\documentclass{aa}
\usepackage{graphicx}
\usepackage{lipsum}
\usepackage{txfonts}
\usepackage{amsmath}
\usepackage{amsfonts}
\usepackage{bm}
\usepackage{empheq}
\usepackage[version=4]{mhchem}
\usepackage[dvipsnames]{xcolor}
\usepackage{subfig}
\usepackage{derivative}

\usepackage{tikz}
\usetikzlibrary{decorations.pathmorphing,patterns,shapes.misc,arrows.meta}

\tikzstyle{box}=[rectangle,very thick, draw = black,rounded corners=3pt,scale=1,minimum width = 2cm, minimum height = 0.8cm,align=center]
\tikzstyle{secondary}=[dash pattern=on 5mm off 2mm]
\tikzstyle{Arrow}=[draw,->,>=Latex,rounded corners=5pt, very thick,double,double distance=1pt]
\tikzstyle{SecondaryArrow}=[draw,->,>=latex,rounded corners=5pt, very thick]

\newcommand\nsigma{\raisebox{.2ex}{$\sigma$}}

\newcommand{\red}[1]{\textcolor{red}{#1}}

\usepackage{natbib,twoopt}
\usepackage[breaklinks=true]{hyperref} %% to avoid \citeads line fills
\bibpunct{(}{)}{;}{a}{}{,}             %% natbib format for A&A and ApJ
\makeatletter
  \newcommandtwoopt{\citeads}[3][][]{\href{http://adsabs.harvard.edu/abs/#3}%
    {\def\hyper@linkstart##1##2{}%
     \let\hyper@linkend\@empty\citealp[#1][#2]{#3}}}
  \newcommandtwoopt{\citepads}[3][][]{\href{http://adsabs.harvard.edu/abs/#3}%
    {\def\hyper@linkstart##1##2{}%
     \let\hyper@linkend\@empty\citep[#1][#2]{#3}}}
  \newcommandtwoopt{\citetads}[3][][]{\href{http://adsabs.harvard.edu/abs/#3}%
    {\def\hyper@linkstart##1##2{}%
     \let\hyper@linkend\@empty\citet[#1][#2]{#3}}}
  \newcommandtwoopt{\citeyearads}[3][][]%
    {\href{http://adsabs.harvard.edu/abs/#3}
    {\def\hyper@linkstart##1##2{}%
     \let\hyper@linkend\@empty\citeyear[#1][#2]{#3}}}
\makeatother
\begin{document}

   \title{The Pedersen and Hall Conductances in the Jovian Polar Regions:\\ New Maps based on a Broadband Electron Energy Distribution}
   \titlerunning{Conductance Maps}

   \subtitle{}

   \author{G. Sicorello\inst{1}
           \and
           D. Grodent\inst{1}
           \and
           B. Bonfond\inst{1}
           \and
           J.-C. Gérard\inst{1}
           \and
           B. Benmahi\inst{1,2}
           \and
           A. Salveter\inst{3}
           \and
           A. Moirano\inst{1,4}
           \and
           L. A. Head\inst{1}
           \and
           J. Vinesse\inst{1}
           \and
           T. Greathouse\inst{5}
           \and
           G. R. Gladstone\inst{5,6}
           \and
           M. Barthélémy\inst{7}
           }

   \institute{Laboratoire de Physique Atmosphérique et Planétaire (LPAP), University of Liège, Liège, Belgium.\\
              \email{guillaume.sicorello@uliege.be}
              \and
              Aix-Marseille Université, CNRS, CNES, Institut Origines, LAM, Marseille, France.
              \and
              Institute of Geophysics and Meteorology, University of Cologne, Cologne, Germany.
              \and
              Institute for Space Astrophysics and Planetology, National Institute for Astrophysics (INAF-IAPS), Rome, Italy.
              \and
              Southwest Research Institute, San Antonio, Texas, USA.
              \and
              Physics and Astronomy Department, University of Texas at San Antonio, Texas, USA.
              \and
              Univ. Grenoble Alpes, CNRS, IPAG, 38000 Grenoble, France.
              }

   \date{
   Received June 30, 2025
   %; accepted
   }

% \abstract{}{}{}{}{} 
% 5 {} token are mandatory
 
  \abstract
   {The ionospheric Pedersen and Hall conductances play an important role in understanding the coupling by which angular momentum, energy and matter are exchanged between the magnetosphere and the ionosphere/thermosphere at Jupiter, modifying the composition and temperature of the planet. In the high latitude regions, the Pedersen and Hall conductances are enhanced by the auroral electron precipitation.}
   {The effect of a broadband precipitating electron energy distribution, similar to the observed electron distributions through particle measurements, on the Pedersen and Hall conductance values is investigated. The new conductance values are compared to the ones obtained from previous studies, notably when considering a mono-energetic distribution.}
   {The broadband precipitating electron energy distribution is modeled by a kappa distribution, which is used as an input in an electron transport model that computes the density vertical profiles of ionospheric ions. Assuming that the conductivity is mostly governed by the density of $\ce{H3+}$ and $\ce{CH5+}$, the vertical profiles of the Pedersen and Hall conductivities are then evaluated from the ion density vertical profiles. Finally, the Pedersen and Hall conductances are computed by integrating the corresponding conductivities over altitude.
   }
   {The Pedersen and Hall conductance values are globally higher when considering a broadband electron energy distribution rather than a mono-energetic distribution. In addition, the use of the direct outputs of an electron transport model rather than the analytical expression presented in \citet{hiraki_parameterization_2008} as well as a change in the electron collision cross-sections considered in these models also have significant impacts on the conductance values. Comparison between our results and the ones deduced from the corotation enforcement theory suggests that either a physical mechanism limits the field-aligned currents or the auroral electrons precipitating in the atmosphere are also accelerated by processes not associated with the field-aligned currents.}
   {}

   % \keywords{Jupiter --
   %           Aurora --
   %           conductivity --
   %           conductance --
   %           Pedersen -- 
   %           Hall --
   %           broadband
   %           }

   \maketitle
%

%%%%%%%%%%%%%%%%%%%%%%%%%%%%%
%%%%%%%%%% SECTION %%%%%%%%%%
%%%%%%%%%%%%%%%%%%%%%%%%%%%%%
\section{Introduction}
\label{section:introduction}

    Jupiter, the largest planet in the solar system, is also known to host the most intense polar aurorae at its poles, observed at many wavelengths. They mainly result from collisions between precipitating energetic electrons with the atmospheric constituents in the upper atmosphere of the planet, \ce{H} and \ce{H2}. The electron precipitation also considerably enhances the ionization of the neutral species, which feeds the Jovian ionosphere and strenghens its ability to carry electric currents. The intensity of these electric current components flowing perpendicular to the magnetic field are proportional to the Pedersen and Hall conductances.

    The Pedersen conductance has an influence on the transfer of angular momentum by large scale field-aligned-current (FAC; currents flowing along the magnetic field lines) loops circulating between the middle magnetosphere and the high latitude ionosphere \citep[e.g.,][]{nichols_magnetosphere-ionosphere_2004,smith_coupled_2009, tao_neutral_2009}. The loops are closed by radially outward flowing currents in the magnetospheric plasma sheet and by equatorward flowing currents in the ionosphere. In the frame of the corotation enforcement theory, these currents transfer momentum from the ionosphere to the magnetosphere through a $\mathbf{J} \times \mathbf{B}$ torque that maintains part of the constant outward-flowing magnetospheric plasma in corotation with the Jovian magnetic field \citep{hill_inertial_1979} up to a distance varying between 20 R\textsubscript{J} and 40 R\textsubscript{J} (1R\textsubscript{J} = 71,492 km), depending on local time. In this framework, the Pedersen conductance has an influence on the transfer of angular momentum, notably modulating the current intensity and the maximum distance at which the plasma corotates \citep{nichols_magnetosphere-ionosphere_2003, nichols_magnetosphere-ionosphere_2004}. The Pedersen and Hall conductances also play a role in the transport of the magnetospheric plasma itself at distances < 20 $R_J$. Notably, the interchange instabilities and plasma convection are highly dependent on the values and spatial distribution of the conductances \citep{wang_simulation_2023}. In addition, the Pedersen and Hall conductances affect the interaction between Alfv\'{e}n waves, propagating along magnetic field lines, and the ionospheric conductive layer, the layer where the conductances become significant. Such an interaction is depicted in the numerical model developed in \citet{lysak_numerical_2023} describing the propagation of Alfv\'{e}n wings between Io and Jupiter. This model notably shows that a higher Pedersen conductance leads to a stronger reflection of the Alfv\'{e}n waves at the ionosphere. Thus, the Jovian ionospheric Pedersen and Hall conductances are key parameters to understand the coupling between the magnetosphere and the ionosphere, i.e., the exchanges of momentum, energy and matter between the two media.
    
    Numerous studies have attempted to constrain these values. \citet{millward_dynamics_2002} used mono-energetic electron beams to study the variation of the Pedersen and Hall conductances with the electron mean energy, number flux and energy flux. They notably used the Jovian Ionospheric Model (JIM; \citeauthor{achilleos_jim_1998} \citeyear{achilleos_jim_1998}) to infer the ion density in the Jovian ionosphere. They found that both conductances are maximum for a mean electron energy value around 60 keV. Based on these results, \citet{nichols_magnetosphere-ionosphere_2004} modeled the precipitating electron energy distribution as a function of the FAC density in the ionosphere and investigated the impact of such a precipitation on the Pedersen conductance and its resulting positive feedback on the FAC. To understand the angular momentum transfer from the thermosphere to the magnetosphere, \citet{smith_coupled_2009} attempted to refine the effective Pedersen conductance by taking into account the difference of velocities (i.e., slippage) between the thermosphere and the deep interior of Jupiter due to collisions with the charged particles \citep{nichols_magnetosphere-ionosphere_2004}. \citet{tao_neutral_2009} also computed the Pedersen conductance in a attempt to evaluate the effect of neutral winds on the magnetosphere-ionosphere-thermosphere (MIT) coupling. The variation of the conductance due to the diurnal change in the solar extreme ultraviolet (EUV) flux has also been investigated in order to evaluate its effect on the FAC. Results from \citet{tao_jovian_2010} suggested that there is a variation of the Pedersen conductance of the order of 0.1 mho between the nightside and the dayside. 

    \citet{hiraki_parameterization_2008} built an analytical expression reproducing the ionization rate vertical profile resulting from the energy degradation in the atmosphere of the precipitating electrons computed with Monte-Carlo techniques. The availability of such analytical formulation allows to simplify the calculations of the ionospheric ion densities and hence the conductance. They notably computed Pedersen conductance values assuming an \ce{H2} atmosphere and a uniform surface magnetic field of $B = 8.4\times 10^{-4}$ T. In their 3D Jupiter Thermospheric General Circulation Model (JTGCM), \citet{bougher_jupiter_2005} used an electron energy distribution made up of three kappa functions to compute the ion density in the ionosphere. They found conductance values of the order of 10 mho, contrasting with previous studies. According to the authors, the difference might arise from the generation of the $\ce{H3+}$ vertical profile peak at a deeper altitude in the JTGCM model compared to the JIM model.
    
    The above studies provided a wide range of ionospheric conductance values that did not easily match with each other. According to \citet{gerard_spatial_2020}, the lack of observational constraints led the authors to postulate values for important parameters that have a major influence on the conductances, such as the electron total energy flux and mean energy as well as the general shape of the energy distribution. The advent of the Juno mission contributed to better constrain these parameters. The presence of hydrocarbons which control the ion density below the methane homopause, the altitude above which the methane molecular diffusion becomes dominant compared with atmospheric turbulence, was not always addressed. It was notably the case for the JIM model used in some of the studies mentioned above \citep[e.g.,][]{millward_dynamics_2002,nichols_magnetosphere-ionosphere_2004}. This issue was resolved to a certain extent in the more recent studies mentioned above, even if the methane homopause level is still partially unconstrained \citep{sinclair_improved_2025}.

    There are however several aspects that are poorly addressed when studying the conductance in the auroral regions. Most of these studies are restricted to the main auroral emission (ME) region, while this auroral emission represents only approximately 1/3 of the total auroral emission. Moreover, they often consider a precipitating electron energy distribution based on inferred ionospheric FAC density \citep{nichols_magnetosphere-ionosphere_2004}, on the assumption that up-going FACs, corresponding to down-going electrons accelerated towards the Jovian ionosphere by field-aligned potentials predicted by the corotation enforcement theory, are the root cause of the auroral ME \citep{cowley_origin_2001}. However, data collected by Juno and the Hubble Space Telescope (HST) over the past years challenge this theory and reveal growing evidence of a more intricate process at work \citep{bonfond_six_2020}. Among the compelling evidence is the discrepancy between theory and observations that arises when we look at the asymmetry in the dawn and dusk brightness of the aurorae. According to corotation enforcement theory, the dawn part of the ME should be brighter than the dusk part due to a stronger magnetic field bend-back on the dawnside \citep{ray_local_2014}. This prediction is not supported by observations which show the opposite trend \citep{bonfond_far-ultraviolet_2015,groulard_dawn-dusk_2024}. The observations also show that not only the electrons are accelerated bidirectionally by stochastic mechanisms in addition to field-aligned potentials \citep{mauk_diverse_2018,sulaiman_jupiters_2022} but these mechanisms are the main acceleration processes at work \citep{salveter_jovian_2022}.

    In their investigation of the MIT coupling parameters, \citet{wang_preliminary_2021}, and at a later stage \citet{al_saati_magnetosphereionospherethermosphere_2022}, avoided part of the problem by using electron energy distributions that are directly drawn from the data acquired by the JADE and JEDI instruments onboard Juno. However, the electron energy distributions measured by the instruments may differ from the precipitating population if the acceleration region is located mostly between the spacecraft and the planet \citep{gerard_contemporaneous_2019}. \citet{gerard_spatial_2020} computed Pedersen conductance maps of the overall auroral region using Juno-UVS spectral images. From the auroral brightness and a methane color ratio (CR), which is the ratio between two specific UV spectral bands absorbed and unabsorbed by hydrocarbons, they retrieved the total energy flux and the mean energy of the precipitating electrons respectively \citep{gustin_characteristics_2016}. By assuming a mono-energetic precipitating electron energy distribution, they derived the ion densities in the ionosphere using the analytic expression developed in \citet{hiraki_parameterization_2008}. Nevertheless, in-situ Juno observations showed that the shape of the precipitating electron energy distribution is mostly broadband \citep{mauk_diverse_2018, salveter_jovian_2022}. The choice of a realistic electron energy distribution may be relevant for the computation of the conductance as it was already shown at low energy for the aurorae on Earth \citep{germany_determination_1994}.

    With the use of Juno data, we investigated the effect of a broadband precipitating electron energy distribution on the ionospheric Pedersen and Hall conductances. One of the added values of this work is that the same electron transport model and distribution model were applied to retrieve the ionospheric ion densities as well as the precipitating electron mean energy. A description of the instruments and data used in this study is presented in section \ref{section:instruments}. The applied method is described in section \ref{section:method}. In section \ref{section:results}, the generic differences between the conductances computed with a broadband and a mono-energetic distribution are identified and the conductance values for a few Juno perijoves (PJs) are compared to already existing values. Finally, the conclusions of the study are presented in section \ref{section:conclusions}.

%%%%%%%%%%%%%%%%%%%%%%%%%%%%%
%%%%%%%%%% SECTION %%%%%%%%%%
%%%%%%%%%%%%%%%%%%%%%%%%%%%%%
\section{Instruments}
\label{section:instruments}

    This study is based on data collected by two instruments onboard the solar-powered Juno spinning spacecraft \citep{bolton_juno_2017}. Launched in August 2011 for a five year journey, Juno was inserted in orbit around Jupiter on 5th July 2016. It follows an elliptical polar orbit, allowing it to avoid most of the radiation belt regions and perform close flybys of Jupiter. The orbital period of Juno around the planet was initially of 53.5 days but has been gradually reduced since the start of the extended mission campaign (PJ34+) down to a value of about 32 days for the most recent PJs. The spacecraft spins at a rate of 30 seconds.

    \subsection{Juno-JEDI}
    The Jupiter Energetic-Particles Detector (JEDI) \citep{mauk_jupiter_2017} is a set of three sensors measuring the flux of ions (\ce{H+}, \ce{He+}, \ce{O+} and \ce{S+}) and electrons crossing the path of Juno, as a function of their energy. JEDI detects electrons in the energy range 25-1,000 keV. The JEDI sensors geometry allows to instantaneously collect the electrons for almost any pitch angle, which is the angle between the electron velocity and the local magnetic field.

    \subsection{Juno-UVS}
    The Ultraviolet Imager/Spectrograph (UVS) instrument \citep{gladstone_ultraviolet_2014} collects EUV and far ultraviolet (FUV) photons in the 68-210 nm range, covering part of the Lyman and Werner bands of the \ce{H2} spectrum and the Lyman-$\alpha$ line of \ce{H}. When entering the instrument, the UV light passes through an aperture oriented along the spacecraft spin axis, defining the instrument field of view (FOV). It is made up of three rectangular slits, two external, wide slits and a central, narrow slit in the form of a ``dog bone''. When projected onto the sky, the two wide slits have a FOV of $2.55^{\circ} \times 0.2^{\circ}$. The narrow slit has a FOV of $2^{\circ} \times 0.025^{\circ}$ and an increased spectral resolution. After the slits, a grating scatters the UV light onto the detector. A scan mirror is placed at the entrance of the spectrograph that permits the shifting of the FOV by $\pm 30^{\circ}$ away from the Juno spin plane. Close to Jupiter, UVS acquires high spatial resolution spectra of the auroral UV emission in the polar regions.

%%%%%%%%%%%%%%%%%%%%%%%%%%%%%
%%%%%%%%%% SECTION %%%%%%%%%%
%%%%%%%%%%%%%%%%%%%%%%%%%%%%%
\section{Method}
\label{section:method}

    The ionospheric electric conductance is essentially dependent on the density vertical profiles of the different ions present in the ionosphere. Since the main goal of this study was to assess the effect of a more realistic precipitating electron energy distribution on the Pedersen and Hall conductances, we only considered ions generated by the auroral electron precipitation. Photo-ionization of the atmospheric components by solar EUV radiation was neglected. Indeed, this effect contributes to increase the Pedersen and Hall conductances by a maximum of about 0.01 mho \citep{tao_jovian_2010, clement_ionospheric_2025}, which is 10 to 100 times smaller than the conductance values computed in the auroral regions. We also did not take into account the presence of ions produced by meteoric impacts. Nevertheless, for a sufficient meteoric influx, this process is expected to have a non-negligible effect at Jupiter, contributing to especially increase the Hall conductance \citep{nakamura_effect_2022, clement_ionospheric_2025}.
    
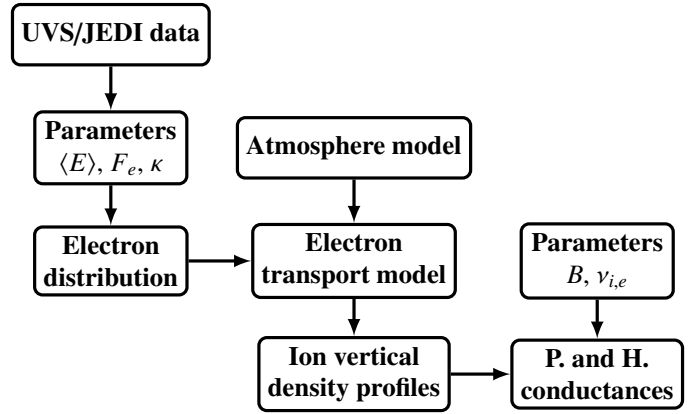
\begin{figure}[h] % Two-columns figure
    \centering
    \captionsetup[subfigure]{width=0.9\linewidth}
    	\subfloat{
        	\begin{tikzpicture}
            \node[box] (A)
                at (-2.2,2) {\textbf{Electron} \\[1pt] \textbf{distribution}};
            \node[box] (B)
                at (1,0.5) {\textbf{Ion vertical} \\[1pt] \textbf{density profiles}};
            \node[box] (C)
                at (4.2,0.5) {\textbf{P. and H.} \\[1pt] \textbf{conductances}};
            \node[box] (AB) 
                at (-2.2,5) {\textbf{UVS/JEDI data}};
            \node[box] (AA) 
                at (-2.2,3.5) {\textbf{Parameters} \\[1pt] $\langle E \rangle$, $F_e$, $\kappa$};
            \node[box] (BA) 
                at (1,3.5) {\textbf{Atmosphere model}};
            \node[box] (BB)
                at (1,2) {\textbf{Electron} \\[1pt] \textbf{transport model}};
            \node[box] (CA) 
                at (4.2,2) {\textbf{Parameters} \\[1pt] $B$, $\nu_{i,e}$};
            \draw[SecondaryArrow] (A) -- (BB);
            \draw[SecondaryArrow] (B) -- (C);
            \draw[SecondaryArrow] (AA) -- (A);
            \draw[SecondaryArrow] (AB) -- (AA);
            \draw[SecondaryArrow] (BA) -- (BB);
            \draw[SecondaryArrow] (BB) -- (B);
            \draw[SecondaryArrow] (CA) -- (C);
            \end{tikzpicture}
        }
    \caption{Schematic representation of the different steps implemented to compute the Pedersen (P.) and Hall (H.) conductances. $\langle E \rangle$ and $F_e$ are the electron mean energy and energy flux, respectively, $\kappa$ is related to the high energy slope of the broadband electron energy distribution, $B$ is the magnetic field intensity and $\nu_{i,e}$ are the ion/electron collision frequency with the neutrals.}
    \label{fig:method}
    \end{figure}

    In Fig. \ref{fig:method} is presented the general flow of the method. The UVS and JEDI data were used to extract the key parameters to model the broadband precipitating electron energy distribution. After setting an atmospheric model, this distribution was input to an electron transport model to obtain the ion density vertical profiles, from which the conductivities and conductances were deduced.

    \subsection{Precipitating auroral electron energy distribution}
    As mentioned in the introduction, the use of the electron energy distribution measured by JEDI might affect the computation of the conductances, since, in some cases, Juno flew above the region where the precipitating electrons were accelerated, expected to extend between about 0.5 and 1.5 R\textsubscript{J} above Jupiter \citep{gerard_contemporaneous_2019}. Instead, we chose to represent the broadband energy spectrum of the precipitating auroral electron population with a kappa distribution model \citep{livadiotis_kappa_2017}. Its two main parameters, the electron mean energy and total energy flux, are deduced from the chosen CR (see equation \eqref{equation:CR}) and the auroral brightness, respectively, which do not depend on the spacecraft altitude.
    
    At low energy, the kappa distribution is similar to a Maxwellian distribution. However, the kappa distribution has the advantage of reproducing the power-low decrease observed at high energy in the JEDI data \citep{mauk_juno_2017}, characterizing collisionless plasma out of thermal equilibrium. A kappa distribution were used to study the particle environment around several celestial bodies. \citet{coumans_electron_2002} employed it to model the proton population generating part of the FUV aurorae at Earth. The energy spectra of electrons responsible for the diffuse aurorae at Earth and Ganymede have also been modeled with a kappa distribution \citep{singhal_diffuse_2016,tripathi_generation_2017,tripathi_diffuse_2023}. At Saturn, the influence of the electron precipitation on the ionosphere was investigated using a kappa distribution \citep{galand_response_2011}. At Jupiter, the energies of the electron population have already been modeled by a kappa distribution to investigate the heating of the Jovian atmosphere resulting from the electron precipitation \citep{grodent_self-consistent_2001}, the Io footprint \citep{bonfond_io_2009} as well as the ME vertical emission profile \citep{bonfond_far-ultraviolet_2015}. 
    
    We used the expression of the kappa distribution from \citet{coumans_electron_2002},
    \begin{equation}
    F(E) = \frac{4\,F_e}{\pi} \, \frac{\kappa \,(\kappa-1)}{(\kappa-2)^2} \, \frac{E}{\langle E \rangle} \, \frac{\langle E \rangle^{\, \kappa-1}}{\left(\frac{2 \,E}{\kappa\,- \,2} + \langle E \rangle\right)^{\, \kappa+1}}.
    \label{equation:kappa}
    \end{equation}
    Here, $E$ [keV] is the electron energy, $\langle E \rangle$ [keV] is the electron mean energy and $F_e$ [mW.m\textsuperscript{-2}] is the total energy flux. The additional parameter $\kappa$ controls the high energy slope. The above function $F(E)$ can also be expressed in terms of the characteristic energy $E_0$ of the distribution. However, we preferred to use the electron mean energy $\langle E \rangle$, which is more meaningful when comparing different distributions \citep{robinson_calculating_1987}. The relation between the mean and characteristic energies is given by
    \begin{empheq}{align}
    \langle E\rangle = 2 \, E_{0} \, \frac{\kappa}{\kappa-2}.
    \label{equation:E-E}
    \end{empheq}

    \subsubsection{Retrieving $F_e$ and $\langle E \rangle$}
    When Juno is at the shortest distance from Jupiter, UVS scans the Jovian aurorae once for each rotation of the spacecraft. The resulting image from each spin only covers a part of the aurorae. In this study, we used complete coverages of the northern and southern aurorae, obtained by averaging the spectral brightness over a stack of 100 spins \citep{bonfond_morphology_2017}. The electron total energy flux $F_e$ was derived from the total \ce{H2} auroral UV brightness, which was obtained by integrating the spectral brightness, measured by UVS, in the wavelength range 145-165 nm and then scaled by a factor of 4.4 in order to obtain the total equivalent \ce{H2} Lyman and Werner bands UV brightness \citep{bonfond_morphology_2017}. Previous studies found that a total brightness of 10 kR roughly corresponds to $F_e$ = 1 mW.m\textsuperscript{-2} \citep{gerard_model_1982, waite_electron_1983, grodent_self-consistent_2001, nichols_relation_2022}.
    
    The electron mean energy $\langle E \rangle$ was retrieved from the CR deduced from UVS observations and computed as
    \begin{empheq}{align}
    CR = \frac{\int_{155}^{162} I(\lambda) \, \text{d}\lambda}
              {\int_{125}^{130} I(\lambda) \, \text{d}\lambda},
    \label{equation:CR}
    \end{empheq}
    where $I(\lambda)$ is the spectral brightness \citep{gerard_contemporaneous_2019}.
    The numerator and the denominator represent parts of the UV spectrum unabsorbed and absorbed by the methane, respectively. This CR is different from the one used when dealing with spectral images from the Space Telescope Imaging Spectrograph (STIS) onboard HST, for which the absorbed part of the spectrum in the denominator starts at 123 nm instead of 125 nm (e.g., \citet{gustin_characteristics_2016}). The spectral range has been adapted to account for the degradation of the UVS detector sensitivity around the Lyman-$\alpha$ line (about 121.6 nm), especially for the wide slits \citep{hue_-flight_2019}. The minimum value of this CR, which is the value without any absorption, is 1.8 \citep{benmahi_energy_2024}.

    We used the $CR-\langle E \rangle$ relationship presented in \citet{benmahi_energy_2024} and updated in \citet{benmahi_auroral_2024}. This relationship is based on a UV emission model of $\ce{H2}$ coupled to the TransPlanet electron transport model \citep{lilensten_ionization_1989, benmahi_etude_2022}. They developed a phenomenological law, assuming a kappa or a mono-energetic distribution for the auroral precipitating electron energy, that gives the CR as a function of the characteristic energy $E_0$ of the distribution and the emission angle $\theta$, which is the angle between the local vertical and the line that joins Juno to the center of the planet,
    \begin{empheq}{align*}
    CR\Big(&E_0,\theta\Big) = \\  
    & A \, C \, \left[ \tanh \left(\frac{E_0 - E_{c}}{B}\right) + 1 \right] \, \ln^{\,\beta}\left( \left(  \frac{E_0}{D} \right)^{\alpha} + e  \right)
    \, \left[  1 + \delta \, \sin^{\gamma}(\theta) \right].
    \end{empheq}
    Here, $A$ is the minimum amplitude of the CR, $E_c$ is a threshold energy and $B$, $C$, $D$, $\alpha$, and $\beta$ are the fit parameters. For a fixed emission angle, the characteristic energy was retrieved from the CR by using the dichotomy method, a root-finding method. The electron mean energy was then deduced from the characteristic energy depending on the chosen distribution. For a mono-energetic distribution, the relation is simply $\langle E \rangle = E_0$. For the kappa distribution, the link is given by equation \eqref{equation:E-E} and depends on the value of the $\kappa$ parameter. It should be kept in mind that, as mentioned in \citet{gustin_characteristics_2016}, the $CR-\langle E \rangle$ relationship is highly dependent on the temperature profile, the $\ce{CH4}$ methane homopause altitude and the electron energy distribution.

    \subsubsection{Retrieving $\kappa$}
\begin{figure}[h] % One-column figure
        \centering
        \includegraphics[scale=0.6]{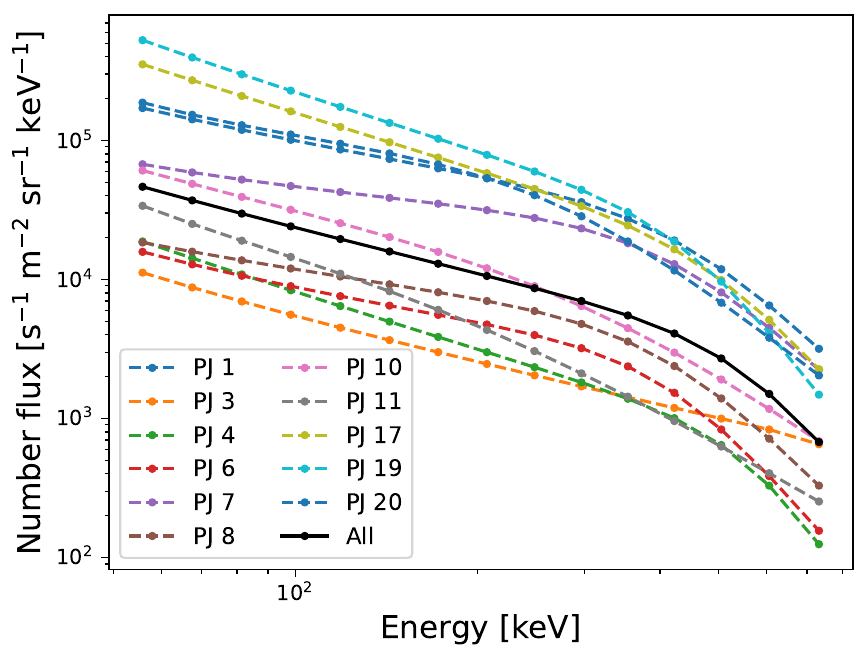}
        \caption{Broadband energy distributions of electrons measured in the loss cone by JEDI for several PJs when Juno was magnetically connected to the ME region. The number flux values are actually median values over a PJ.}
        \label{fig:jedi_median}
    \end{figure}

    A Markov-Chain Monte Carlo method was applied to sample the $\kappa$ parameter providing 11 sets of median JEDI electron energy distribution measurements magnetically linked to the ME crossing. The JEDI data were part of the dataset used in \citet{salveter_jovian_2022} and are plotted in Fig. \ref{fig:jedi_median}. They were corrected from the minimum ionizing bump usually seen around 150 keV and corresponding to an excess of energy left in the detector by high-energy (> 800 keV) electrons fully penetrating the instrument \citep{mauk_diverse_2018}. A value of $\kappa$ = 2.5 was retrieved that maximizes the likelihood with a computed uncertainty of $\delta \kappa$ = 0.48. It is worth noting that this $\kappa$ value may change when considering other auroral regions than the ME region. This possibility will be addressed in a future study.

    \subsection{Ion density vertical profile}
    \begin{figure}[h] % One-column figure
        \centering
        \includegraphics[scale=0.8]{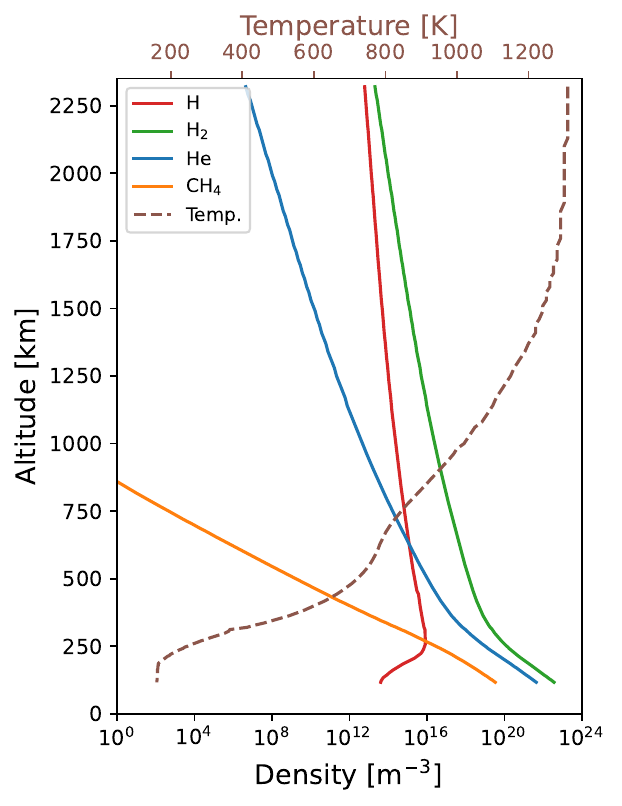}
        \caption{Density and temperature profiles of the neutral atmosphere.}
        \label{fig:dty_neutrals}
    \end{figure}
    
    We now briefly describe the main steps and assumptions applied in this study to obtain the ion densities in the Jovian atmosphere. A complete description of the computations is provided in Appendix \ref{appendix:a}. As explained in the beginning of section \ref{section:method}, we only considered ions generated by collisions of auroral precipitating electrons with the principal Jovian atmospheric constituents, which were assumed to be atomic hydrogen (\ce{H}), dihydrogen (\ce{H2}), helium (\ce{He}) and methane (\ce{CH4}). Their densities were taken from the atmosphere model of \citet{grodent_self-consistent_2001} and plotted in Fig. \ref{fig:dty_neutrals}. The zero reference altitude was defined as the altitude corresponding to the 1 bar pressure level.
    
    We considered the system at steady-state. The contribution of $\ce{H+}$ to the conductance was assumed to be small as this ion is dominant above 2000 km \citep{tao_uv_2011}, way above the conductive layer. The \ce{He+} ions were also neglected, as their relative abundance is low \citep{hiraki_parameterization_2008}. Below the altitude region where $\ce{H+}$ is important and above the methane homopause, the dominant ion was then assumed to be $\ce{H3+}$\citep{millward_dynamics_2002}. Its production and loss are described by
    \begin{empheq}{align}
    \ce{H2+} + \ce{H2} & \ce{->} \ce{H3+} + \ce{H},
    \label{equation:chemical_reaction_h3+}\\
    \ce{H3+} + \ce{e-} & \ce{->} \ce{H2} + \ce{H}.
    \end{empheq}
    
    Close to and below the homopause, \ce{H3+} reacts with \ce{CH4} to produce hydrocarbon ions, which become the main contributors to the conductance at these altitudes. According to \citet{perry_chemistry_1999}, the main hydrocarbon ions are \ce{CH5+} and $\ce{C3H_n+}$-class ions. The production and loss of $\ce{CH5+}$ is given by
    \begin{empheq}{align}
    \ce{H3+} + \ce{CH4} &\ce{->} \ce{CH5+} + \ce{H2},
    \\
    \ce{CH5+} + \ce{e-} &\ce{->} \text{products}.
    \end{empheq}
    The $\ce{C3H_n+}$ ions result from a chain of ion-molecule reactions of the $\ce{C}$ and $\ce{C2}$ hydrocarbon molecules. As in \citet{wang_preliminary_2021} and \citet{clement_ionospheric_2025}, we decided to limit our study to the $\ce{CH5+}$ ions. This simplification allowed us to analytically solve the reaction equations in order to obtain the ion densities. As the $\ce{C3H_n+}$ ion layer is located deeper in the atmosphere, this assumption mainly impact the conductance generated by higher-energy electrons, as they penetrate deeper into the atmosphere.

    \subsubsection{Electron transport model: TransPlanet}
    The $\ce{H2+}$ production rate, from which the $\ce{H3+}$ ion density was derived, was retrieved using the one dimensional TransPlanet multi-stream electron transport model \citep{lilensten_ionization_1989, benmahi_etude_2022}.
    
    TransPlanet is designed to compute the degradation of energy of the auroral electrons precipitating in the Jovian atmosphere. In practice, after setting an initial electron energy distribution, the model computes the evolution of the energy distribution of the auroral electrons as they travel deeper into the atmosphere and interact with their surroundings through various elastic and inelastic collisions. The energy exchanged between the electrons and the neutral atmospheric constituents is assessed by solving the dissipative Boltzmann radiative transfer equation, detailed in \citet{benmahi_etude_2022}. The values of the semi-relativistic cross-sections relative to the accounted elastic and inelastic collisions between the auroral particles and the atmosphere, come from the Atomic and Molecular Cross section for Ionization and Aurora Database (ATMOCIAD) \citep{gronoff_atmociad_2011}.

    In this study, the initial precipitating electron energy distribution input in TransPlanet was either broadband or mono-energetic. A lower and upper energy cutoffs were applied in the case of an initial broadband energy distribution. Indeed, electrons with a kinetic energy less than 1 eV were considered as thermalized and hence irrelevant. In addition, it was assumed that the fluxes and collision cross-sections of the electrons with a kinetic energy higher than $10^6$ eV were too small to efficiently produce conductances. Therefore, these energetic electrons were set to 0.
    
    \subsection{Pedersen and Hall conductances}
    The Pedersen and Hall conductivities are defined as
    \begin{empheq}{align}
        \nsigma_{\!P} &= \frac{e}{B}
        \left(- \, n_e\,
        \frac{\omega_e \, \nu_{en}}{\nu^2_{en} + \omega^2_e}
        +
        \sum_i n_i \, \frac{\omega_i \, \nu_{in}}{\nu^2_{in} + \omega^2_i}
        \right),
        \label{equation:ctyp}\\
        \nsigma_{\!H} &= \frac{e}{B}
        \left(n_e \,
        \frac{\omega^2_e}{\nu^2_{en} + \omega^2_e}
        -
        \sum_i n_i \, \frac{\omega^2_i}{\nu^2_{in} + \omega^2_i}
        \right).
        \label{equation:ctyh}
    \end{empheq}
    Here, $e$ is the electron charge, $\nu_{en}$ $(\nu_{in})$ is the electron (ion)-neutral collision frequency, $m_e$ $(m_i)$ is the electron (ion) mass, $\omega_e$ $(\omega_i)$ is the electron (ion) gyrofrequency and $B$ is the magnetic field intensity. For the conductance maps, the value of the local Jovian ionospheric magnetic field is obtained from the most recent JRM33 internal magnetic field model \citep{connerney_new_2022,wilson_internal_2023}. Here, the electron and ion gyrofrequencies are defined as $\omega_{e} = - \frac{e\, B}{m_e}$ and $\omega_{i} = \frac{e \, B}{m_i}$.
    The electron-neutral and ion-neutral collision frequencies are given by
    \begin{empheq}{align*}
        \nu_{en} &= 4.5\times 10^{-9} n_n\, (1-1.35\times 10^{-4} \,T_e) \, T_e^{\frac{1}{2}}\\
        \nu_{in} &= 2.6\times 10^{-9} n_n \sqrt{\frac{\alpha_0}{\mu_{in}}},
    \end{empheq}
    where $n_n$ [cm\textsuperscript{-3}] and $\alpha_0$ [$10^{-24}$ cm\textsuperscript{3}] are the number density and the polarisability of the considered neutral, $\mu_{in}$ is the reduced mass between the considered ion and neutral and $T_e$ [K] is the electron temperature \citep{banks_aeronomy_1973,schunk_ionospheres_2009}.
    Here, we considered \ce{H2} as the dominant neutral specie ($\alpha_{0,\ce{H2}} = 0.82$). Typically, the interactions between the suprathermal auroral electrons with the thermalised ionospheric electrons contribute to increase the electron temperature, resulting in a difference between the electron and neutral temperatures. However, results from a thermosphere-ionosphere model applied at Saturn \citep{galand_response_2011} showed that, at low altitude, the neutral density is sufficiently high to effectively cool the electrons and maintain the electron temperature equal to the neutral temperature. By transposing these results to Jupiter, we found that this assumption is valid up to 800 km which is above the altitude range of the conductive layer.

    The Pedersen and Hall conductances are the corresponding conductivity integrated over the altitude
    \begin{empheq}{align}
        \Sigma_P &= \int \nsigma_{\!P} \, \text{d}z,\\
        \Sigma_H &= \int \nsigma_{\!H} \, \text{d}z.
    \end{empheq}
    Since the production rate $q$ is directly proportional to the total energy flux $F_e$ (see Appendix \ref{appendix:a}), the conductances increase as the square root of the electron total energy flux
    \begin{empheq}{align*}
        \Sigma_{P,H} 
        \propto \left[\ce{H3+}\right] \propto \sqrt{q} \propto \sqrt{F_e}.
    \end{empheq}

    The slippage of the thermosphere is not included in those definitions. As we ultimately used conductance ratios to investigate the variations resulting from different energy distributions and since this effect is independent from the chosen precipitating electron energy distribution, we did not include it in the conductances presented in this study. Nevertheless, effective Pedersen conductance values, which take the slippage of the thermosphere into account, were computed at the end of section \ref{section:results} to compare our results to previous work.

    \subsection{Maps}
\begin{figure} % One-column figure
    \centering
    \captionsetup[subfigure]{width=1\textwidth}
        \subfloat[ UV brightness.]{\includegraphics[width=0.5\linewidth]{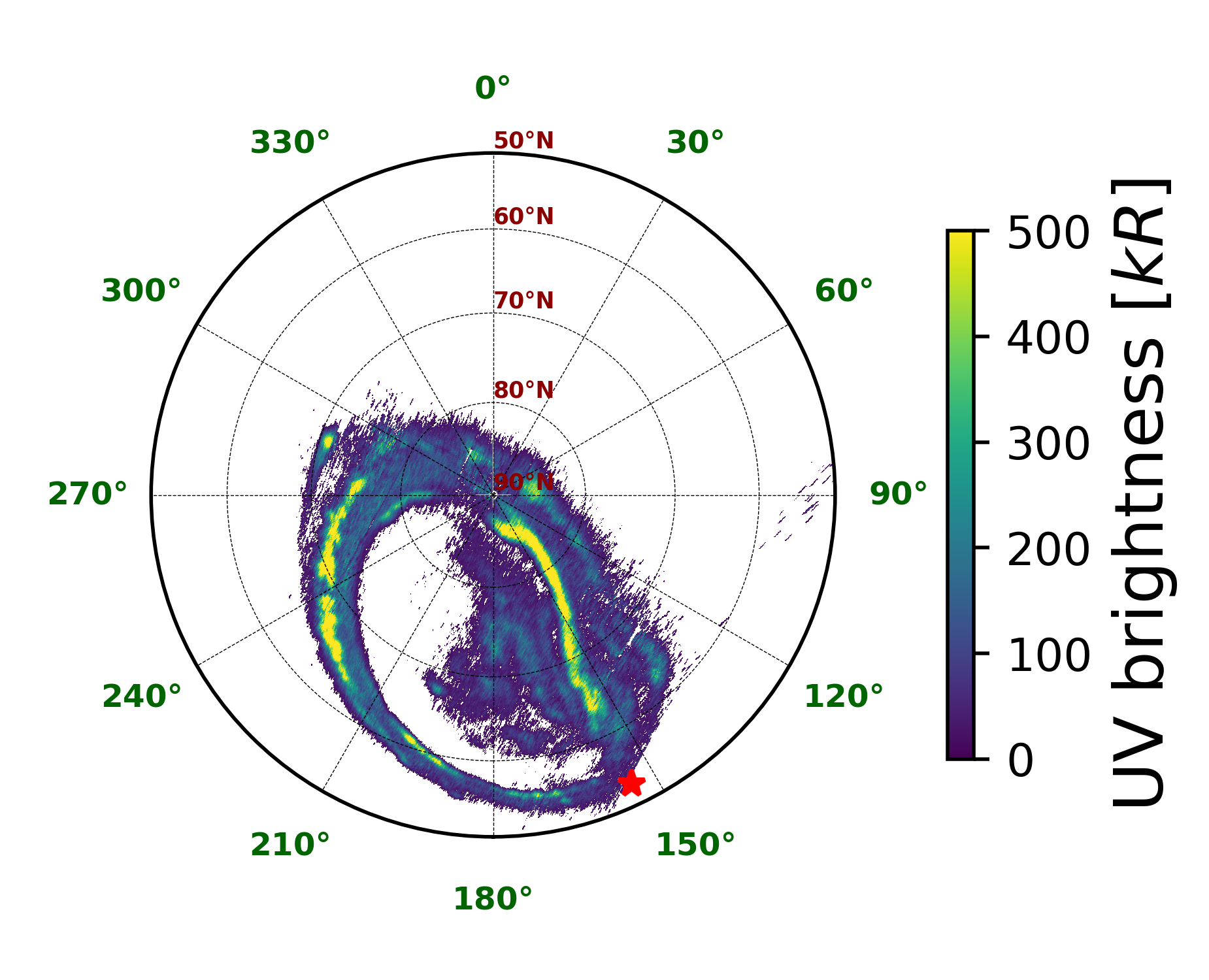}
        \label{fig:maps_inputs_1}
        }
        \subfloat[ UV brightness - ME only.]{\includegraphics[width=0.5\linewidth]{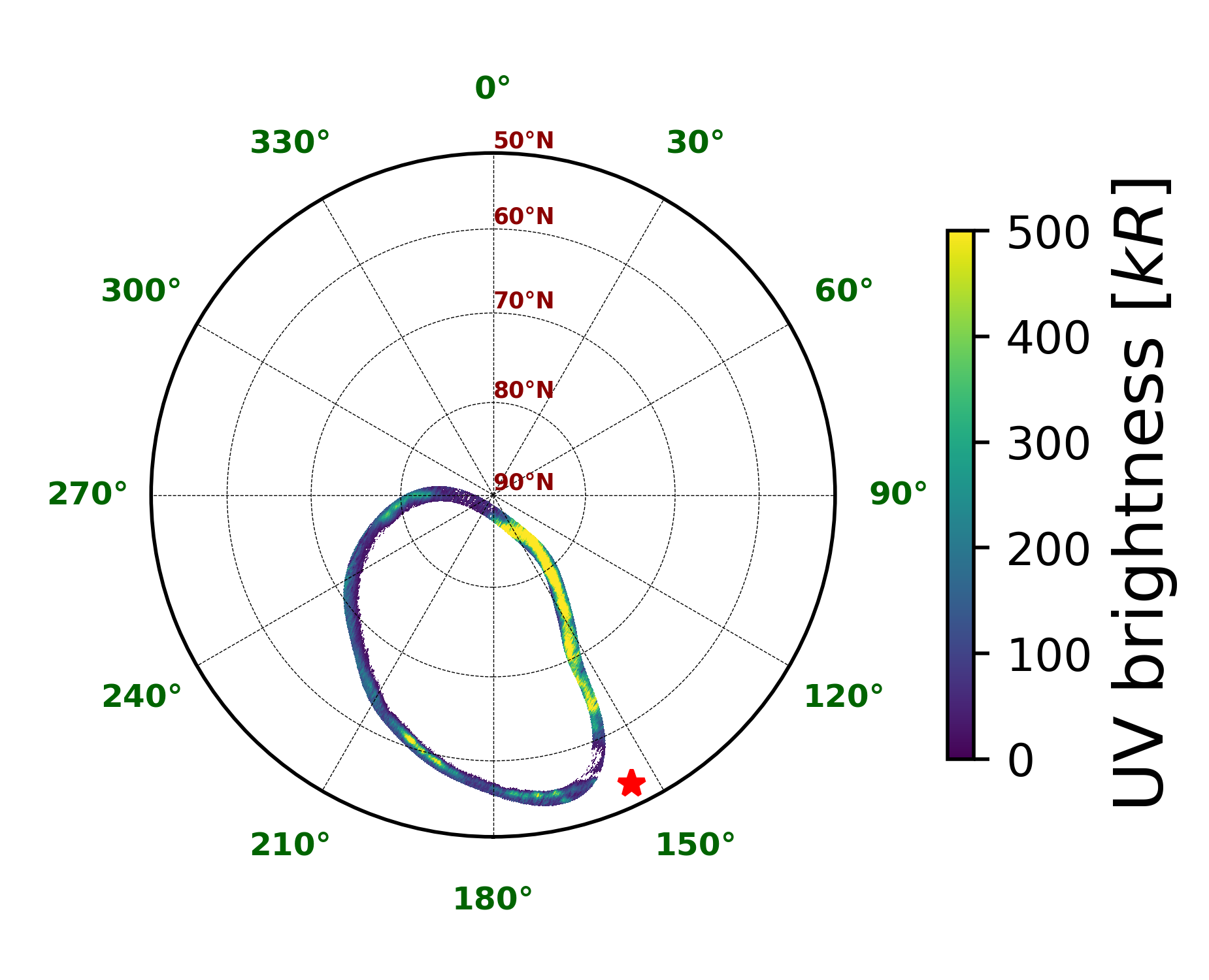}
        \label{fig:maps_inputs_2}
        }
    \caption{
    (a) UV brightness map for PJ1 north. (b) UV brightness map for PJ1 north after we applied a mask that hides the regions poleward and equatorward of the ME. On each map, the red star represents the subsolar longitude.
    }
    \label{fig:maps_inputs}
    \end{figure}

    For each PJ, the physical quantities used in this study (e.g., the CR, the total auroral UV brightness of $\ce{H2}$, the Pedersen and Hall conductances,...) were organized in the form of longitude-latitude maps, spanning from $0^{\circ}$ to $360^{\circ}$ in longitude and from $-90^{\circ}$ to $90^{\circ}$ in latitude with a $0.1^{\circ}$ interval ($\approx$ 100 km). For convenience, the maps presented in this study are polar-projected. For the north pole, the projection represents the view an observer would have from above the pole. For the south pole, the maps are built to represent the view an observer would have through the planet from above the north pole. The $0^{\circ}$ longitude in System-III is set on top of the map and the longitude increases clockwise. As an illustration, the map of the total UV brightness of $\ce{H2}$ for PJ1 is plotted in Fig. \ref{fig:maps_inputs_1}.
    
    For a given auroral feature, most of the emission is assumed to come from the emission peak altitude. Because Juno is not always right above the auroral emission, this peak altitude is needed to infer the location of the aurora on Jupiter. The peak altitude depends on the type of auroral emission. For example, most of the emission coming from the ME region is assumed to take place between 200 km and 400 km \citep{vasavada_jupiters_1999,bonfond_far-ultraviolet_2015,benmahi_auroral_2024}. Instead, the Io footprint emission rather peaks at a higher altitude, around 900 km \citep{bonfond_io_2009}. Here, we assumed the peak emission altitude to be at 400 km above the 1 bar level. As explained in \citet{head_effect_2024}, this assumption introduces an error that could shift the actual position of the aurorae up to about 100 km, depending on the position of Juno with respect to the aurorae. This error is comparable to the map resolution.

    In this study, we also focused on the ME region alone. To retrieve information from this region only, we applied masks on the polar maps, that hide the auroral regions equatorward and poleward of the ME \citep{groulard_dawn-dusk_2024}. As an illustration, we applied a mask on the total UV brightness map of PJ1 north and plotted the result in Fig. \ref{fig:maps_inputs_2}.  

    Concerning the CR maps, the value on each pixel is the mean CR computed in a circle of 500 km. To avoid low SNR, the CR was only computed with the spectral brightness coming from UVS wide slits and if the total auroral \ce{H2} UV brightness was higher than 2 kR \citep{bonfond_morphology_2017}.

%%%%%%%%%%%%%%%%%%%%%%%%%%%%%
%%%%%%%%%% SECTION %%%%%%%%%%
%%%%%%%%%%%%%%%%%%%%%%%%%%%%%
\section{Results and discussion}
\label{section:results}
    
    In this section, we present a comparison between the Pedersen and Hall conductivities and conductances computed using a kappa and a mono-energetic precipitating electron energy distributions. In addition, we present and discuss the conductance maps for several Juno's PJs. Hereafter, the conductivity/conductance computed using a kappa (mono-energetic) distribution is called the kappa (mono-energetic) conductivity/conductance.

    \subsection{Conductivity vertical profiles}
    \begin{figure*} % Two-columns figure
    \centering
    \captionsetup[subfigure]{width=1\textwidth}
        % Panel 1-4
        \subfloat[ $\langle E \rangle$ = 10 keV.]{\includegraphics[width=0.25\linewidth]{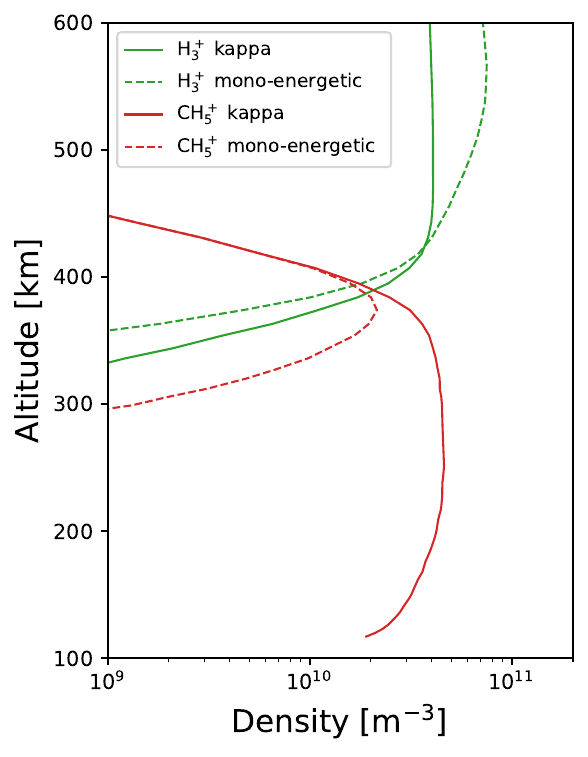}
        \label{fig:panel_a}
        }
        \subfloat[ $\langle E \rangle$ = 30 keV.]{\includegraphics[width=0.25\linewidth]{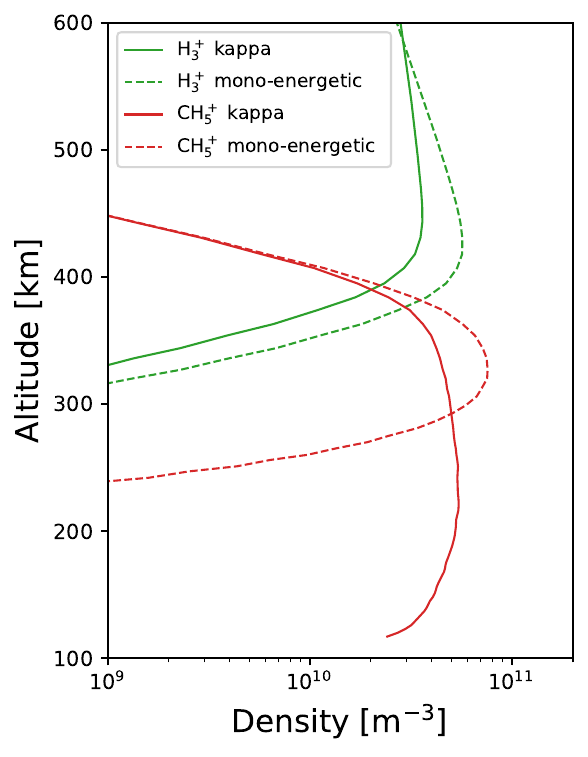}
        \label{fig:panel_b}
        }
        \subfloat[ $\langle E \rangle$ = 100 keV.]{\includegraphics[width=0.25\linewidth]{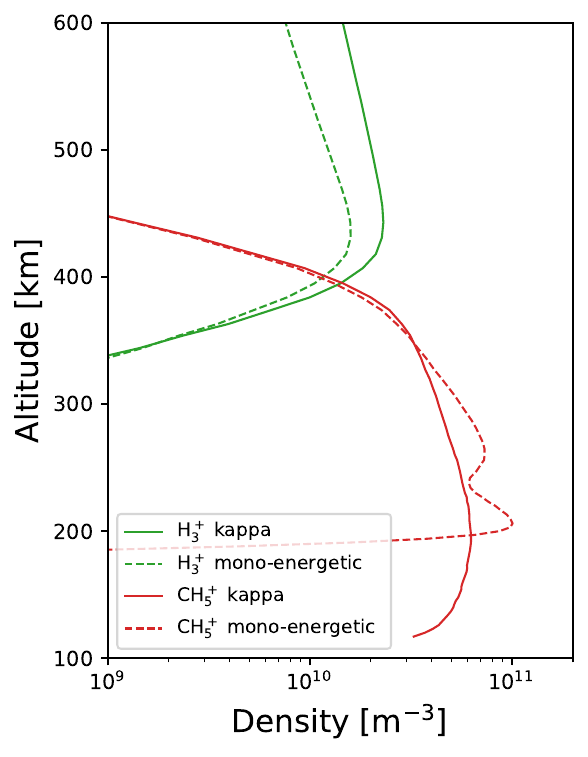}
        \label{fig:panel_c}
        }
        \subfloat[ $\langle E \rangle$ = 500 keV.]{\includegraphics[width=0.25\linewidth]{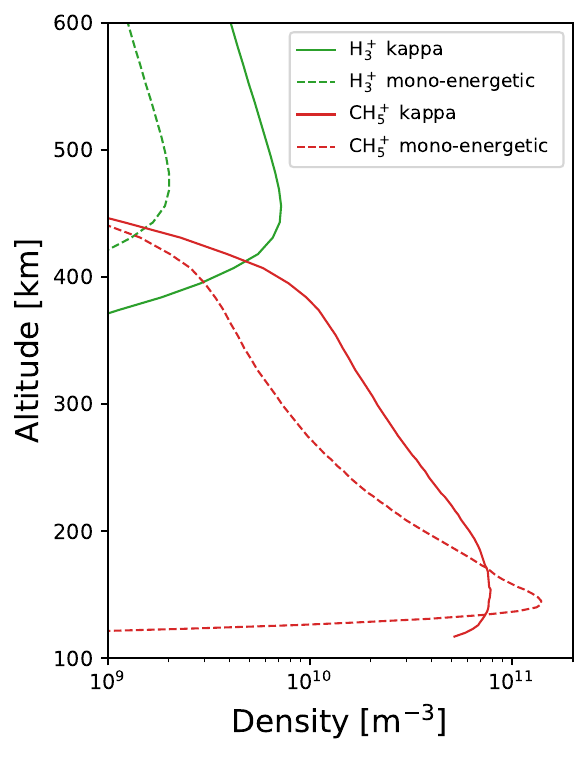}
        \label{fig:panel_d}
        }

        % Panel 5-8
        \subfloat[ $\langle E \rangle$ = 10 keV.]{\includegraphics[width=0.25\linewidth]{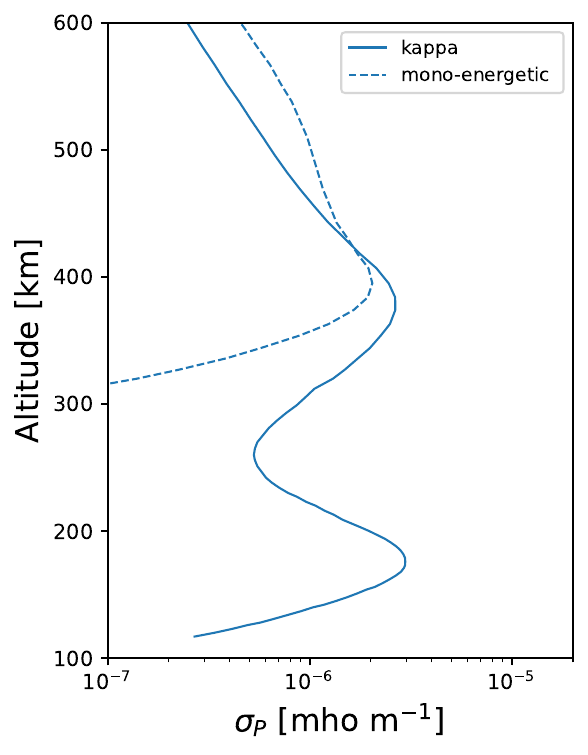}
        \label{fig:panel_e}
        }
        \subfloat[ $\langle E \rangle$ = 30 keV.]{\includegraphics[width=0.25\linewidth]{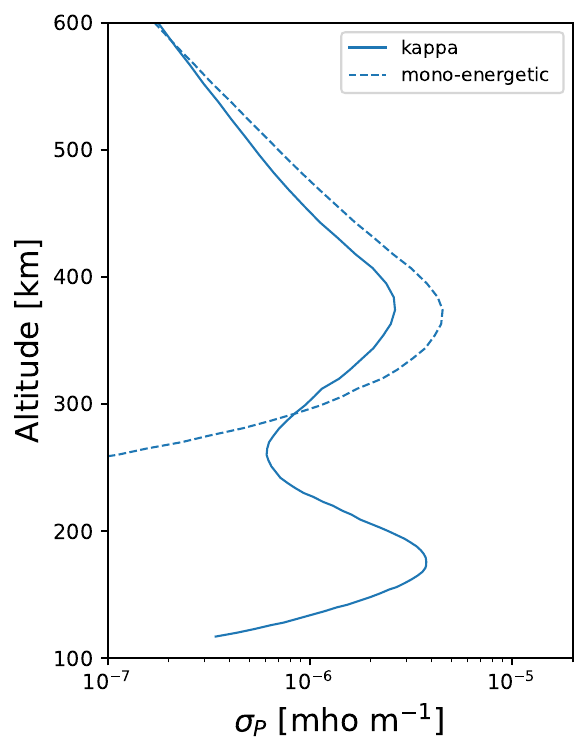}
        \label{fig:panel_f}
        }
        \subfloat[ $\langle E \rangle$ = 100 keV.]{\includegraphics[width=0.25\linewidth]{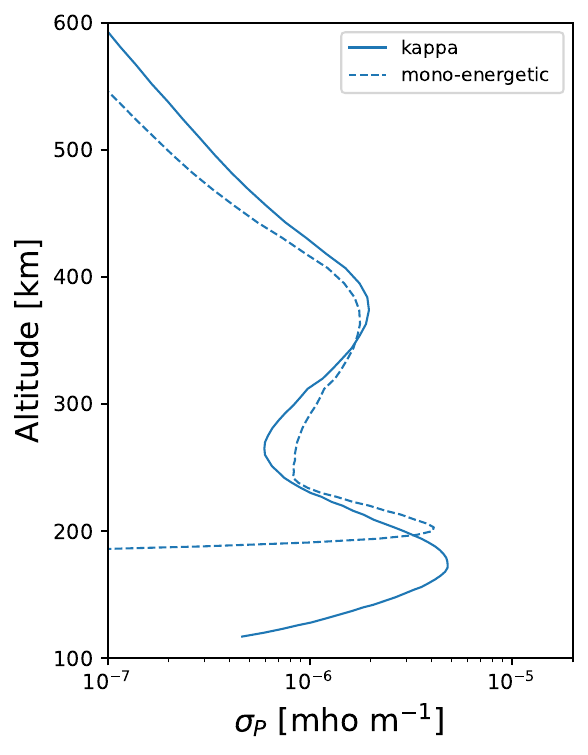}
        \label{fig:panel_g}
        }
        \subfloat[ $\langle E \rangle$ = 500 keV.]{\includegraphics[width=0.25\linewidth]{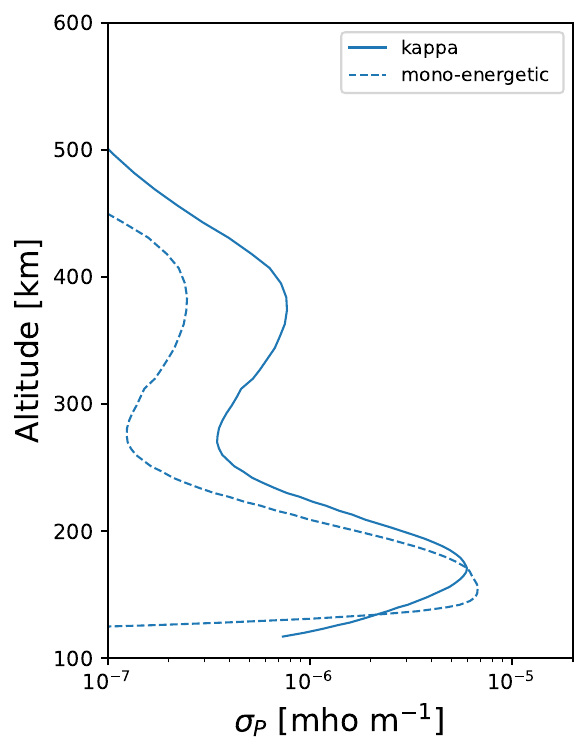}
        \label{fig:panel_h}
        }

        % Panel 9-12
        \subfloat[ $\langle E \rangle$ = 10 keV.]{\includegraphics[width=0.25\linewidth]{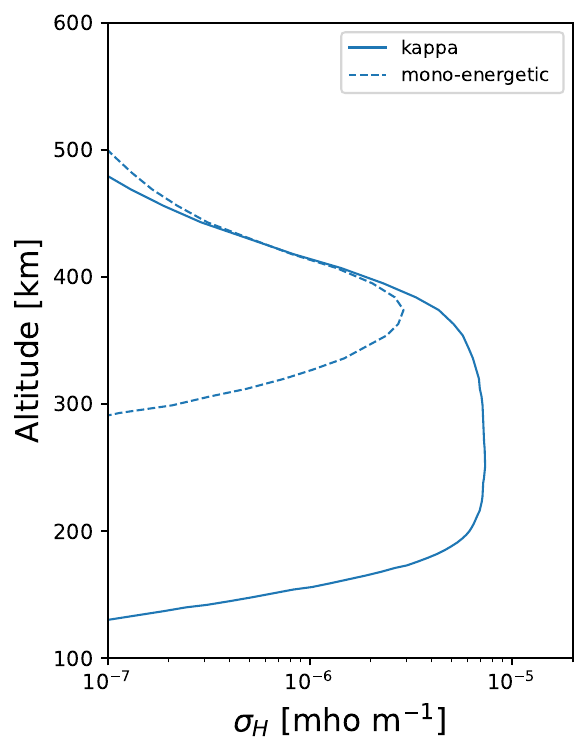}
        \label{fig:panel_i}
        }
        \subfloat[ $\langle E \rangle$ = 30 keV.]{\includegraphics[width=0.25\linewidth]{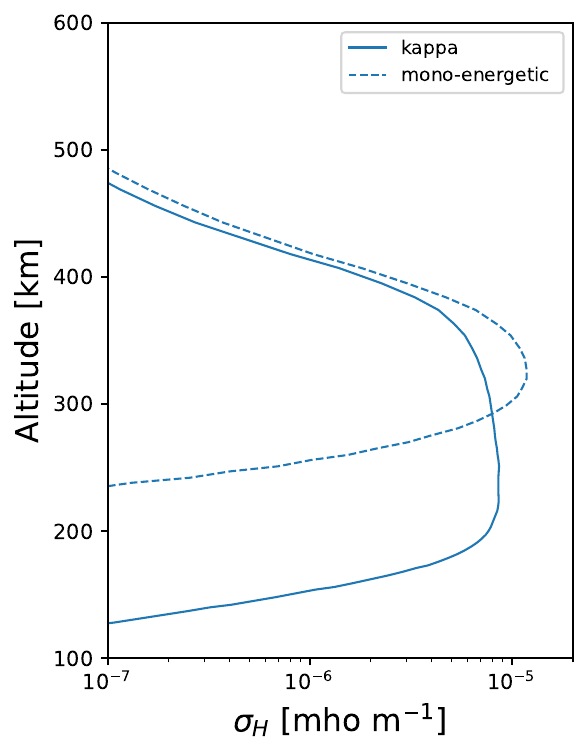}
        \label{fig:panel_j}
        }
        \subfloat[ $\langle E \rangle$ = 100 keV.]{\includegraphics[width=0.25\linewidth]{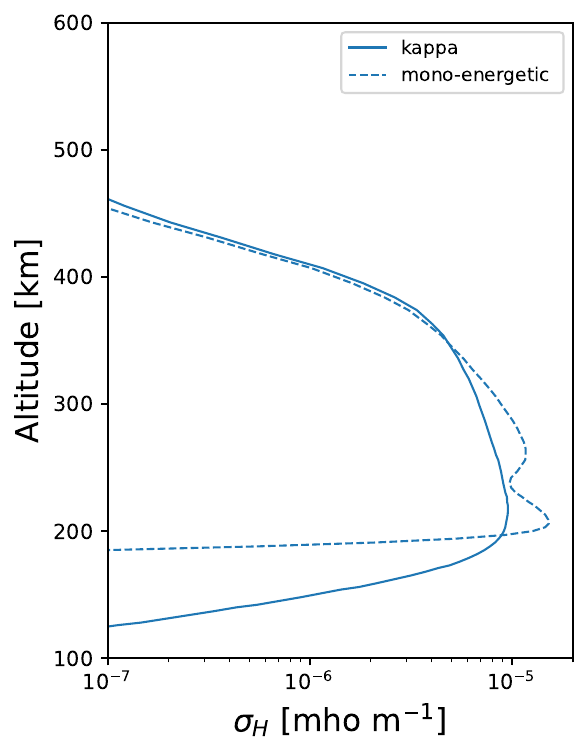}
        \label{fig:panel_k}
        }
        \subfloat[ $\langle E \rangle$ = 500 keV.]{\includegraphics[width=0.25\linewidth]{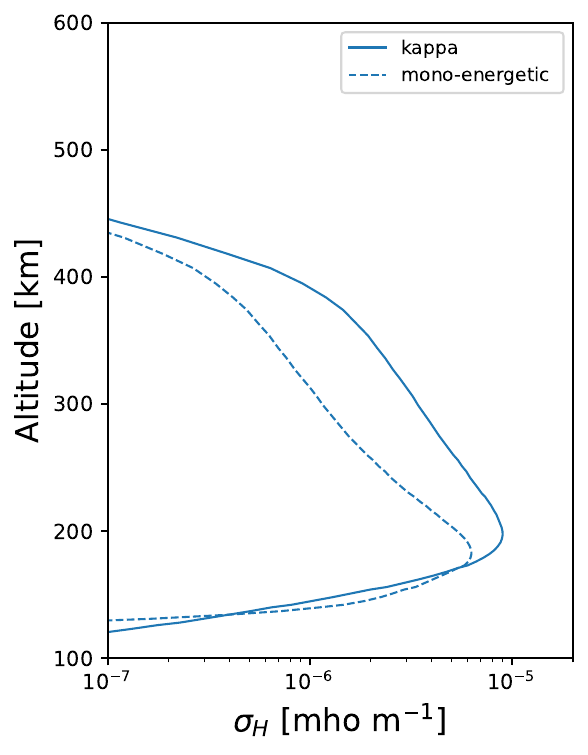}
        \label{fig:panel_l}
        }
    \caption{Vertical profiles of (a)-(d) the densities of $\ce{H3+}$ and $\ce{CH5+}$, (e)-(h) the Pedersen conductivity and (i)-(l) the Hall conductivity, computed assuming either a kappa or a mono-energetic electron energy distribution and for $\langle E \rangle$ = 10, 30, 100 and 500 keV. The electron total energy flux is 1 mW.m\textsuperscript{-2}.}
    \label{fig:vertical_profiles}
    \end{figure*}

    \begin{figure} % One-column figure
    \centering
    \captionsetup[subfigure]{width=0.5\linewidth}
        \subfloat[ Pedersen.]{\includegraphics[width=0.939\linewidth]{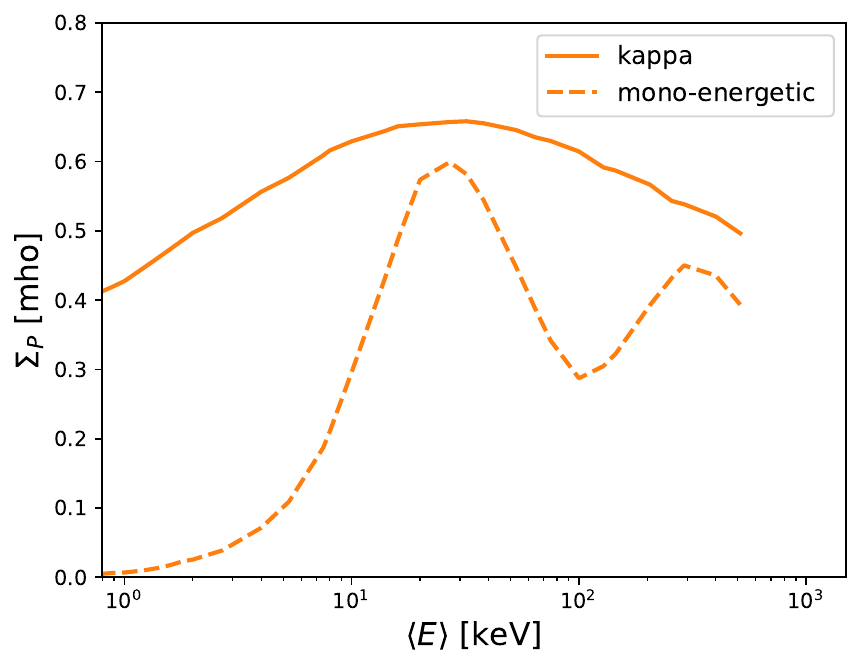}
        }

        \subfloat[ Hall.]{\includegraphics[width=0.939\linewidth]{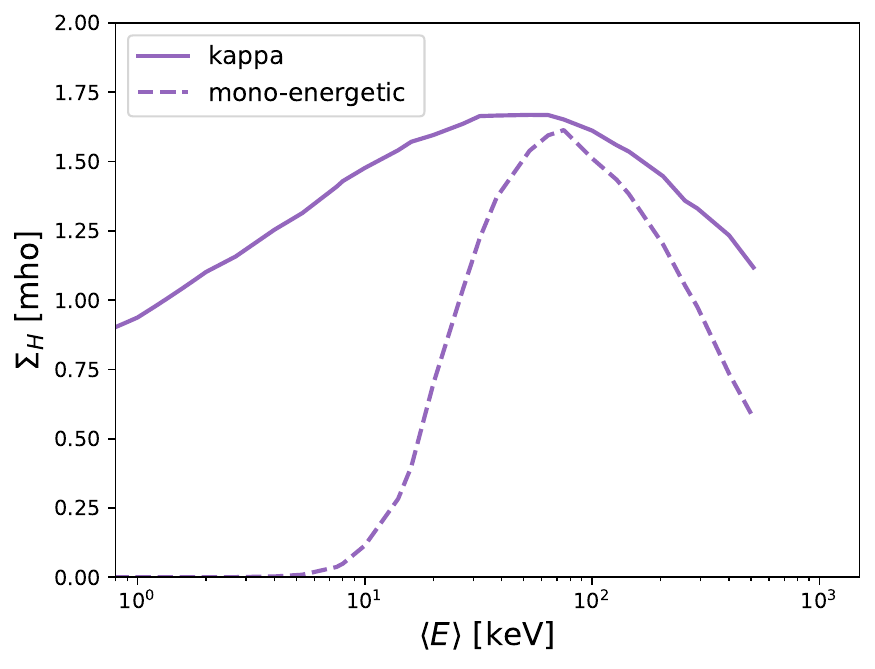}
        }
    \caption{
    Evolution of the (a) Pedersen and (b) Hall conductances as a function of the mean energy of the precipitating electrons $\langle E \rangle$, assuming a kappa (solid line) or a mono-energetic (dotted line) energy distribution ($F_e = $ 1 mW.m\textsuperscript{-2}, $ B = 10^{-3}$ T).
    }
    \label{fig:cVSmean_flux1}
    \end{figure}
    
    \begin{figure} % One-column figure
    \centering
    \includegraphics[width=0.939\linewidth]{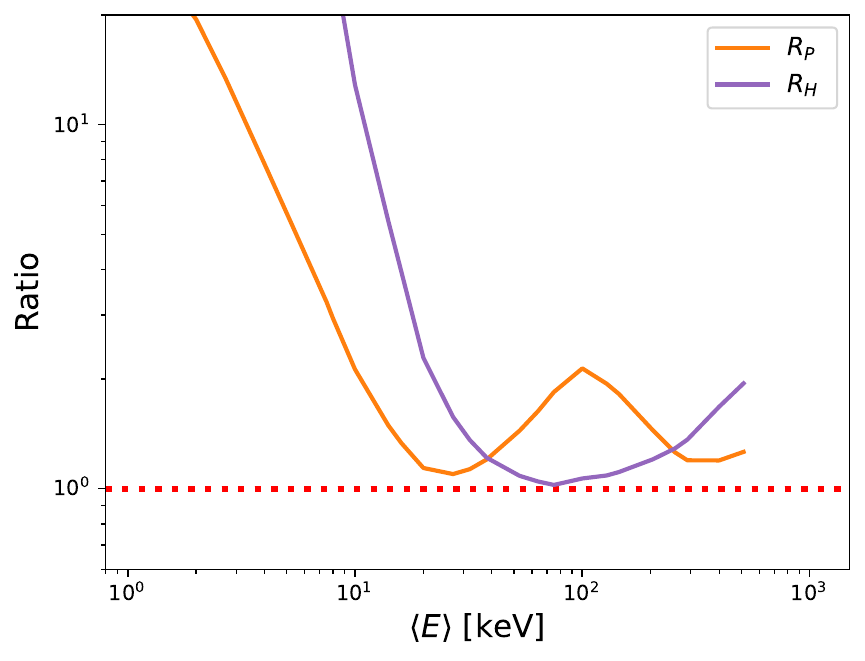}
    \caption{Ratio kappa/mono-energetic Pedersen ($R_P$) and Hall ($R_H$) conductances as a function of the electron mean energy $\langle E \rangle$ ($F_e = $ 1 mW.m\textsuperscript{-2}, $ B = 10^{-3}$ T).}
    \label{fig:ctance_ratio}
    \end{figure}

    Following the method described in section \ref{section:method}, we computed generic density vertical profiles of \ce{H3+} and \ce{CH5+} as well as the Pedersen and Hall conductivity profiles using either a kappa or a mono-energetic precipitating electron energy distribution, for $\langle E \rangle$ = 10, 30, 100 and 500 keV. We assumed an electron total energy flux of 1 mW.m\textsuperscript{-2}. While this value is associated with a dim aurora, the reader can easily scale the conductances $\Sigma_{P,H}$ to another energy flux $F_e$ since $\Sigma_{P,H} \propto \sqrt{F_e}$ (see section \ref{section:method}). We also assumed a constant magnetic field intensity of $10^{-3}$ T. 

    In the case of a mono-energetic distribution with a specific mean energy and total energy flux, all the electrons have a kinetic energy equal to the mean energy of the distribution, while, in the case of a kappa distribution with the same mean energy and total energy flux, the electron energies are distributed between 1 eV and 1 MeV. This difference has an impact on the density vertical profiles of $\ce{H3+}$ and $\ce{CH5+}$ shown in Fig. \ref{fig:panel_a} to \ref{fig:panel_d}. The mono-energetic electrons deposit most of their energy in a limited range of altitudes depending on the average electron energy, resulting in the observed ion density vertical profiles with sharp peaks. The main part of the ion production is thus confined around these peaks. For a kappa distribution having the same mean energy and total energy flux, the electrons with different kinetic energy lose most of their energy at different altitudes, leading to the observed wider and less peaked ion density vertical profiles. In particular, the high-energy electrons always reach lower altitudes down to a minimum altitude sets by the chosen high-energy cutoff (1 MeV, see section \ref{section:method}). In addition, the ion density vertical profiles computed with a kappa distribution depend less on the mean energy.
    
    The Pedersen and Hall conductivity vertical profiles are displayed in Fig. \ref{fig:panel_e} to \ref{fig:panel_l} for both energy distributions and for the same set of mean energies as for the ion density profiles. For the kappa Pedersen conductivity profiles, we distinguish an upper and lower conductivity peaks present around 180 km and 350 km, respectively, for all mean energy values considered. In the case of the mono-energetic Pedersen conductivity profiles, no sharp peak appears for $\langle E \rangle = 10$ keV. For $\langle E \rangle = 30$ keV, only one peak is present around 350 km, becoming less sharp as the mean energy increases. A second peak appears at a lower altitude in the case of $\langle E \rangle = 100$ keV, becoming more prominent and moving to lower altitudes as the mean energy increases. The kappa Hall conductivity is significant between about 150 km and 380 km for all the considered mean energy values. In contrast, the mono-energetic Hall profile displays a small peak around 380 km for $\langle E \rangle = 10$ keV. The Hall conductivity increases and the peak moves to lower altitudes as the mean energy increases. The conductivities are likely to maximize around the altitudes where the electron and ion gyro-frequencies match their corresponding collision frequencies with the neutrals. As mentioned in the introduction, this altitude region is often called the ionospheric conductive layer \citep[e.g.,][]{nakamura_effect_2022, clement_ionospheric_2025}. To be precise, the Pedersen conductivity maximizes around the upper and lower boundaries of the conductive layer and the Hall conductivity maximizes inside this layer. A clear definition of the conductive layer used in this study can be found in Appendix \ref{appendix:b}.
    
    The difference in behavior between the kappa and mono-energetic conductivities directly arises from the ion vertical profiles. Consider the conductivity profiles computed with a mono-energetic distribution. For $\langle E \rangle = 10$ keV, most of the ions are created above the upper boundary of the conductive layer, resulting in a poor contribution to the Pedersen and Hall conductivities. For $\langle E \rangle = 30$ keV, the peaks of the $\ce{H3+}$ and $\ce{CH5+}$ profiles more or less match the altitude of the upper boundary of the conductive layer. It results in sharp Pedersen conductivity peaks around 350 km in Fig. \ref{fig:panel_f}. Here, no peak appears around the lower boundary of the conductive layer because the mean energy is too low for the electrons to reach this altitude. As the electrons penetrate deeper into the atmosphere with $\langle E \rangle = 100$ keV, the density peak of $\ce{CH5+}$ comes closer to the lower boundary of the conductive layer, resulting in a second peak of Pedersen conductivity. Finally, for $\langle E \rangle = 500$ keV, the peak of the $\ce{CH5+}$ density matches the lower boundary of the conductive layer, giving rise to a sharp second, lower peak for the Pedersen conductivity. However, fewer ions are produced at the upper boundary of the conductive layer, leading to a large difference between the two peaks. The Hall conductivity becomes also non-negligible for $\langle E \rangle =$ 30, 100 and 500 keV since a significant amount of electrons can reach the altitudes of the conductive layer. Understandably, the altitude of the Hall conductivity peak follows the altitude of the $\ce{CH5+}$ density peak. If we now consider a kappa electron energy distribution, we see that a significant number of electrons are formed from 600 km down to an altitude of about 120 km for all the considered mean energies. As a consequence, all the kappa Pedersen conductivity profiles similarly display two peaks and all the Hall profiles are extended over the altitude range of the conductive layer. As the electrons penetrate deeper in altitude with the increase of the mean energy, the second, lower, peak of the Pedersen conductivity becomes prominent compared to the first, upper one and the Hall conductivity profile slowly deforms itself into a peaked profile.

    We then integrated the Pedersen and Hall conductivity profiles to obtain the corresponding conductances. In Fig. \ref{fig:cVSmean_flux1}, the two graphs display the evolution of the Pedersen and Hall conductances, for a kappa and a mono-energetic electron energy distribution, respectively, as a function of the electron mean energy. As in \citet{gerard_spatial_2020}, the mono-energetic Pedersen conductance presents a maximum value around 30 keV, previously predicted in \citet{millward_dynamics_2002} around 60 keV. The mono-energetic Pedersen conductance values obtained in this study are in accordance with the ones presented in \citet{millward_dynamics_2002}. However, their mono-energetic Hall conductance values are lower than ours by an order of magnitude. It is most likely due to the fact that they did not take into account the presence of hydrocarbon ions. It resulted in a depletion of electric charges, notably electrons, inside the ionospheric conductive layer, where the Hall conductivity is likely to maximize, resulting in the low computed Hall conductance.
    
    The kappa Pedersen and Hall conductances present maximum values around 60 keV and 70 keV, respectively. We see that the mono-energetic conductance is always less than the kappa conductance. The conductance plots follow directly the evolution of the conductivity profiles with the mean energy. Notably, the very low values of the mono-energetic conductance for $\langle E \rangle < 10$ keV reflects the fact that most of the ions are formed above the conductive layer and the mono-energetic Pedersen conductance peak, around 30 keV, results from the concordance between the ion density peak altitude and the upper boundary of the conductive layer. As mentioned before for the conductivities, the kappa conductances vary less with the mean energy than the corresponding mono-energetic conductances. Thus, the polar conductance maps generated assuming a kappa electron energy distribution would appear more uniform than the ones based on a mono-energetic distribution.

    Finally, we computed the Pedersen ratio $R_P = \Sigma_{P,k}/\Sigma_{P,m}$ and Hall ratio $R_H = \Sigma_{H,k}/\Sigma_{H,m}$. These ratios are plotted in Fig. \ref{fig:ctance_ratio} as a function of the electron mean energy $\langle E \rangle$. In the considered energy range, $R_P$, $R_H$ $\geq$ 1. At low energy ($\langle E \rangle$ $\leq$ 4 keV), both ratios are very high ($\geq$ 10) and decrease as the mean energy increases. The Pedersen ratio reaches down to $R_P \approx 1.2$ around $\langle E \rangle = 30$ keV before increasing up to $R_P \approx 2$ around $\langle E \rangle = 130$ keV. It then decreases again toward $R_P = 1$ with the increase of mean energy. The Hall ratio decreases down to $R_H = 1$ around 70 keV before increasing again. From these plots, we see that the Pedersen and Hall conductances computed assuming a mono-energetic electron energy distribution are generally less inside the 1-1000 keV energy range compared to those computed assuming a kappa electron energy distribution spanning the same range of energies. 

    In the auroral regions, the conductance in the low electron mean energy zones should then be the most impacted by a change in the electron energy distribution. For the Hall conductance, since the difference is significant for $\langle E \rangle <$ 20 keV and for 300 keV $< \langle E \rangle$, it should be impacted by a change in the electron energy distribution either in the low and high mean energy zones in the auroral regions.

    \subsection{Conductance maps}
    \begin{figure*} % One-column figure
    \centering
    \captionsetup[subfigure]{width=1\textwidth}
        \subfloat[ $\Sigma_P$.]
        {\includegraphics[width=0.5\linewidth]{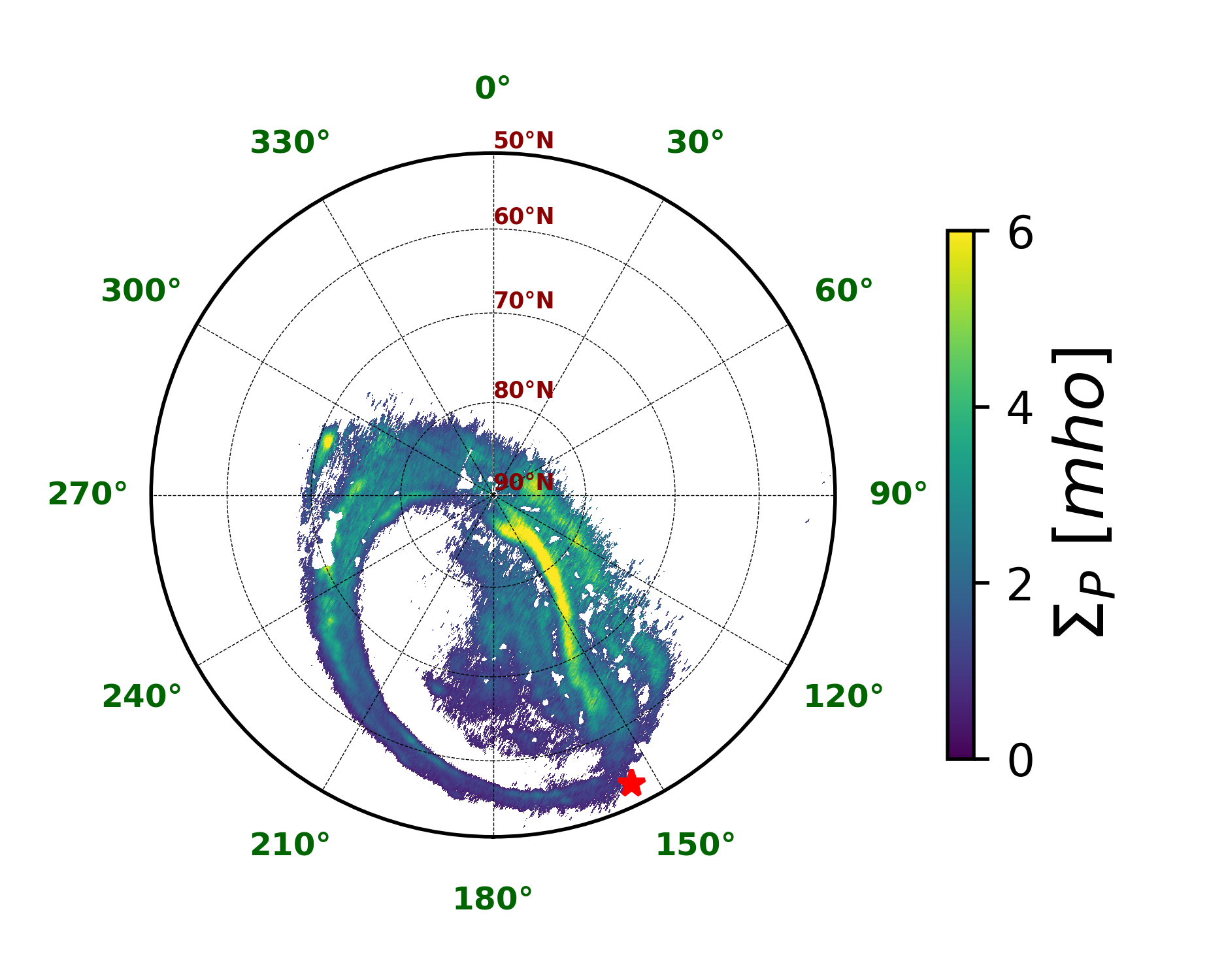}
        \label{fig:maps_conductnorth_pj1_p}
        }
        \subfloat[ $\Sigma_H$.]
        {\includegraphics[width=0.5\linewidth]{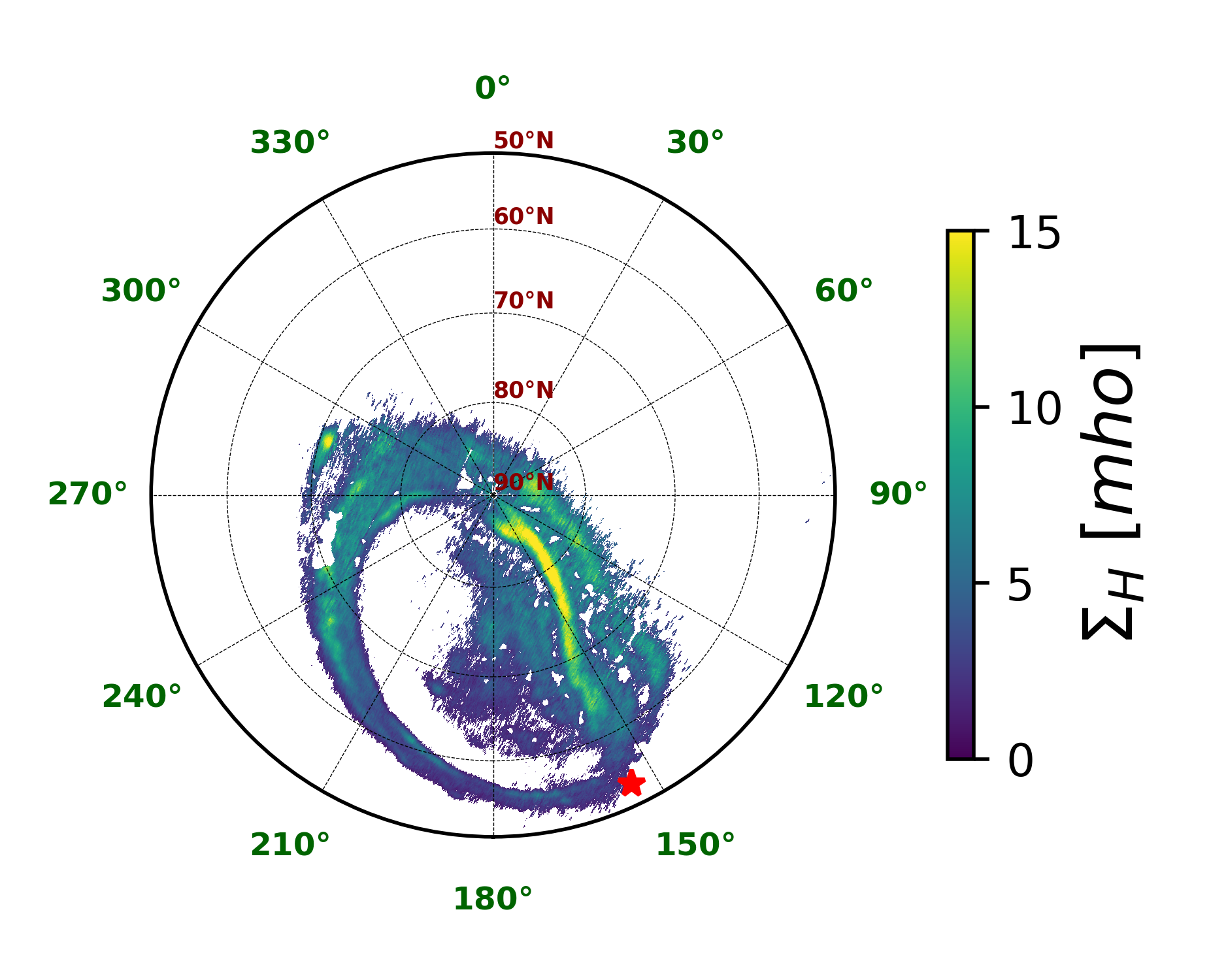}
        \label{fig:maps_conductnorth_pj1_h}
        }
    \caption{
    The (a) Pedersen and (b) Hall conductance maps for PJ1 north. On each map, the red star represents the subsolar longitude.
    }
    \label{fig:maps_conductnorth_pj1}
    \end{figure*}

    \begin{figure*} % One-column figure
    \centering
    \captionsetup[subfigure]{width=1\textwidth}
        \subfloat[ Ratio Pedersen.]
        {\includegraphics[width=0.5\linewidth]{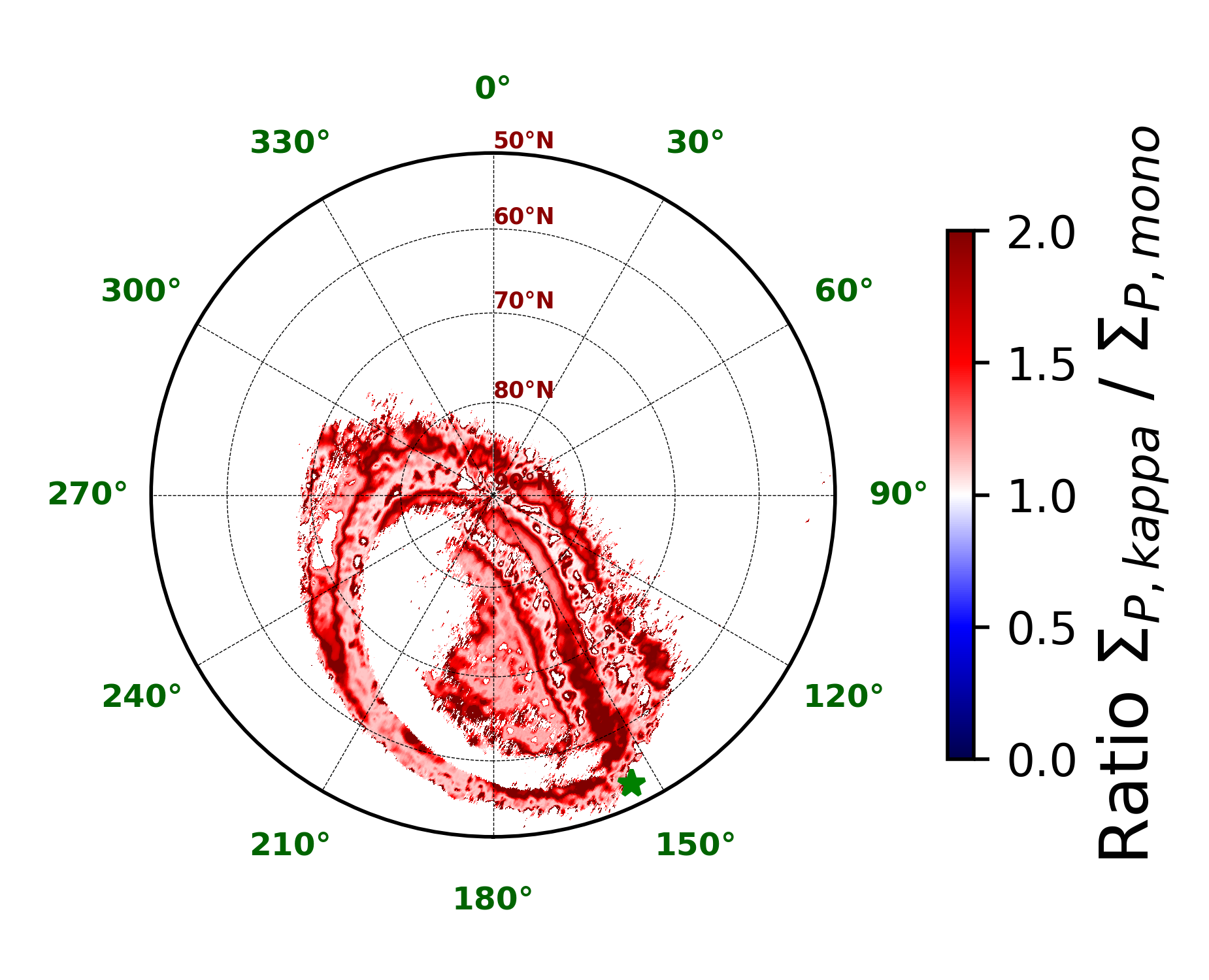}
        \label{fig:maps_conductnorth_pj1_rp}
        }
        \subfloat[ Ratio Hall.]
        {\includegraphics[width=0.5\linewidth]{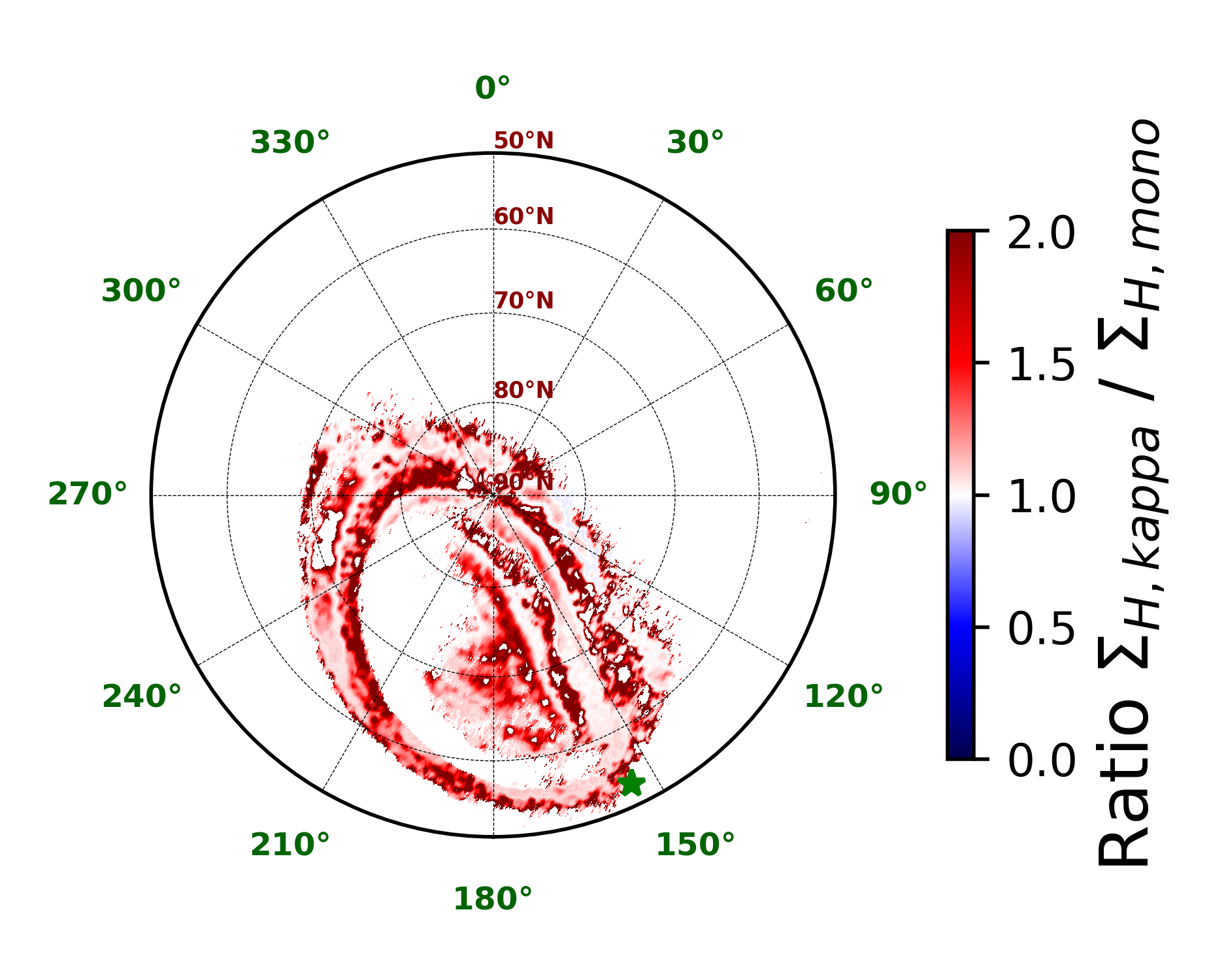}
        \label{fig:maps_conductnorth_pj1_rh}
        }
    \caption{
    Ratio of the Pedersen (Hall) conductance map computed with a kappa distribution over the Pedersen (Hall) conductance map computed with a mono-energetic distribution. The scale goes up to 2 but the ratio can be locally higher. On each map, the green star represents the subsolar longitude.
    }
    \label{fig:maps_conductnorth_pj1_ratio}
    \end{figure*}

    \begin{table*}
        \centering
        \begin{tabular}{|c|c||c|c|c|c||c|c||c|c|c|c||c|c|}
             \hline
             \multicolumn{2}{|c||}{}
             & \multicolumn{6}{c||}{\textbf{North}} 
             & \multicolumn{6}{c|}{\textbf{South}}\\
             \hline
             \textbf{Zone} & \textbf{Value type} 
             & \textbf{PJ1} & \textbf{PJ6} 
             & \textbf{PJ12} & \textbf{PJ22} 
             & \textbf{Mean} & \textbf{Med.} 
             & \textbf{PJ15} & \textbf{PJ27} 
             & \textbf{PJ34} & \textbf{PJ35} 
             & \textbf{Mean} & \textbf{Med.} \\
             \hline
             \hline
             Whole aurora & < 95\% 
             & 4.7 & 3.2 & 3.9 & 3.4 
             & \textbf{3.8} & \textbf{3.7} 
             & 4.0 & 4.0 & 3.0 & 3.0 
             & \textbf{3.5} & \textbf{3.5}\\
             Whole aurora & Mean 
             & 2.4 & 1.8 & 1.7 & 1.8 
             & \textbf{1.9} & \textbf{/} 
             & 2.0 & 2.0 & 1.6 & 1.6 
             & \textbf{1.8} & \textbf{/}\\
             Whole aurora & Med. 
             & 2.0 & 1.7 & 1.5 & 1.6 
             & \textbf{/} & \textbf{1.7} 
             & 1.7 & 1.7 & 1.5 & 1.4 
             & \textbf{/} & \textbf{1.6}\\
             \hline
             ME & < 95\% 
             & 6.9 & 4.8 & 4.7 & 3.3 
             & \textbf{4.9} & \textbf{4.8} 
             & 4.6 & 4.9 & 3.9 & 4.2 
             & \textbf{4.4} & \textbf{4.4}\\
             ME & Mean 
             & 2.9 & 2.4 & 2.6 & 1.9 
             & \textbf{2.4} & \textbf{/} 
             & 2.8 & 3.0 & 2.3 & 2.4 
             & \textbf{2.6} & \textbf{/}\\
             ME & Med. 
             & 2.1 & 2.1 & 2.4 & 1.7 
             & \textbf{/} & \textbf{2.1} 
             & 2.7 & 2.9 & 2.2 & 2.4 
             & \textbf{/} & \textbf{2.5}\\
             \hline
        \end{tabular}
        \caption{A few figures relative to the kappa Pedersen conductance computed for the whole polar aurora (including the ME) and for the ME only (units: mho). The lines entitled "<95\%" specify the boundary value below which we find 95\% of the conductance values. The lines and columns entitled "Med." designates median values.}
        \label{tab:condP}
    \end{table*}

    \begin{table*}
        \centering
        \begin{tabular}{|c|c||c|c|c|c|c|c||c|c|c|c||c|c|}
             \hline
             \multicolumn{2}{|c||}{}
             & \multicolumn{6}{c||}{\textbf{North}} 
             & \multicolumn{6}{c|}{\textbf{South}}\\
             \hline
             \textbf{Zone} & \textbf{Value type} 
             & \textbf{PJ1} & \textbf{PJ6} 
             & \textbf{PJ12} & \textbf{PJ22} 
             & \textbf{Mean} & \textbf{Med.} 
             & \textbf{PJ15} & \textbf{PJ27} 
             & \textbf{PJ34} & \textbf{PJ35} 
             & \textbf{Mean} & \textbf{Med.} \\
             \hline
             \hline
             Whole aurora & < 95\% 
             & 10.8 & 7.5 & 9.2 & 7.9 
             & \textbf{8.9} & \textbf{8.6} 
             & 9.7 & 10.0 & 7.6 & 7.8 
             & \textbf{8.8} & \textbf{8.8}\\
             Whole aurora & Mean 
             & 5.6 & 4.3 & 4.2 & 4.4 
             & \textbf{4.6} & \textbf{/}
             & 4.9 & 4.9 & 4.0 & 4.0 
             & \textbf{4.5} & \textbf{/}\\
             Whole aurora & Med. 
             & 5.1 & 4.0 & 3.6 & 4.0 
             & \textbf{/} & \textbf{4.0} 
             & 4.3 & 4.1 & 3.7 & 3.6 
             & \textbf{/} & \textbf{3.9}\\
             \hline
             ME & < 95\% 
             & 15.7 & 11.2 & 11.0 & 7.8 
             & \textbf{11.4} & \textbf{11.1} 
             & 11.0 & 12.0 & 9.8 & 10.3
             & \textbf{10.8} & \textbf{10.8}\\
             ME & Mean 
             & 6.9 & 5.7 & 6.3 & 4.5 
             & \textbf{5.8} & \textbf{/} 
             & 6.8 & 7.5 & 5.6 & 6.0 
             & \textbf{6.5} & \textbf{/}\\
             ME & Med. 
             & 5.2 & 5.1 & 5.9 & 4.2 
             & \textbf{/} & \textbf{5.1}
             & 6.7 & 7.4 & 5.3 & 5.8 
             & \textbf{/} & \textbf{6.2}\\
             \hline
        \end{tabular}
        \caption{Same description as Table \ref{tab:condP} for the kappa Hall conductance (units: mho).}
        \label{tab:condH}
    \end{table*}
    
    We carried out a case study by investigating the Pedersen and Hall conductances along the northern and southern polar aurorae for specific PJs. We selected PJs with a wide range of auroral characteristics. For the north pole, we chose PJ1, PJ6, PJ12 and PJ22. PJ1 and PJ6 are characterized by bright equatorward and poleward auroral emissions, respectively. Concerning the two other PJs, the ME zones present on PJ12 and PJ22 are contracted and expanded, respectively. Other features like polar auroral filaments \citep{nichols_observations_2009} are also present on PJ12. For the southern hemisphere, we selected PJ15, PJ27, PJ34 and PJ35. PJ27 displays a clear auroral signature of plasma injection \citep{dumont_evolution_2018}. Finally, PJ34 and PJ35 were chosen because they display similar ME morphologies \citep{head_effect_2024} and sub-solar latitudes.
    
    For illustration purposes, Fig. \ref{fig:maps_conductnorth_pj1} displays the Pedersen and Hall conductance maps for PJ1 built with a kappa distribution. The maps of the other PJs can be found in Appendix \ref{appendix:c}. At first glance, the Pedersen and Hall conductances present similar spatial distributions for each PJ, even if the Hall conductance is 2-3 times higher than the Pedersen conductance. In the north, the spatial variations in the ionospheric magnetic field give rise to variations in the conductances, notably along the ME. As pointed out in Appendix \ref{appendix:b}, such discrepancies are expected since the Pedersen and Hall conductances increase as the magnetic field intensity decreases. Both Pedersen and Hall conductances are larger in the ME and in the outer emission regions than in the polar emission region. 
 
    We also derived the ratio of the Pedersen (Hall) conductance maps computed with a kappa distribution over the Pedersen (Hall) conductance maps computed with a mono-energetic distribution. These ratio maps are plotted in Fig. \ref{fig:maps_conductnorth_pj1_ratio} for PJ1 (see Appendix \ref{appendix:c} for the maps relative to the other PJs). As anticipated in the previous subsection, the auroral zones that are less affected by a change of distribution seem to be associated with a high CR (i.e. a high electron mean energy). In contrast, the local zones with strong ratio values are matching with very low CR (i.e., low electron mean energies). Concerning the Hall conductance, part of the auroral zones has a Hall ratio strongly greater than 1, often matching with low CR.
    
    For each hemisphere, the conductance maps computed with a kappa precipitating electron energy distribution display substantial spatial variations, depending on the considered auroral zones. Locally, the conductances can be of the order of $50$ mho, such as in the auroral zone linked to the injection event that took place during PJ27. However, as we read in Tables \ref{tab:condP} and \ref{tab:condH}, more than 95\% of the kappa Pedersen and Hall conductance values remain lower than about 5 mho and 11 mho, respectively. These tables also display the median kappa conductance values computed for each PJ for the whole polar aurora, including the ME, as well as for the ME only. In our study, the values drawn from the Pedersen conductance maps contrast with the ones presented in \citet{gerard_spatial_2020}, for which the conductance did not exceed 10 mho. In addition, our mean Pedersen conductance value computed for each PJ are about 3-5 times higher than the mean conductance value obtained in \citet{gerard_variability_2021}. Here, the difference in the electron energy distribution used (kappa distribution in this study VS mono-energetic distribution in \citet{gerard_spatial_2020, gerard_variability_2021}) only partially explain this discrepancy. It seems that the origin of the remaining difference lies within the method used. In \citet{gerard_spatial_2020, gerard_variability_2021}, the $\ce{H2+}$ production rate is computed using the analytical expression presented in \citet{hiraki_parameterization_2008}, built from the results of an electron transport model based on a Monte-Carlo simulation. In this study, the $\ce{H2+}$ production rate is instead the direct output of the chosen electron transport model. In Appendix \ref{appendix:ctris} is presented a comparison between the mono-energetic Pedersen conductance values obtained with the analytical expression presented in \citet{hiraki_parameterization_2008}, with TransPlanet and with a Monte-Carlo simulation. It appears that there is a discrepancy between the use of the \citet{hiraki_parameterization_2008}'s analytical expression and direct outputs of models. This discrepancy is partially, but not only, due to a difference in the electron collision cross-sections.
    
    The values present in Tables \ref{tab:condP} and \ref{tab:condH} seem to indicate that, on average, slightly more conductance is produced in the north than in the south when considering the whole polar aurora. The opposite tendency is observed when taking only the ME into account. This latter statement could partially explain the observed difference in the mean ionospheric current density between the north and the south poles \citep{kotsiaros_birkeland_2019}. These asymmetries could also be linked to the already observed brightness asymmetries between the northern and southern hemispheres for which the emissions poleward of the ME are brighter in the north and the ME is brighter in the south \citep{bonfond_north_2024}. As no north-south conductance asymmetry has yet been reported \citep{gerard_variability_2021}, a statistical study of the northern and southern Pedersen and Hall conductances should be performed over the whole set of available PJs before drawing any robust conclusion.

    Concerning the ME region, we see that the mean Pedersen and Hall conductance values, displayed in Tables \ref{tab:condP} and \ref{tab:condH} for the northern and southern aurorae, globally match the ranges presented by \citet{al_saati_magnetosphereionospherethermosphere_2022}, although the mean values are slightly higher. We suggest several reasons for this discrepancy. Firstly, they only have access to the conductances along the path of Juno whereas, with our method, we reach the conductances all along the ME. Second, the region where the electrons are accelerated might be below the path of Juno, as highlighted earlier in this paper. This acceleration would increase the auroral UV brightness, which is used to infer the electron total energy flux in this study. Finally, part of the discrepancy might be explained by the use of the analytic formula presented in \citet{hiraki_parameterization_2008} to retrieve the $\ce{H2+}$ production rate, also invoked to explain the discrepancy between our results and the ones derived by \citet{gerard_spatial_2020}.
    
    The Pedersen conductances obtained for the ME also strongly contrast with the ones presented in \citet{rutala_variation_2024}. By incorporating the auroral brightness observed by HST and the magnetospheric plasma flow measured by the Galileo spacecraft in the equations relative to the corotation enforcement theory, they derived median effective Pedersen conductance values relative to the ME region of $\Sigma^*_P$ = $0.14^{+0.31}_{-0.08}$ mho for the north pole and $\Sigma^*_P$ = $0.14^{+0.34}_{-0.09}$ for the south pole. The effective conductance is related to the actual conductance through a coefficient $k$ representing the slippage of the neutral thermosphere: $\Sigma^*_P = (1-k) \Sigma_P$. We took $k$ = 0.4 - 0.7 \citep{millward_dynamics_2005} and computed the median effective Pedersen conductances corresponding to the actual conductance values obtained in this study. Without taking their error bars into account, we obtained values 5-9 times and 6-12 times higher than 0.14 mho for the north and south poles, respectively. Even if we use their maximal estimate, our effective conductance remains at least 1.5 times higher than theirs. In the corotation enforcement theory, the effective Pedersen conductance is directly proportional to the FACs per radian of azimuth (equation 3 in \citet{rutala_variation_2024}). The effective conductance values computed in this study would then lead to higher FAC values that would not be in accordance with the ones deduced from the Juno magnetometer measurements \citep{nichols_relation_2022}. Therefore, we suggest that the discrepancy between the conductance values computed in \citet{rutala_variation_2024} and the ones computed in this study may arise from one of the following two causes or a combination of them: either the factor limiting the FACs is not the ionospheric conductance, but a phenomenon taking place elsewhere; or the ionospheric conductance is increased by additional precipitating electrons accelerated by processes which are not associated with the FACs. In any case, the actual relation between the FAC per radian of azimut and the effective Pedersen conductance would then differ from the one obtained from the corotation enforcement theory.

%%%%%%%%%%%%%%%%%%%%%%%%%%%%%
%%%%%%%%%% SECTION %%%%%%%%%%
%%%%%%%%%%%%%%%%%%%%%%%%%%%%%
\section{Conclusions}
\label{section:conclusions}
    We used a kappa distribution to model the energy distribution of the auroral electrons that precipitate at the poles in the Jovian atmosphere and compute the ionospheric Pedersen and Hall conductances resulting from this precipitation. We used data from UVS and JEDI to retrieve the distribution parameters. We assumed an atmosphere made up of $\ce{H}$, $\ce{H2}$, $\ce{He}$ and $\ce{CH4}$ with densities taken from \citet{grodent_self-consistent_2001} and used an electron transport model (TransPlanet) to infer the ion density vertical profiles. The conductance maps were computed with the most advanced internal magnetic field model, JRM33, currently available for Jupiter \citep{connerney_new_2022}. From this study, we observed that the Pedersen and Hall conductivities are more broadly distributed in altitudes when computed with a kappa electron energy distribution than with a mono-energetic distribution. As a consequence, the Pedersen and Hall conductance values are higher when computed with a kappa electron energy distribution rather than with a mono-energetic distribution. The importance of this underestimation varies between 1 and $10^3$ keV, decreasing when increasing the electron mean energy. We then computed the northern and southern conductance maps for a series of PJs and observed that the conductances are higher and more uniform on the maps associated with a kappa electron energy distribution. We also noticed that the Pedersen conductance values obtained in this study are in general higher than the ones derived in previous studies. The difference is particularly large when comparing our values with those computed through the corotation enforcement theory \citep{cowley_origin_2001}. Finally, the case study highlights the possible existence of north-south conductance asymmetries. From these observations, we drew the following conclusions:
    \begin{enumerate}
        \item Considering a broadband electron energy distribution rather than a mono-energetic distribution has a significant impact on the determination of the Pedersen and Hall conductances. This impact is especially large at low electron mean energy for the Pedersen conductance and at low and high electron mean energy for the Hall conductance.
        \item The high values obtained for the Pedersen conductance are only partially due to the consideration of a more realistic precipitating electron energy distribution. The use of the analytical expression presented in \citet{hiraki_parameterization_2008} also seems to give lower conductance values than the direct outputs of electron transport models. This difference is partially, but not only, due to a difference in the electron collision cross-sections.
        \item The disagreement between our values and the ones deduced from the corotation enforcement theory could indicate that either a physical mechanism taking place elsewhere contributes to limit the FACs and/or precipitating electrons accelerated by processes not associated with the FACs increase the ionospheric conductance.
    \end{enumerate}
    This study also highlighted the possible existence of a north-south asymmetry in the Pedersen and Hall conductances. This asymmetry will be further investigated in a future statistical analysis of the conductance over the whole set of available PJs.

\begin{acknowledgements}
This work was supported by the Fonds de la Recherche Scientifique - FNRS under Grant(s) No T003524F. T. Greathouse was funded by the NASA's New Frontiers Program for Juno via contract NNM06AA75C with the Southwest Research Institute.
\end{acknowledgements}

%%%%%%%%%%%%%%%%%%%%%%%%%%%%%
%%%%%%%%%%% BIBLIO %%%%%%%%%%
\bibliographystyle{bibtex/aa}
\bibliography{bibtex/biblio}

\begin{appendix}
\section{Densities of $\ce{H3+}$ and $\ce{CH5+}$} %First appendix
\begin{figure} % One-column figure
    \centering
    \captionsetup[subfigure]{width=1\textwidth}
        \subfloat[$\ce{H2+}$ production rates ($q$).]
        {\includegraphics[width=0.5\linewidth]{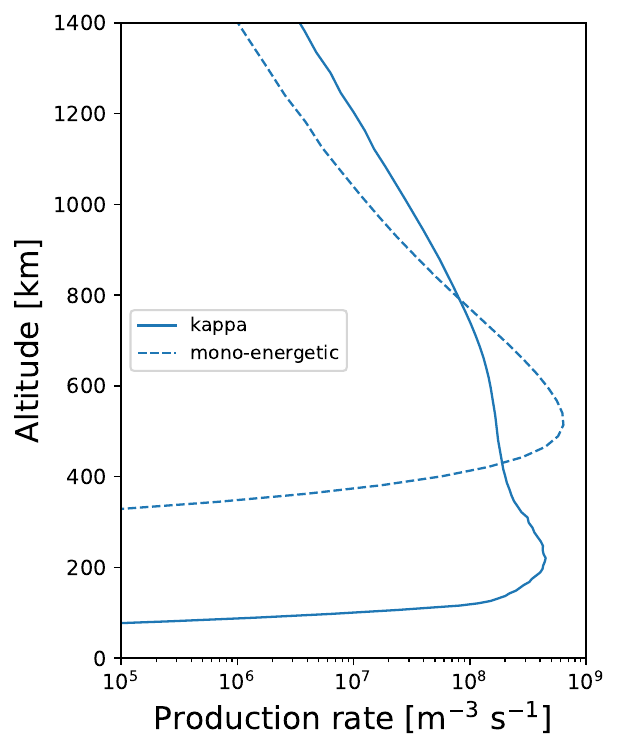}
        \label{fig:ph2plus}
        }
        \subfloat[Ion density vertical profiles.]
        {\includegraphics[width=0.505\linewidth]{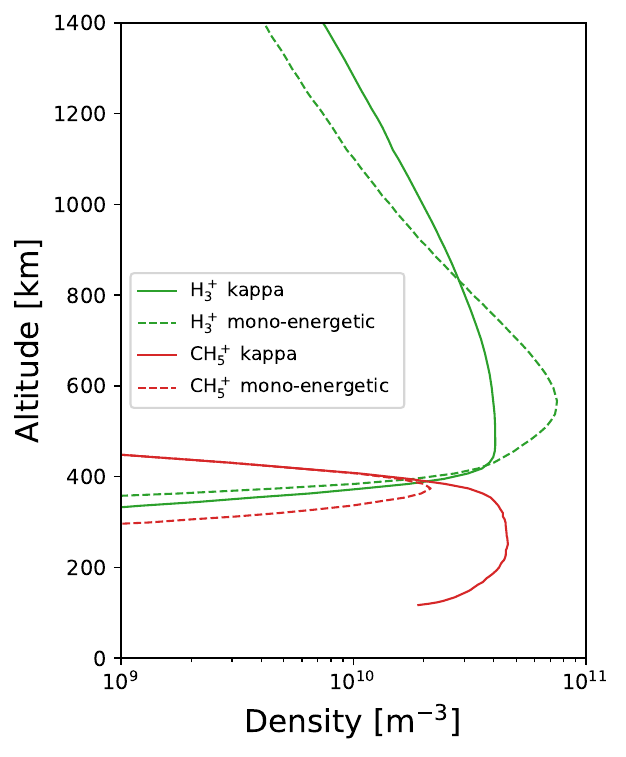}
        \label{fig:dtyions2}
        }
    \caption{(a) Production rate vertical profiles of $\ce{H2+}$ ($q$) and (b) density vertical profiles of $\ce{H3+}$ and $\ce{CH5+}$ computed with a kappa and a mono-energetic electron energy distribution ($\langle E \rangle$ = 10 keV, $F_e$ = 1 mW.m\textsuperscript{-2}).
    }
    \label{fig:appendix1}
\end{figure}

\begin{figure} % One-column figure
        \centering
        \includegraphics[scale=0.6]{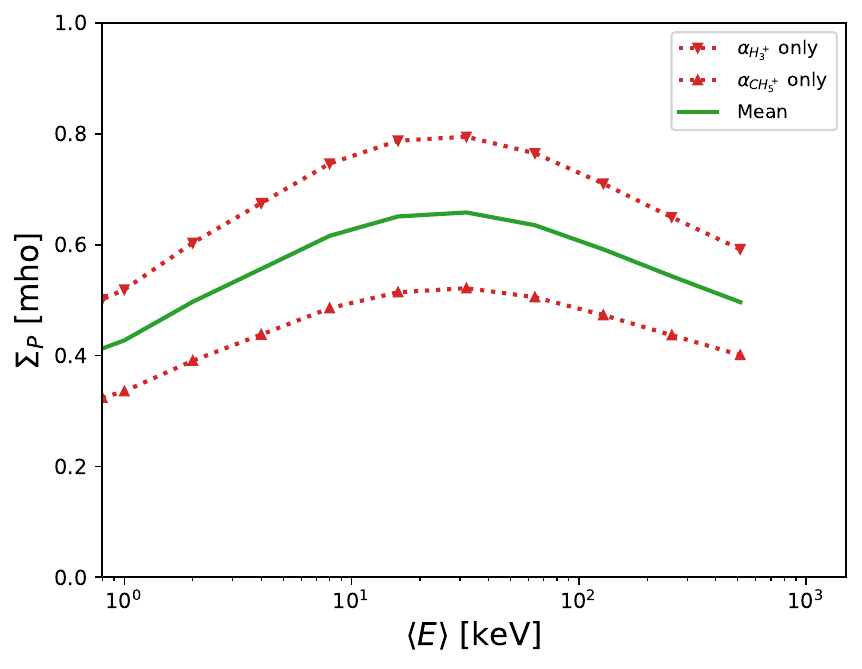}
        \caption{Pedersen conductance computed using a kappa electron energy distribution. The upper (lower) red curve represents the conductance when using the value of $\alpha_{\ce{H3+}}$ ($\alpha_{\ce{CH5+}}$). The green curve represents the mean conductance deduced from the corresponding values on the red curves.}
        \label{fig:appendix:a_boundary}
\end{figure}

\label{appendix:a}
In the Jovian atmosphere, a fraction of the \ce{H2} molecules is ionized through collisions with the auroral precipitating electrons
\begin{empheq}{align}
    \ce{H2} + \ce{e-}_{\text{au}} & \ce{->} \ce{H2+} + 2 \ce{e-}.
\end{empheq}
The newly formed $\ce{H2+}$ ions rapidly react with $\ce{H2}$ to form $\ce{H3+}$
\begin{empheq}{align}
    \ce{ H2+} + \ce{H2} & \ce{->} \ce{H3+} + \ce{H}.
\end{empheq}
The $\ce{H3+}$ ions are essentially destroyed by dissociative recombinations and reactions with methane
\begin{empheq}{align}
    \ce{H3+} + \ce{e-} &\ce{->[\alpha_{\ce{H3+}}]} \ce{H2} + \ce{H},
    \\
    \ce{H3+} + \ce{CH4} &\ce{->[k]} \ce{CH5+} + \ce{H2}.\label{equation:chemical_reaction_ch5+}
\end{empheq}
Here, $\alpha_{\ce{H3+}}$ and $k$ are the reaction rate coefficients. We take $\alpha_{\ce{H3+}} = 1.15\times 10^{-13} \, \left(\frac{300}{T_e}\right)^{0.65}$ m\textsuperscript{3} s\textsuperscript{-1} \citep{sundstrom_destruction_1994} and $k=2.9\times 10{-15}$ m\textsuperscript{3} s\textsuperscript{-1} \citep{perry_chemistry_1999}. The quantity $T_e$ is the electron temperature that we consider equal to the neutral temperature (see text). The equation \ref{equation:chemical_reaction_ch5+} also depicts the primary reaction that produces \ce{CH5+}. These hydrocarbon ions are mainly destroyed through electron recombinations
\begin{empheq}{align}
    \ce{CH5+} + \ce{e-} &\ce{->[\alpha_{\ce{CH5+}}]} \text{products},
\end{empheq}
with $\alpha_{\ce{CH5+}} = 2.78\times 10^{-13} \, \left(\frac{300}{T_e}\right)^{0.52}$ m\textsuperscript{3} s\textsuperscript{-1} \citep{perry_chemistry_1999}. The evolutions of the ion densities [$\ce{H3+}$] and [$\ce{CH5+}$] with time are governed by
\begin{empheq}{align}
    \odv{\left[\ce{H3+}\right]}{t}
    &= P(\ce{H3+}) - D(\ce{H3+}) + T(\ce{H3+}),
    \label{equation:evolution_H3+}\\
    \odv{\left[\ce{CH5+}\right]}{t}
    &= P(\ce{CH5+}) - D(\ce{CH5+}) + T(\ce{CH5+}).
    \label{equation:evolution_CH5+}
\end{empheq}
Here, $P$, $D$ and $T$ are the production, loss and transport rates, respectively. We assumed that the atmosphere was in photochemical equilibrium, implying that the transport terms $T(\ce{H3+})$ and $T(\ce{CH5+})$ are equal to zero, and that the ion densities are at steady-states, giving $\odv{\left[\ce{H3+}\right]}{t} = \odv{\left[\ce{CH5+}\right]}{t} = 0$. The $\ce{H3+}$ and $\ce{H2+}$ production rates are almost equal since $\ce{H2+}$ rapidly reacts with $\ce{H2}$ and we write $P(\ce{H3+}) \approx P(\ce{H2+}) \equiv q$. Typical vertical profiles of $q$ for the two kinds of distribution considered in this study are plotted in Fig. \ref{fig:ph2plus}. The equations \ref{equation:evolution_H3+} and \ref{equation:evolution_CH5+} become
\begin{empheq}{align}
    0 &= q - \alpha_{\ce{H3+}} \, \left[\ce{H3+}\right] \, \left[\ce{e-}\right] - k \, \left[\ce{H3+}\right] \, \left[\ce{CH4}\right],\\
    0 &= k \, \left[\ce{H3+}\right] \, \left[\ce{CH4}\right] - \alpha_{\ce{CH5+}} \, \left[\ce{CH5+}\right] \, \left[\ce{e-}\right].
\end{empheq}
We assume charge equilibrium $\left[\ce{e-}\right] = \left[\ce{H3+}\right] + \left[\ce{CH5+}\right]$ and inject it in the above equations. We obtain a set of two equations with the two ion densities as unknowns
\begin{empheq}{align}
    0 &= q - \alpha_{\ce{H3+}} \left[\ce{H3+}\right] \, \Big(\left[\ce{H3+}\right] + \left[\ce{CH5+}\right]\Big) - k \, \left[\ce{H3+}\right]\left[\ce{CH4}\right],
    \label{equ:appendix_a_1}\\
    0 &= k \, \left[\ce{H3+}\right]\left[\ce{CH4}\right] - \alpha_{\ce{CH5+}} \left[\ce{CH5+}\right] \, \Big(\left[\ce{H3+}\right] + \left[\ce{CH5+}\right]\Big).
    \label{equ:appendix_a_2}
\end{empheq}
To analytically solve this system, we make the additional assumption that the loss rate coefficients $\alpha_{\ce{H3+}}$ and $\alpha_{\ce{CH5+}}$ are equal: $\alpha_{\ce{H3+}} = \alpha_{\ce{CH5+}} =\alpha$. By choosing the value of $\alpha_{\ce{H3+}}$ ($\alpha_{\ce{CH5+}}$) for $\alpha$, we shift the equilibrium towards more (fewer) ions in the atmosphere and compute an upper (lower) limit for the conductance. For a chosen electron mean energy and total energy flux, the conductance is taken as the mean value of the corresponding upper and lower boundary values. The boundary and mean values are illustrated in Fig. \ref{fig:appendix:a_boundary} for the Pedersen conductance computed using a kappa electron energy distribution. Consequently, the system of equations \eqref{equ:appendix_a_1}-\eqref{equ:appendix_a_2} yields the following expressions for the ion densities
\begin{empheq}{align}
    \left[\ce{CH5+}\right] &= \frac{k \, \left[\ce{CH4}\right] \sqrt{\frac{q}{\alpha}}}{k \, \left[\ce{CH4}\right] + \alpha \sqrt{\frac{q}{\alpha}}}, \\
    \left[\ce{H3+}\right] &= \sqrt{\frac{q}{\alpha}} \, - \, \left[\ce{CH5+}\right].
\end{empheq}
As an illustration, the density vertical profiles of these ions are plotted on Fig. \ref{fig:dtyions2} for $\langle E \rangle$ = 10 keV. 

\section{The conductive layer} %Second appendix
\label{appendix:b}

\begin{figure} % One-column figure
        \centering
        \includegraphics[scale=0.6]{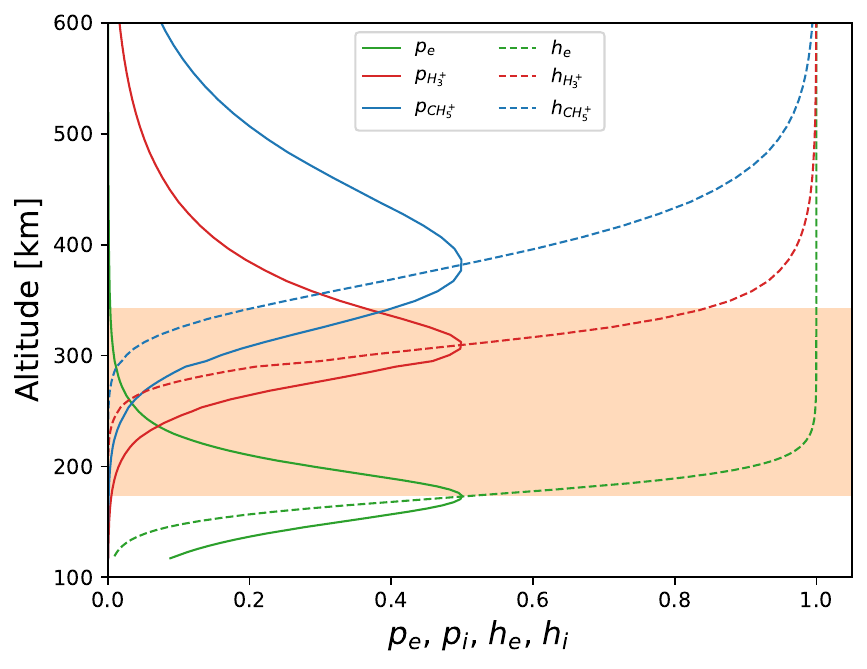}
        \caption{Vertical profiles of $p_e$, $p_i$, $h_e$ and $h_i$ for $B=10^{-3}$ T. The orange rectangle represents the location of the conductive layer.}
        \label{fig:ctygenVSalt}
    \end{figure}

    The electron and ion dynamics determine the altitudes where the Pedersen and Hall conductivities are likely to be the highest. We can highlight this dynamics in the expressions of the conductivities \eqref{equation:ctyp} and \eqref{equation:ctyh} by grouping together the ion (electron) collision frequencies with the neutrals and the ion (electron) gyrofrequencies
    {\small
    \begin{empheq}{align}
    \small
        \nsigma_{\!P} &= \frac{e}{B}
        \left(\, n_e\,
        \red{\frac{- \, \omega_e \, \nu_{en}}{\nu^2_{en} + \omega^2_e}}
        +
        \sum_i n_i \, \red{\frac{\omega_i \, \nu_{in}}{\nu^2_{in} + \omega^2_i}}
        \right)
        =
        \frac{e}{B}
        \left(n_e \, \red{p_e}
        +
        \sum_i n_i \, \red{p_i}
        \right),
        \label{equation:ctyp2}\\
        \nsigma_{\!H} &= \frac{e}{B}
        \left(n_e \,
        \red{\frac{\omega^2_e}{\nu^2_{en} + \omega^2_e}}
        -
        \sum_i n_i \, \red{\frac{\omega^2_i}{\nu^2_{in} + \omega^2_i}}
        \right)
        =
        \frac{e}{B}
        \left(n_e \, \red{h_e}
        -
        \sum_i n_i \, \red{h_i}
        \right).
        \label{equation:ctyh2}
    \end{empheq}
    }
    The vertical profiles of $p_e$, $p_i$, $h_e$ and $h_i$, plotted in Fig. \ref{fig:ctygenVSalt} for $B = 10^{-3}$ T, set the altitudes where the Pedersen and Hall conductivities are likely to maximise. At high altitude, there are almost no collision. We have then $\nu_{en} = \nu_{in} \approx 0$ leading to $p_e = p_i \approx 0$ and $h_e = h_i \approx 1$. As a result, there is almost no conductivity. As we go deeper into the atmosphere, the increase of density results in more collisions, leading to an increase of $\nu_{en}$ and $\nu_{in}$ which modify the values of $p_e$, $p_i$, $h_e$ and $h_i$. The vertical distributions of $p_e$ and $p_i$ present a peak value at the altitudes where $\nu_{en} = \omega_e$ and $\nu_{in} = \omega_i$. Thus, the Pedersen conductivity is likely to maximize around these altitudes. Concerning the vertical distributions of $h_e$ and $h_i$, they drop at different altitudes. Since the electrons are much lighter than the ions, we have $\omega_e >> \omega_i$. As a result, $h_e$ decreases at a deeper altitude than $h_i$. Due to the presence of the minus sign in equation \eqref{equation:ctyh2}, the Hall conductivity is then likely to maximize in the altitude region where $h_e \approx 1$ and $h_i \approx 0$ (i.e., $\nu_{en} << \nu_{in}$).
    
    The ionosphere layer present between the altitudes where $\nu_{in}=\omega_i$ and $\nu_{en}=\omega_e$ is often called the conductive layer \citep{nakamura_effect_2022, clement_ionospheric_2025}. Since we consider two kinds of ions (\ce{H3+} and \ce{CH5+}), we define the altitude of the conductive layer upper boundary halfway between the altitudes where we have $\nu_{CH5+} = \omega_{CH5+}$ and $\nu_{H3+} = \omega_{H3+}$ (see Fig. \ref{fig:ctygenVSalt}). The Hall and Pedersen conductivities are likely to maximize inside the conductive layer and around its boundaries, respectively.
    
\begin{figure} % One-column figure
    \centering
    \captionsetup[subfigure]{width=0.5\linewidth}
        \subfloat[ $B = 10^{-2}$ T.]
        {\includegraphics[width=0.5\linewidth]{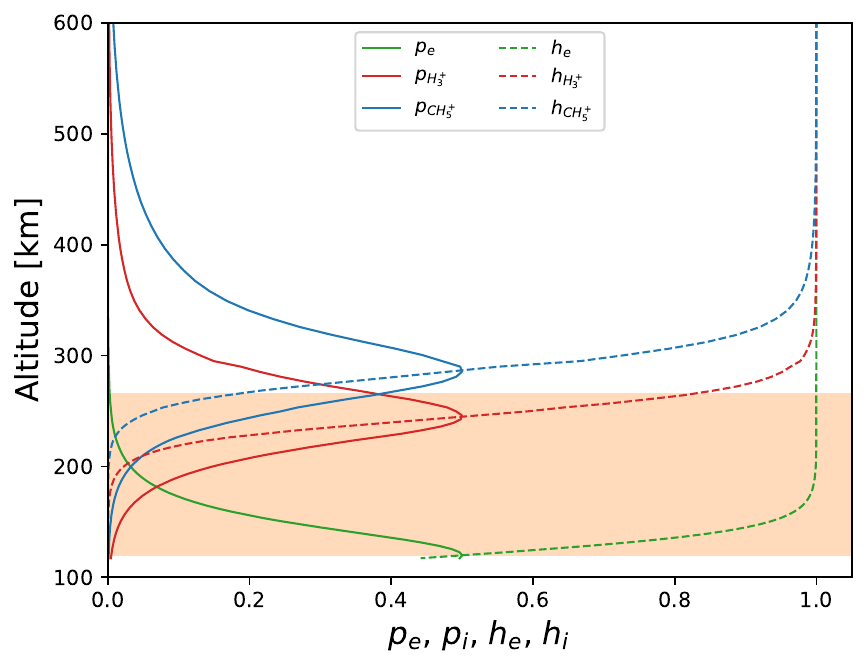}
        }
        \subfloat[ $B = 10^{-4}$ T.]
        {\includegraphics[width=0.5\linewidth]{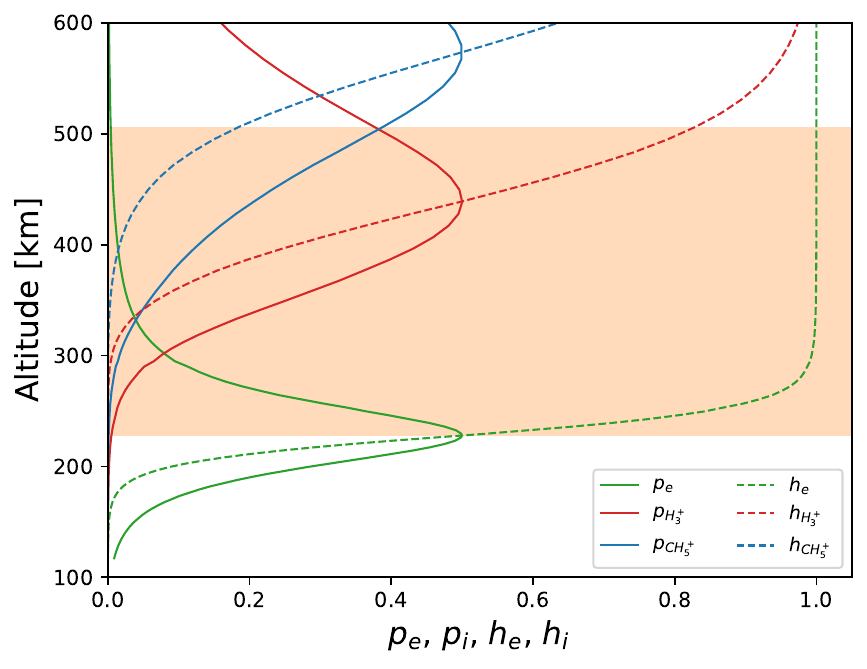}
        }
    \caption{Effect of the magnetic field intensity on the vertical profiles of $p_e$, $p_i$, $h_e$ and $h_i$. The orange rectangle represents the location of the conductive layer.}
    \label{fig:ctygenVSaltBmodified}
    \end{figure}

\begin{figure} % One-column figure
        \centering
        \includegraphics[scale=0.6]{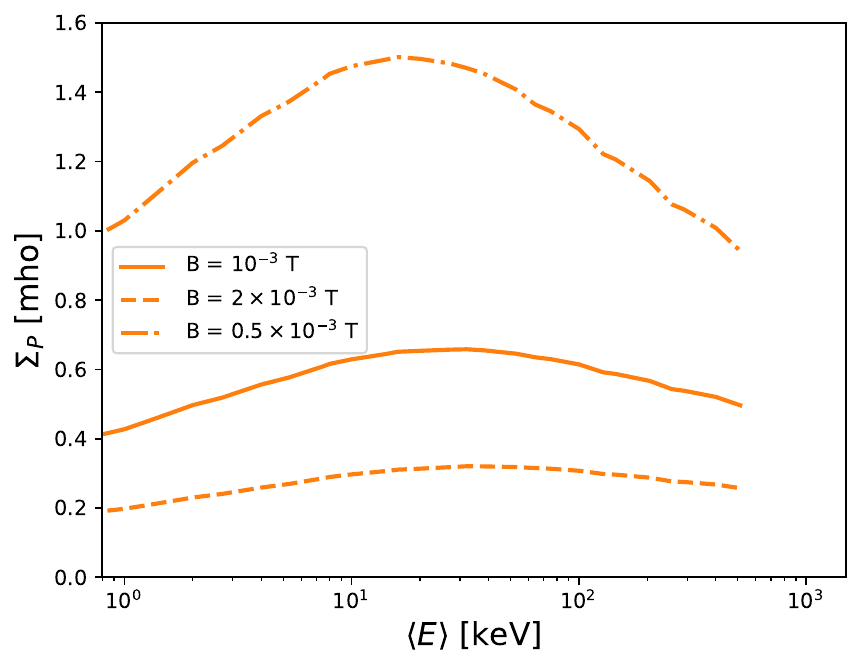}
        \caption{Evolutions of the Pedersen conductance $\Sigma_P$ with the electron mean energy $\langle E \rangle$ for different values of the the magnetic field intensity $B$. In addition to a change in the magnitude of $\Sigma_P$, the mean energy value that maximizes conductance slightly shifts toward lower (higher) values with an increase (decrease) of $B$.}
        \label{fig:cpVSmean_variousBmag}
    \end{figure}

    The conductive layer location depends on the magnetic field intensity \citep{nakamura_effect_2022}. We illustrated the effect of a change in the magnetic field intensity on the factors $p_e$, $p_i$, $h_e$ and $h_i$ present in the expressions \eqref{equation:ctyp2} and \eqref{equation:ctyh2}. On Fig. \ref{fig:ctygenVSaltBmodified}, we plotted their vertical profiles with different magnetic field intensity. We see that an increase (decrease) of the magnetic field intensity shifts the peaks of the vertical profiles towards lower (higher) altitudes and contributes to move down (up) the conductive layer. As a consequence, a change in the magnetic field intensity changes the evolution of the conductance with the electron mean energy, as illustrated in Fig. \ref{fig:cpVSmean_variousBmag} for the Pedersen conductance. As shown in \citet{gerard_spatial_2020}, this effect has a greater impact on the conductances in the northern hemisphere due to the magnetic anomaly.

\section{Impact of a change in the $\ce{CH4}$ density vertical profile} % Third appendix
\label{appendix:cbis}
    \begin{figure*} % Two-column figure
    \centering
    \captionsetup[subfigure]{width=1\linewidth}
        \subfloat[ $\ce{CH4}$ profiles.]{\includegraphics[width=0.25\linewidth]{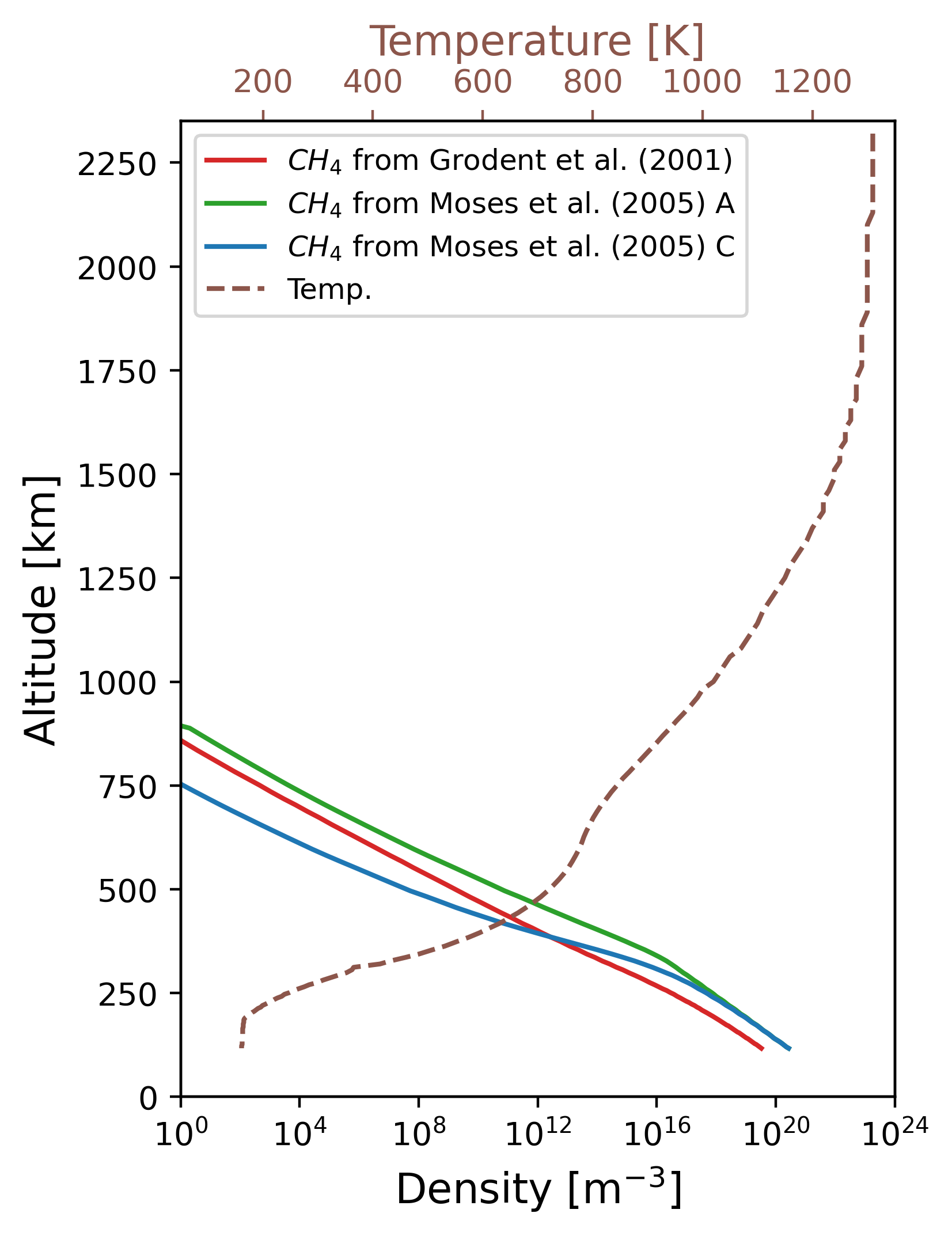}
        \label{fig:appendix_cbis_ch4}
        }
        \subfloat[ Pedersen.]{\includegraphics[width=0.36\linewidth]{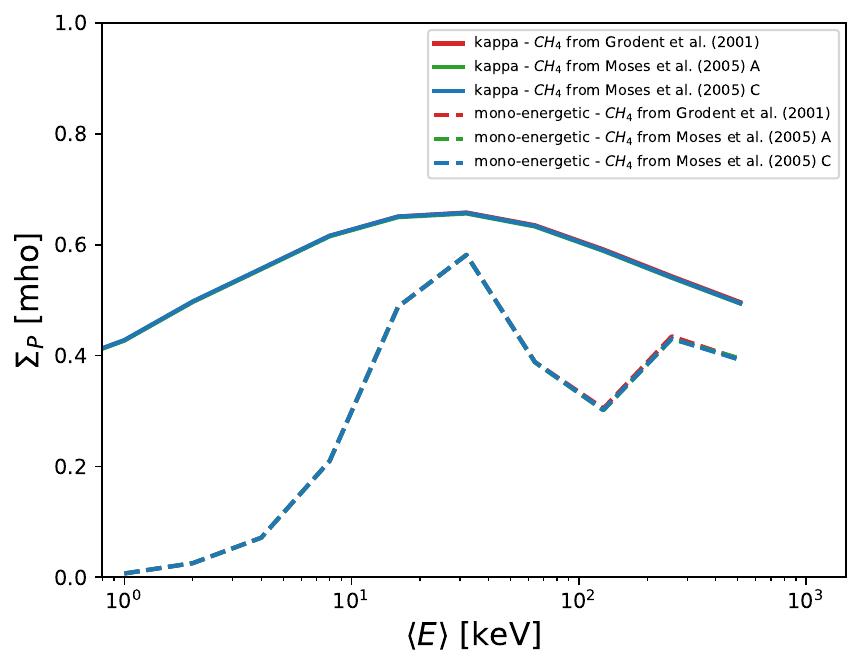}
         \label{fig:appendix_cbis_conductP}
        }
        \subfloat[ Hall.]{\includegraphics[width=0.36\linewidth]{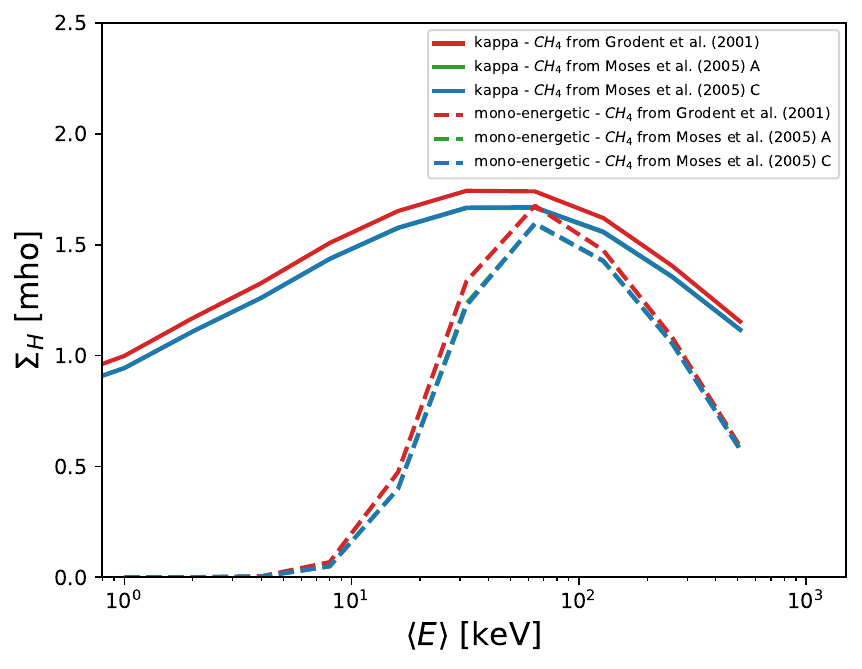}
        \label{fig:appendix_cbis_conductH}
        }
        
    \caption{(a) Different $\ce{CH4}$ density vertical profiles. The temperature is taken from \citet{grodent_self-consistent_2001}. (b)-(c) Evolutions of the Pedersen and Hall conductances with the electron mean energy computed with the $\ce{CH4}$ vertical profiles plotted in (a).}
    \end{figure*}
    
    We evaluated the impact of a change in the $\ce{CH4}$ density profile on the Pedersen and Hall conductances. In Fig. \ref{fig:appendix_cbis_ch4} are plotted the $\ce{CH4}$ abundance profile used in this study as well as two other profiles computed from the $A$ and $C$ eddy diffusion models of \citet{moses_photochemistry_2005} and \citet{hue_photochemistry_2018}. The $\ce{CH4}$ profiles mainly differ by their homopause level. 
    
    The evolution of the Pedersen and Hall conductances, computed using either a kappa or a mono-energetic electron energy distribution, as a function of the electron mean energy, is plotted in Fig. \ref{fig:appendix_cbis_conductP}     and \ref{fig:appendix_cbis_conductH} for the different $\ce{CH4}$ density profiles considered. We see that the Pedersen conductance is almost insensitive to a change in the $\ce{CH4}$ profiles. Concerning the Hall conductance, there is a difference of maximum 0.1 mho between the conductance computed with the $\ce{CH4}$ profile from \citet{grodent_self-consistent_2001} and the other profiles.

\section{Influence of the electron transport model} % Fourth appendix
\label{appendix:ctris}
    The electron transport model used to determine the vertical distribution of charged particles resulting from the auroral electron precipitation, may play a role in the computation of the Pedersen and Hall conductances. In Fig. \label{fig:appendix:ctris} are plotted the evolutions of the mono-energetic Pedersen conductance with the electron mean energy, computed with TransPlanet, a Monte-Carlo simulation and the analytical expression taken from \citet{hiraki_parameterization_2008}, built on the results of a Monte-Carlo simulation. For a part of the results, we also modified the ionization cross-sections of $\ce{H2}$, which are among the most significant cross-sections, to see the impact on the conductance. The cross-sections used in the Monte-Carlo simulation were identical as the ones used to compute the results of this study with TransPlanet. We also computed the conductance values using TransPlanet in which we incorporated the ionization cross-sections of $\ce{H2}$ from \citet{hiraki_parameterization_2008}. To better compare our results, the conductance values were computed assuming an atmosphere only composed of $\ce{H2}$.

    The likeliness between the red and green curves in Fig. \label{fig:appendix:ctris} leads to the comforting conclusion that different types of different electron transport model give similar results when they use similar electron collision cross-sections. In addition, the noticeable differences beyond 10 keV between the green and orange curves mean that a change in the collision cross-section, in particular the ionization cross-sections of $\ce{H2}$, has a significant impact on the computation of the conductance. Finally, the fact that the orange and blue curves diverge seems to indicate that the analytical expression from \citet{hiraki_parameterization_2008} has to be used with caution.

    \begin{figure} % One-column figure
        \centering
        \includegraphics[scale=0.6]{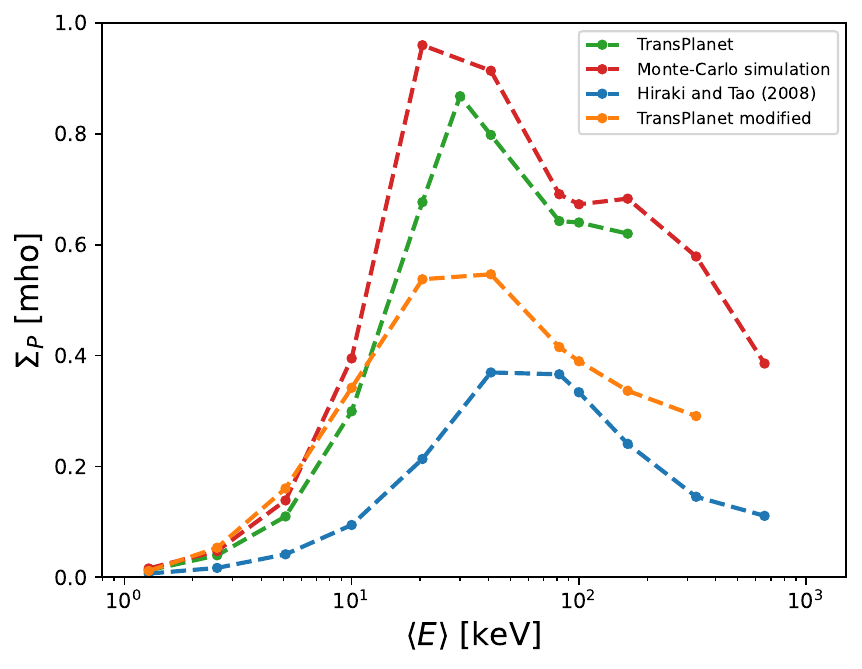}
        \caption{Evolution of the mono-energetic Pedersen conductance computed with TransPlanet (green and orange curves), a Monte-Carlo simulation (red curve) and the analytical expression from \citet{hiraki_parameterization_2008} giving the ion production rate (blue curve). The green and red curves are computed with the same ionization cross-sections of $\ce{H2}$, which are the cross-sections used in this study, and the blue and orange curves are computed with the cross-sections from \citet{hiraki_parameterization_2008}.}
        \label{fig:appendix:ctris}
    \end{figure}

\section{Additional maps} % Fifth appendix
\label{appendix:c}

Here, we present the conductance maps for all the PJs considered in this study.

    \begin{figure} % One-column figure
    \centering
    \captionsetup[subfigure]{width=1\textwidth}
        \subfloat[ $\Sigma_P$ - PJ1 north.]
        {\includegraphics[width=0.62\linewidth]{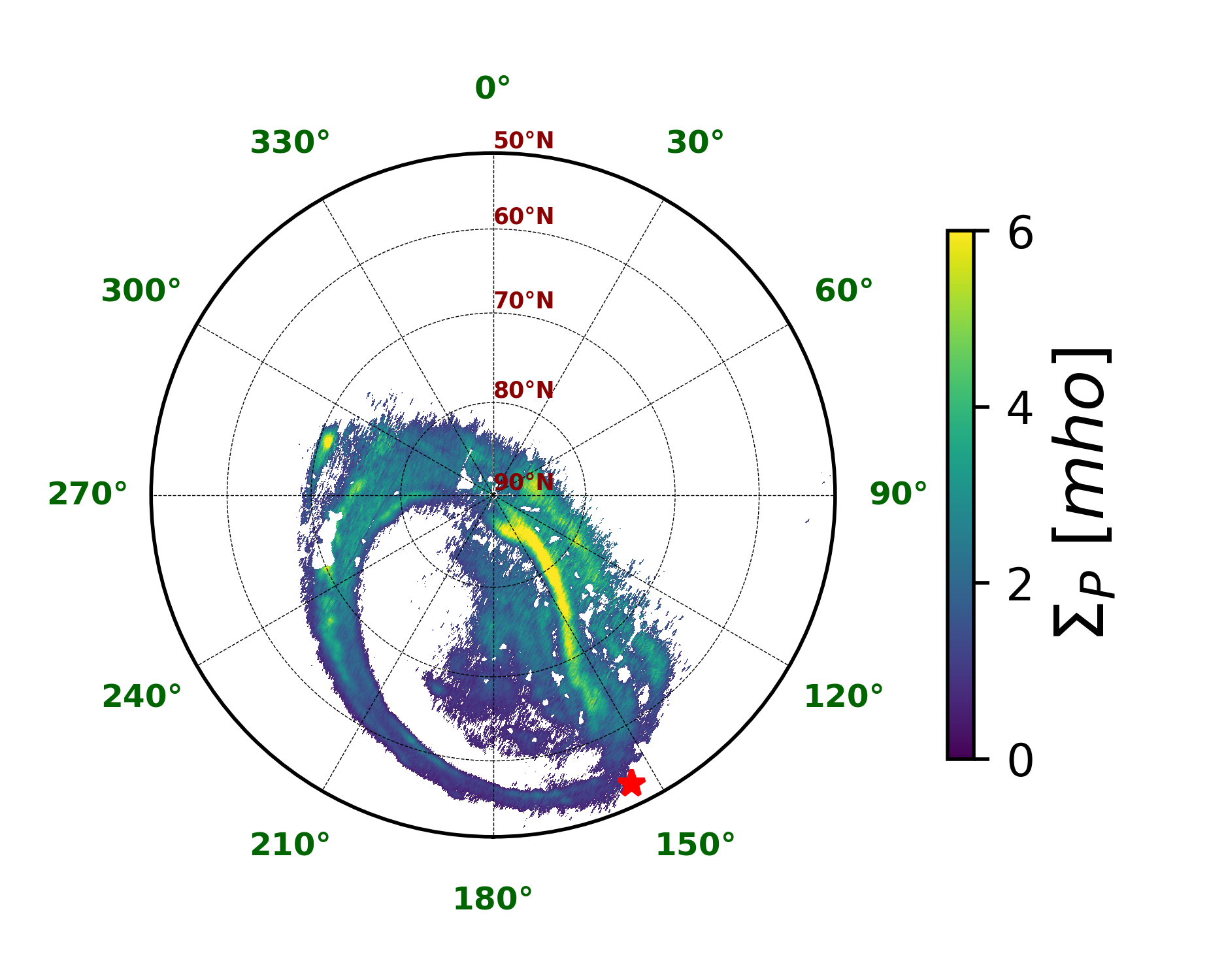}
        }
        
        \subfloat[ $\Sigma_H$ - PJ1 north.]
        {\includegraphics[width=0.62\linewidth]{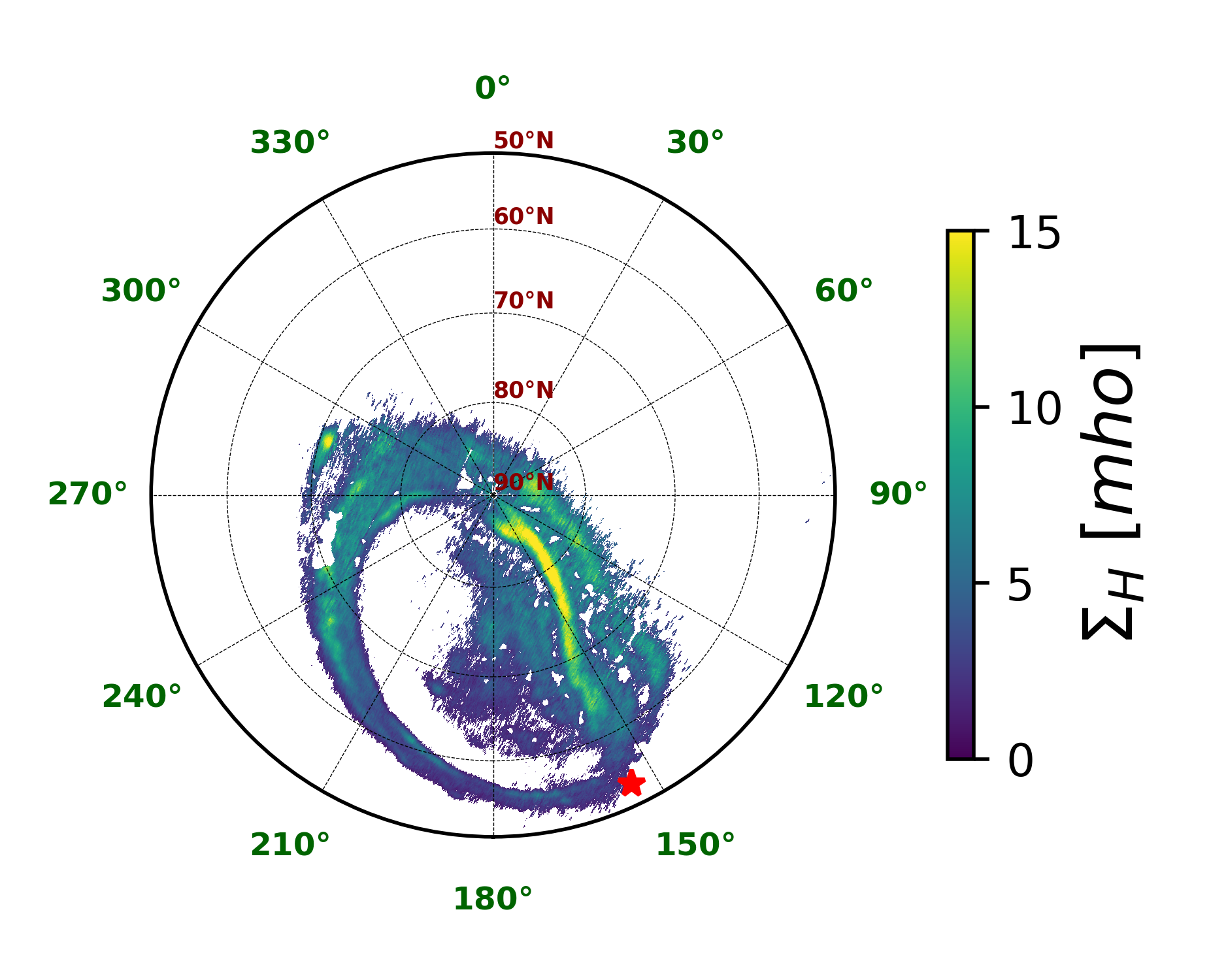}
        }
        
        \subfloat[ Ratio P - PJ1 north.]
        {\includegraphics[width=0.62\linewidth]{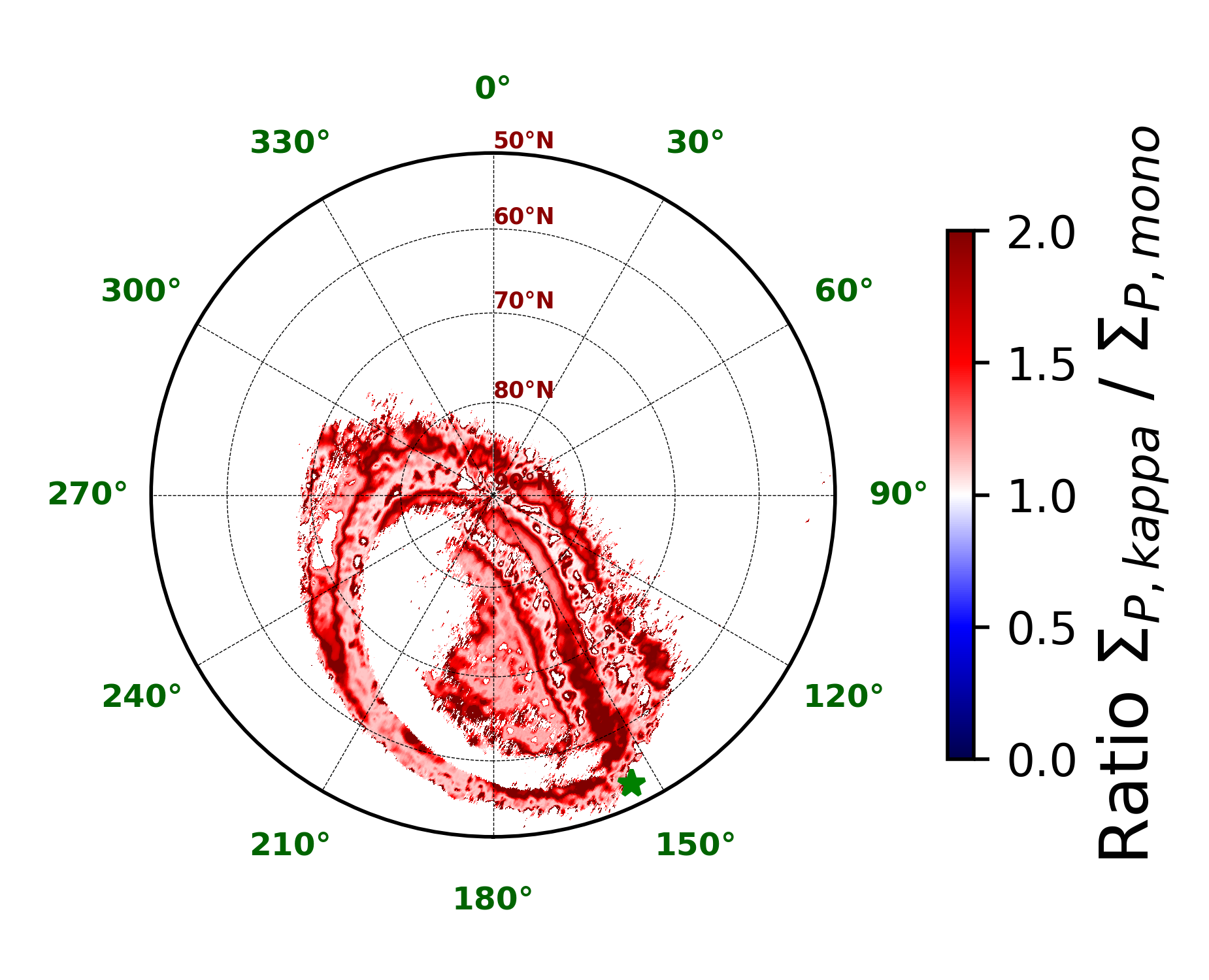}
        }
        
        \subfloat[ Ratio H - PJ1 north.]
        {\includegraphics[width=0.62\linewidth]{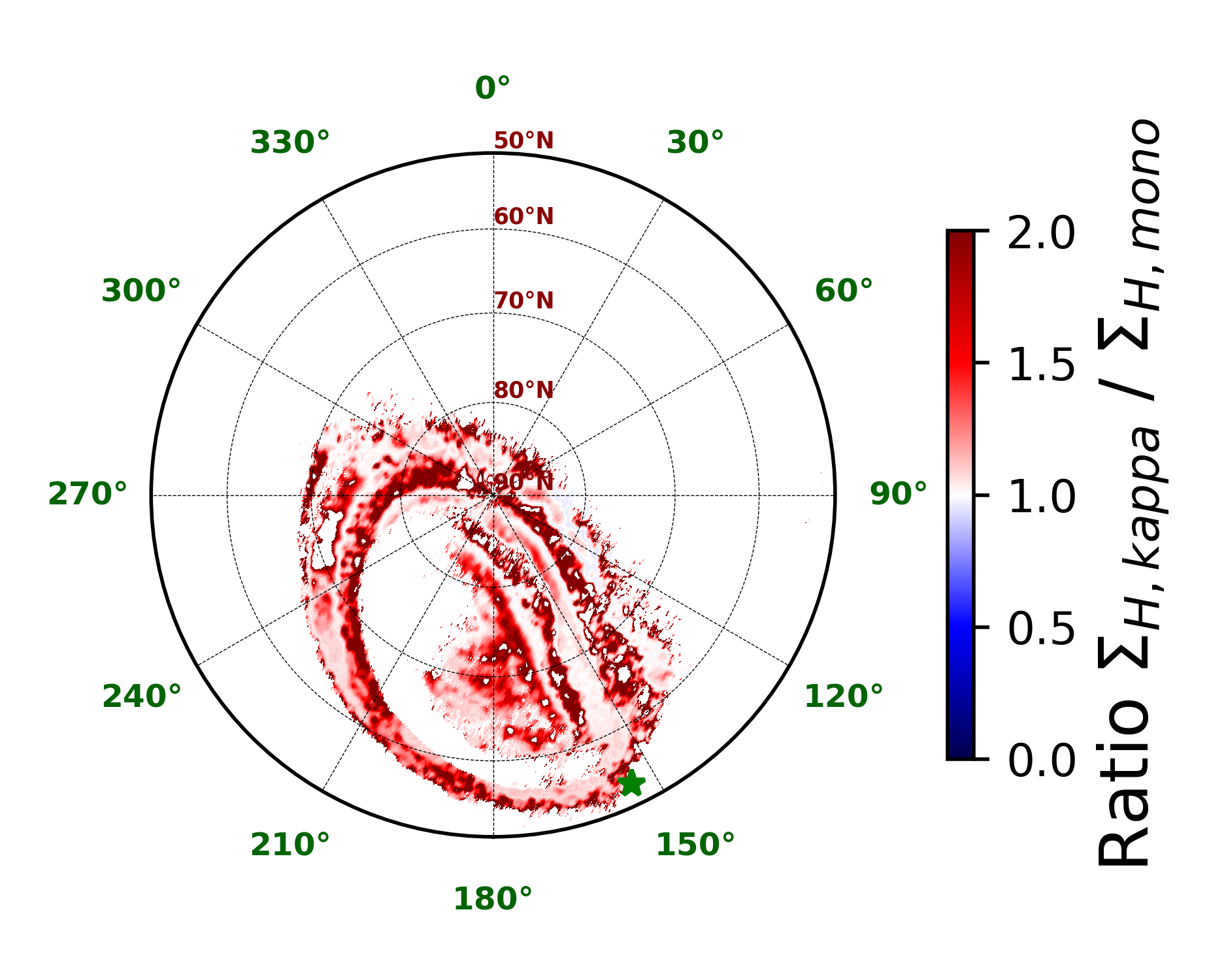}
        }
    \caption{
    Conductance maps of the northern auroral region during PJ1. (a) and (b) represent the Pedersen and Hall conductance maps, respectively. The ratios of the conductance map computed with a kappa electron energy distribution over the conductance map computed with a mono-energetic electron energy distribution are also represented for the (c) Pedersen and (d) Hall conductances. For the ratio maps, the scale goes up to 2 but the ratio can be locally higher. On each map, the red or green star represents the subsolar longitude.
    }
    \label{fig:appendix_maps_conductnorth_pj1}
    \end{figure}
    
    \begin{figure} % One-column figure
    \centering
    \captionsetup[subfigure]{width=1\textwidth}
        \subfloat[ $\Sigma_P$ - PJ6 north.]
        {\includegraphics[width=0.7\linewidth]{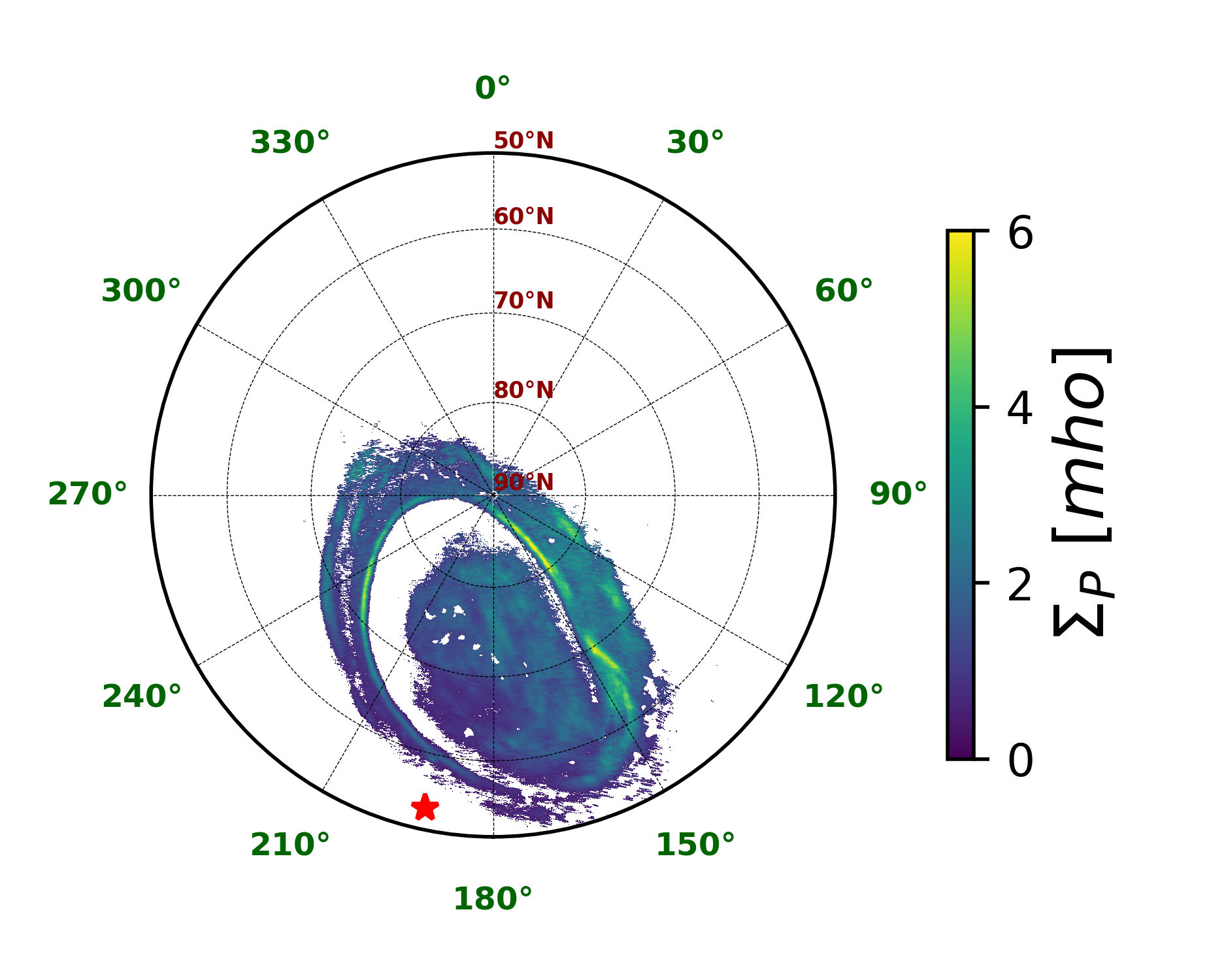}
        }
        
        \subfloat[ $\Sigma_H$ - PJ6 north.]
        {\includegraphics[width=0.7\linewidth]{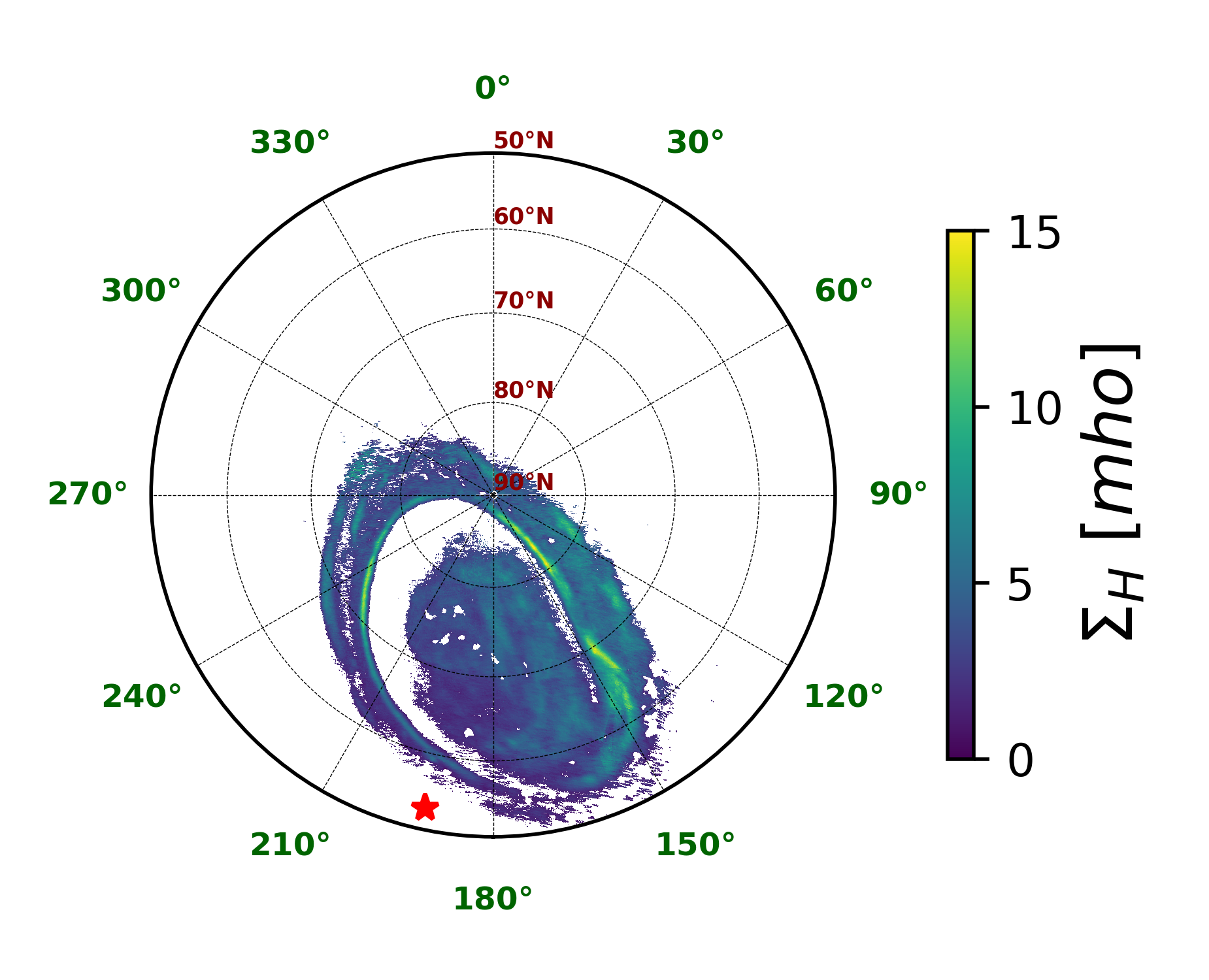}
        }
        
        \subfloat[ Ratio P - PJ6 north.]
        {\includegraphics[width=0.7\linewidth]{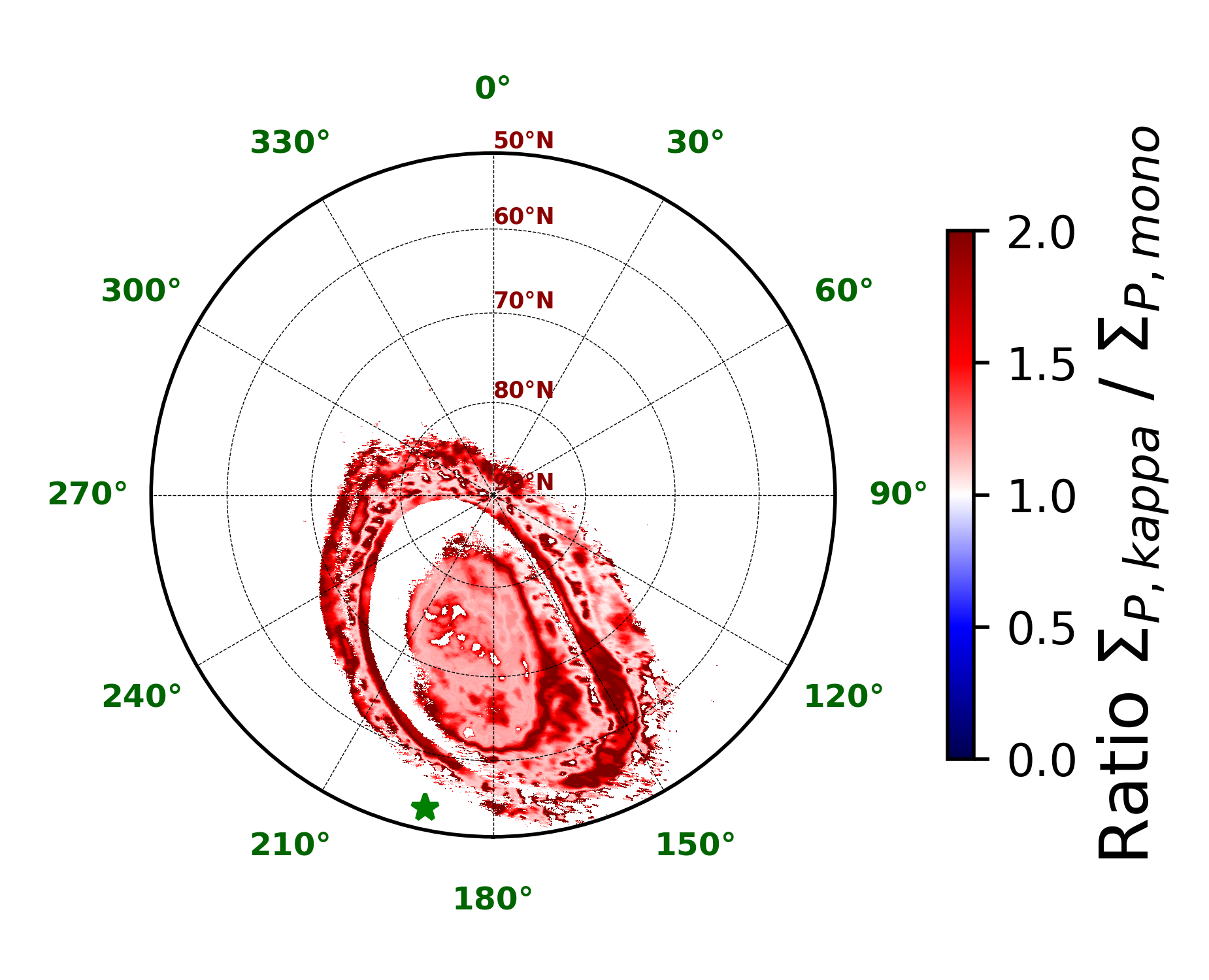}
        }
        
        \subfloat[ Ratio H - PJ6 north.]
        {\includegraphics[width=0.7\linewidth]{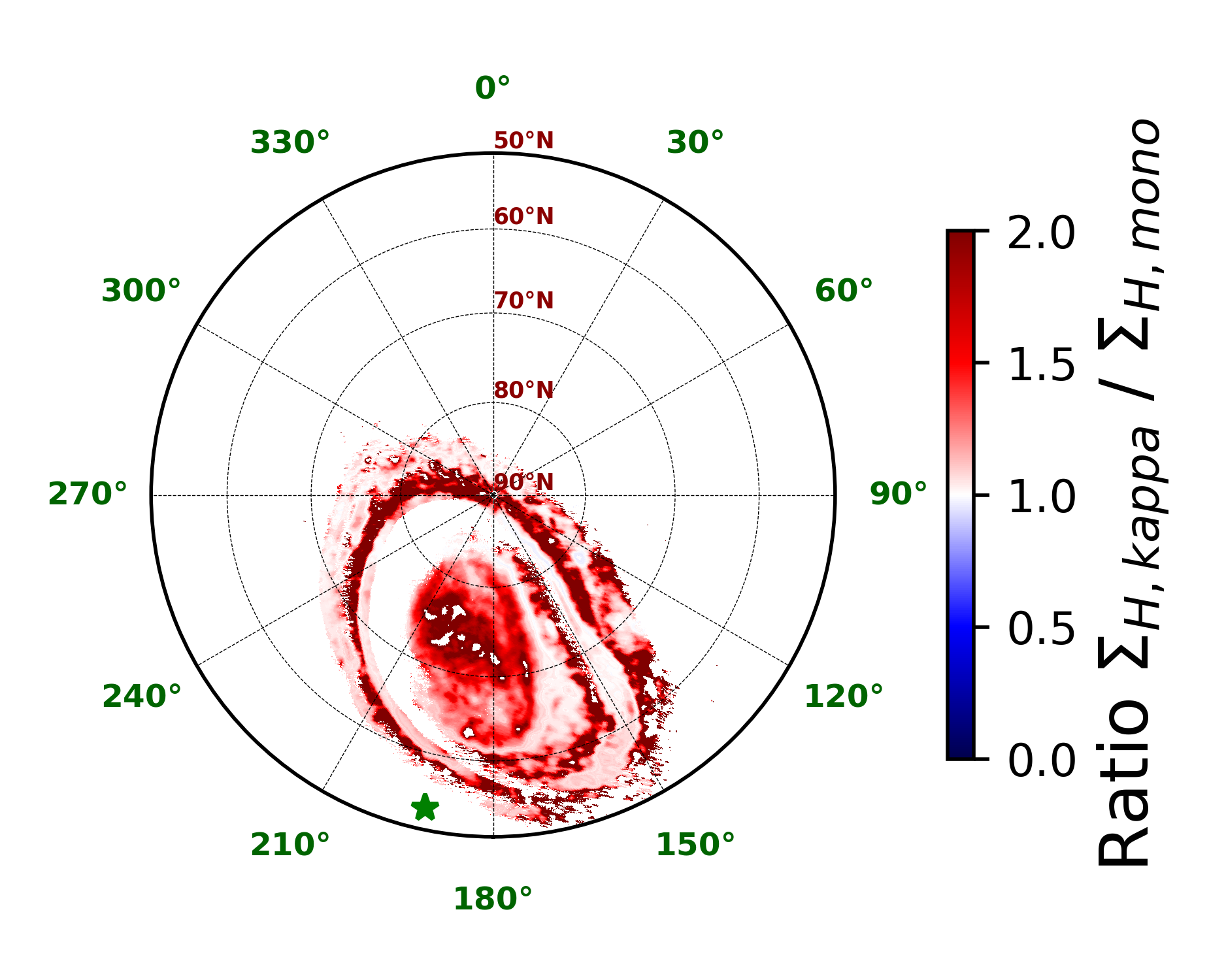}
        }
    \caption{
    Same description as Fig. \ref{fig:appendix_maps_conductnorth_pj1} for PJ6 north.
    }
    \label{fig:appendix_maps_conductnorth_pj6}
    \end{figure}

    \begin{figure} % One-column figure
    \centering
    \captionsetup[subfigure]{width=1\textwidth}
        \subfloat[ $\Sigma_P$ - PJ12 north.]
        {\includegraphics[width=0.7\linewidth]{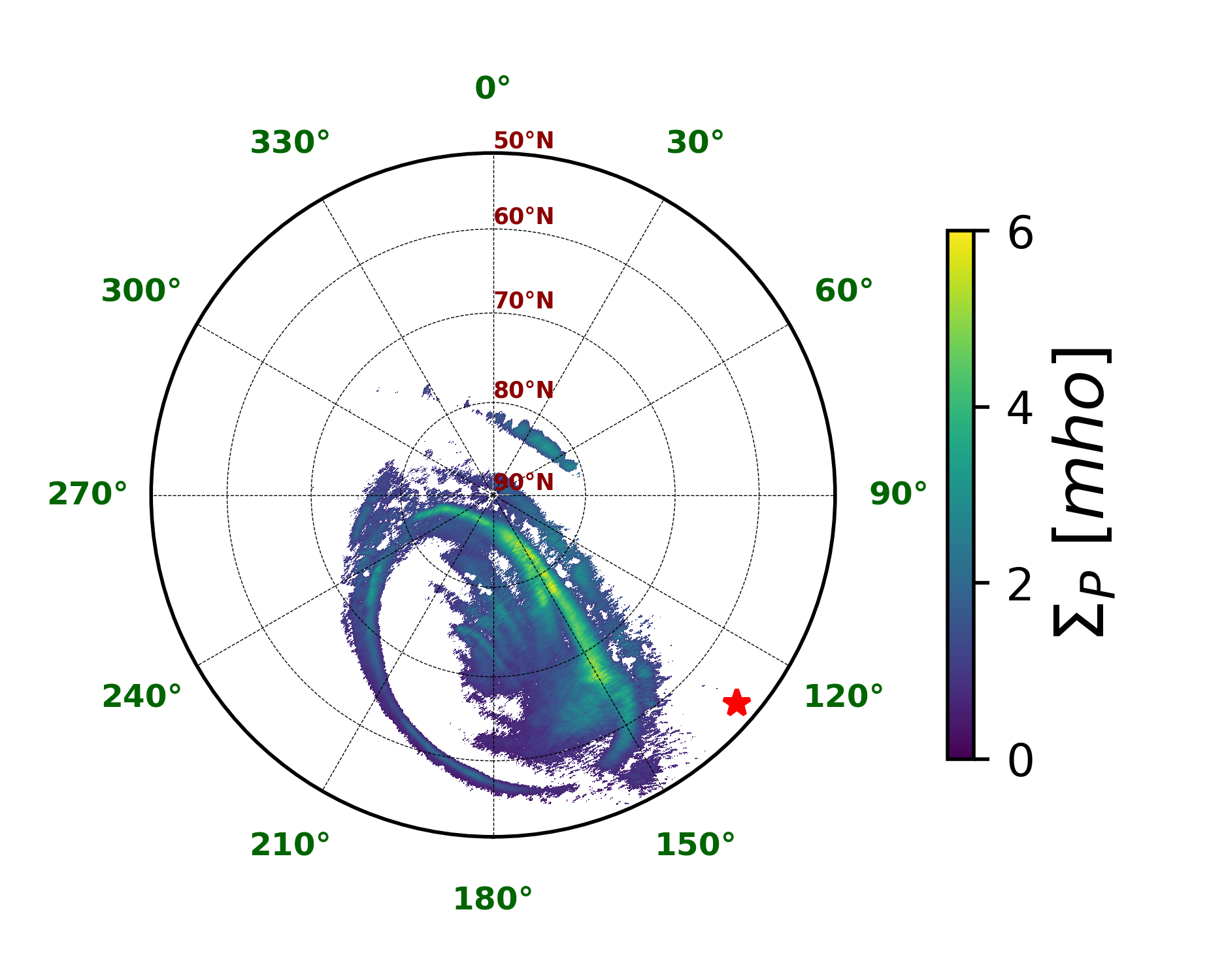}
        }
        
        \subfloat[ $\Sigma_H$ - PJ12 north.]
        {\includegraphics[width=0.7\linewidth]{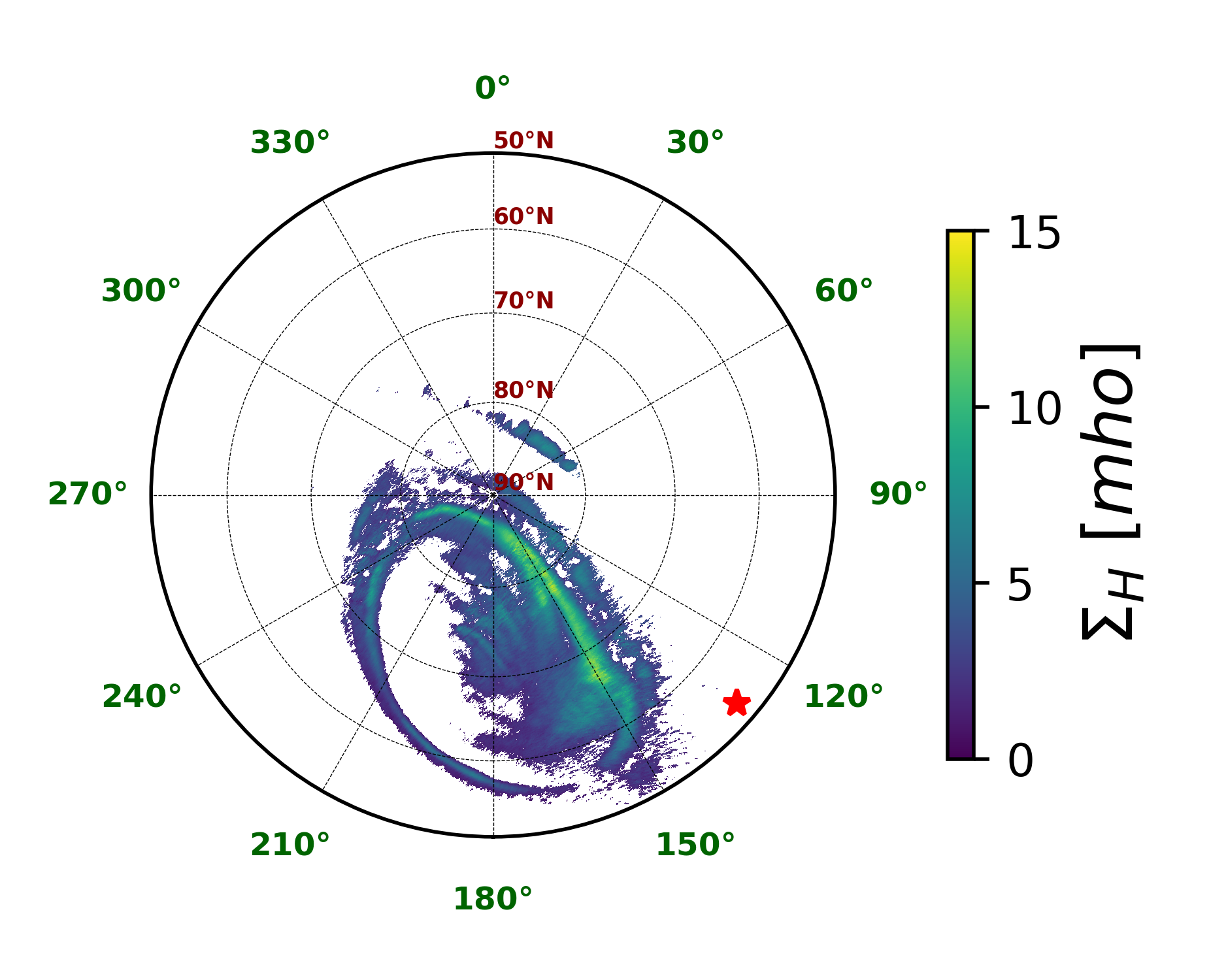}
        }
        
        \subfloat[ Ratio P - PJ12 north.]
        {\includegraphics[width=0.7\linewidth]{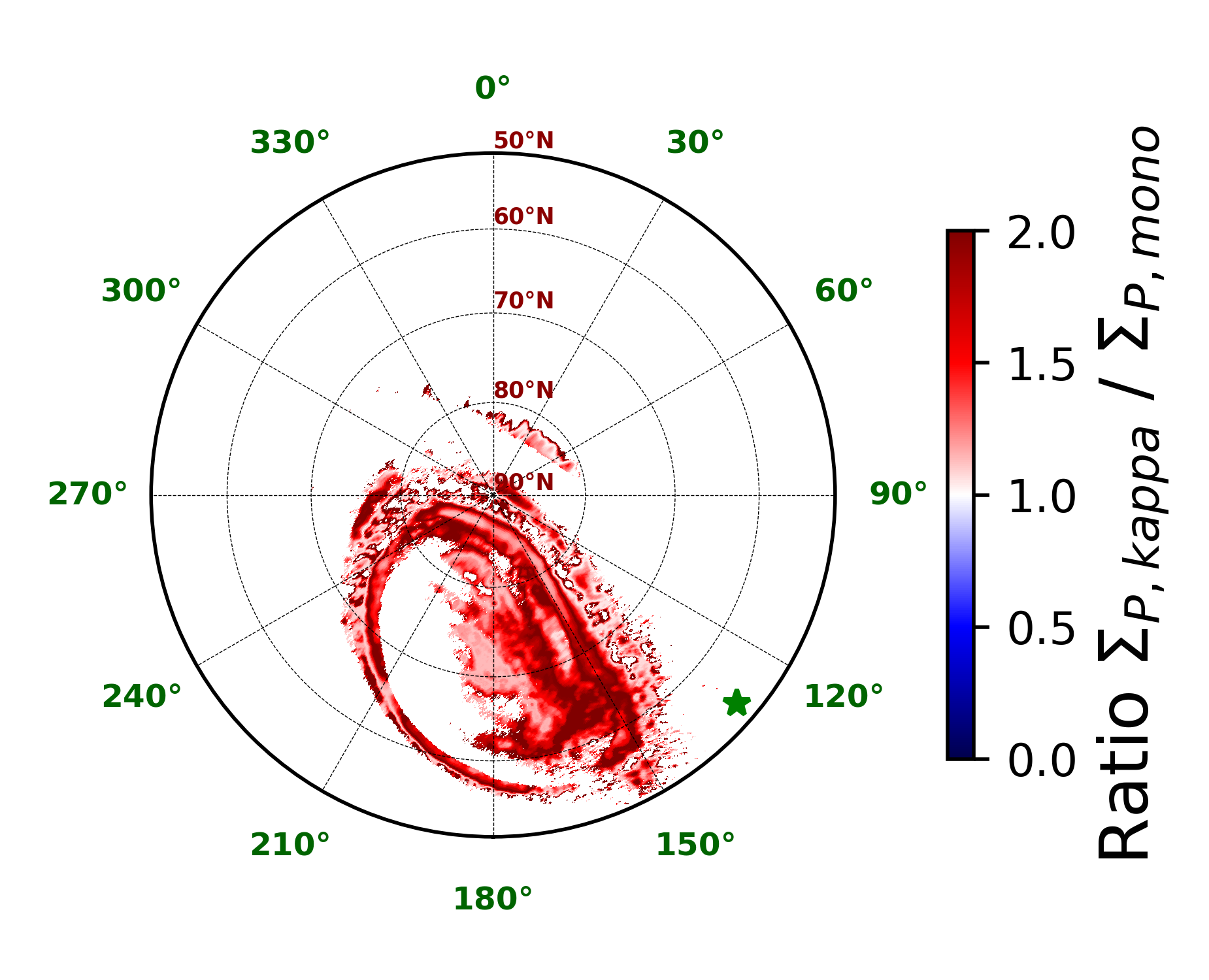}
        }
        
        \subfloat[ Ratio H - PJ12 north.]
        {\includegraphics[width=0.7\linewidth]{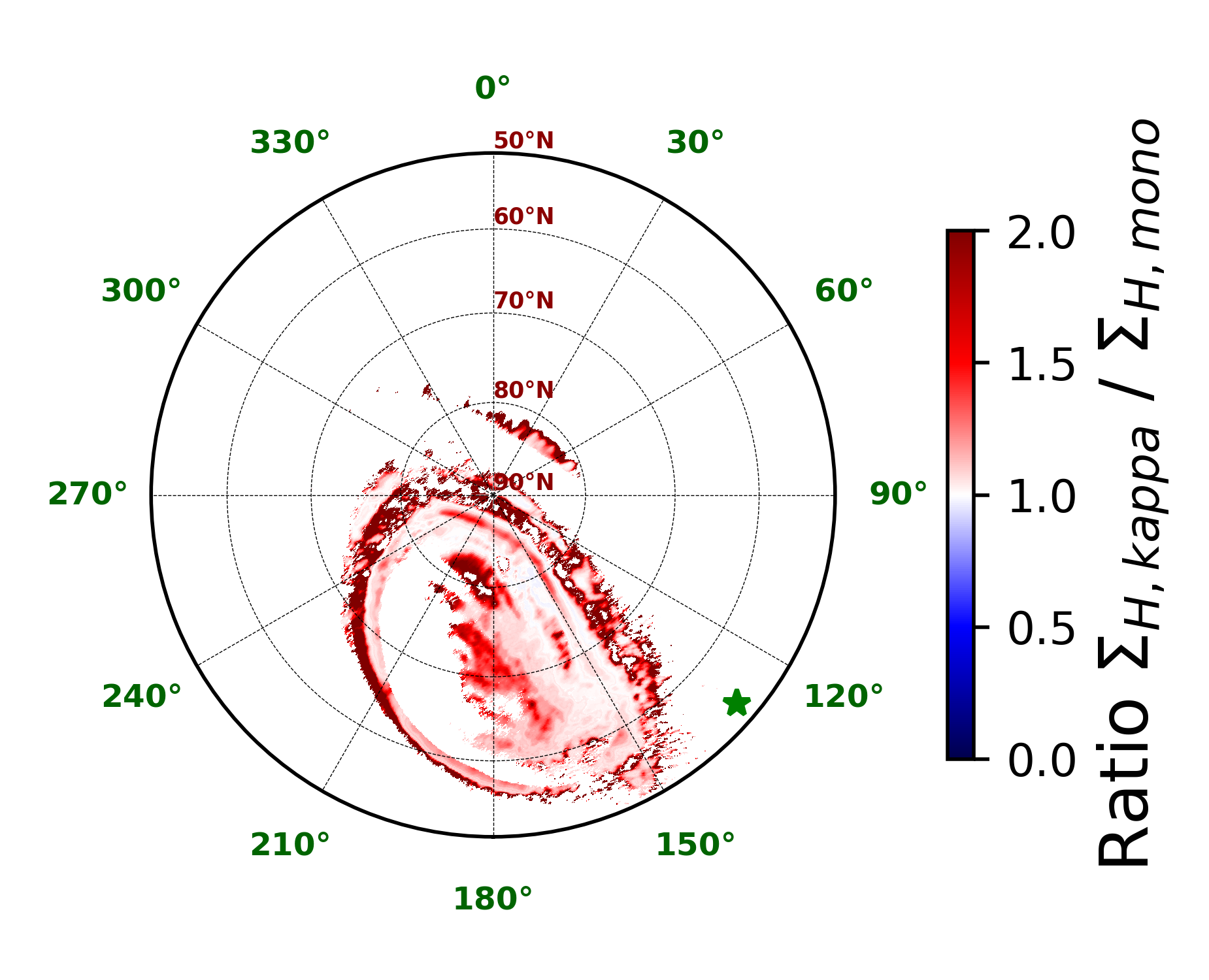}
        }
    \caption{
    Same description as Fig. \ref{fig:appendix_maps_conductnorth_pj1} for PJ12 north.
    }
    \label{fig:appendix_maps_conductnorth_pj12}
    \end{figure}

    \begin{figure} % One-column figure
    \centering
    \captionsetup[subfigure]{width=1\textwidth}
        \subfloat[ $\Sigma_P$ - PJ22 north.]
        {\includegraphics[width=0.7\linewidth]{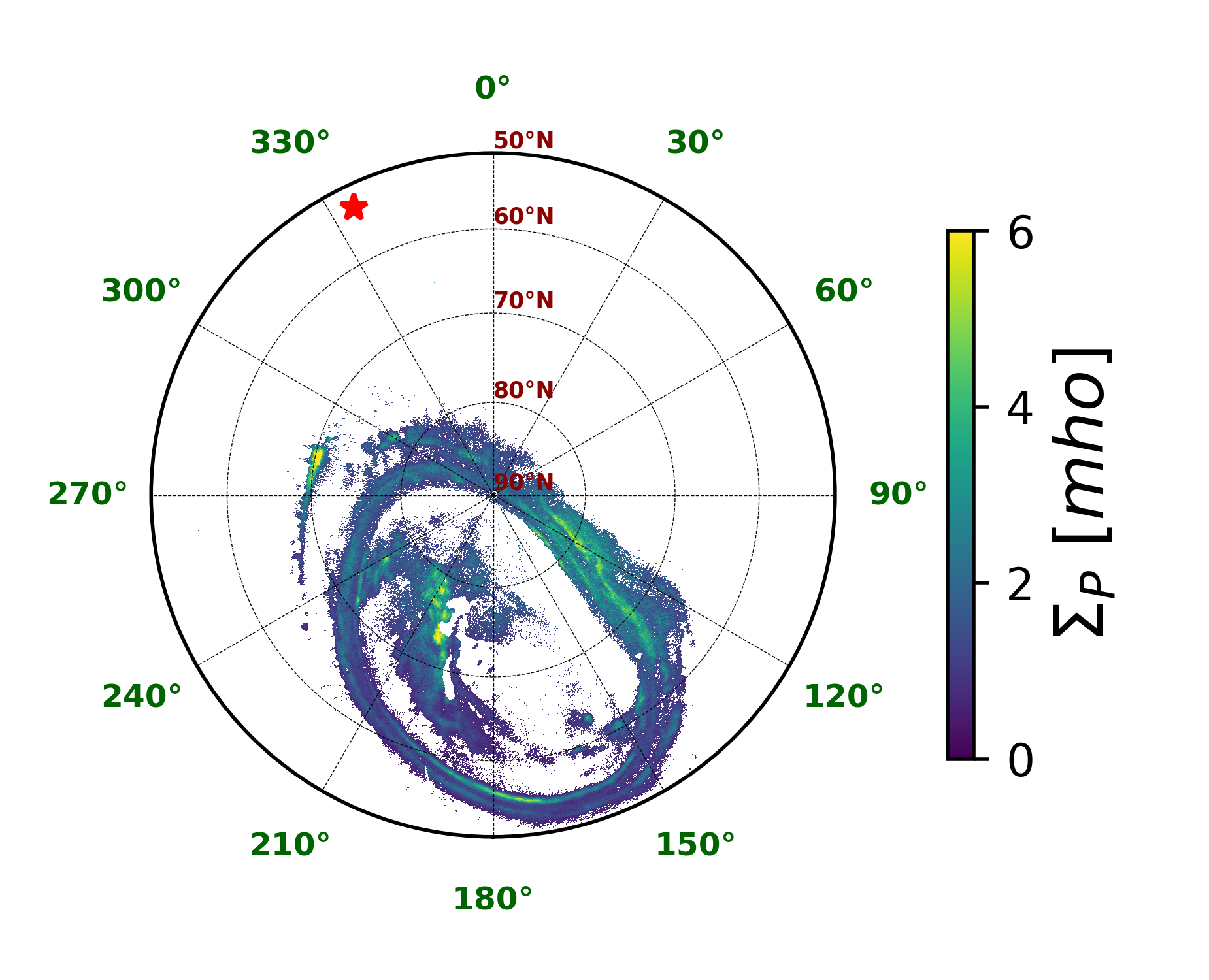}
        }
        
        \subfloat[ $\Sigma_H$ - PJ22 north.]
        {\includegraphics[width=0.7\linewidth]{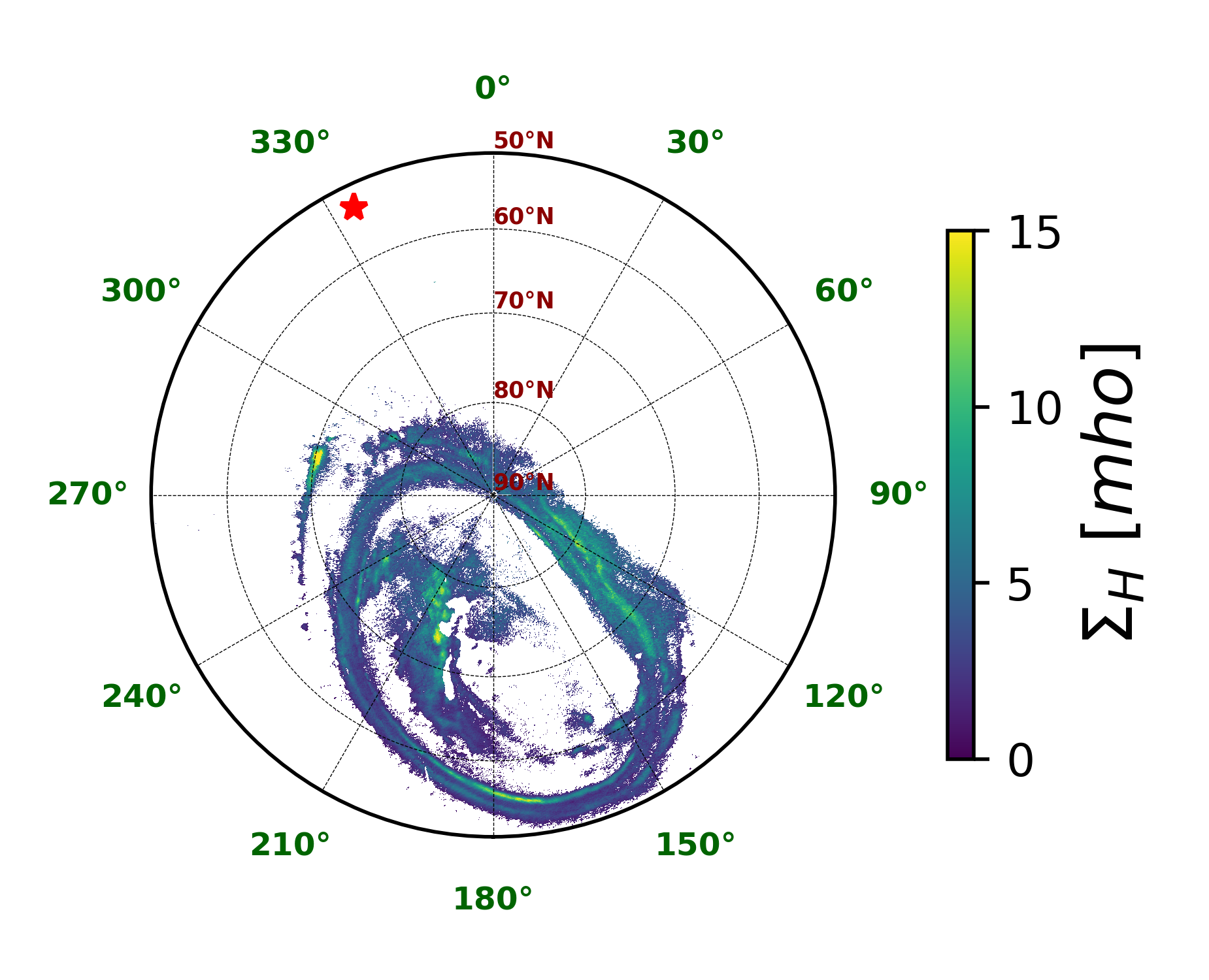}
        }
        
        \subfloat[ Ratio P - PJ22 north.]
        {\includegraphics[width=0.7\linewidth]{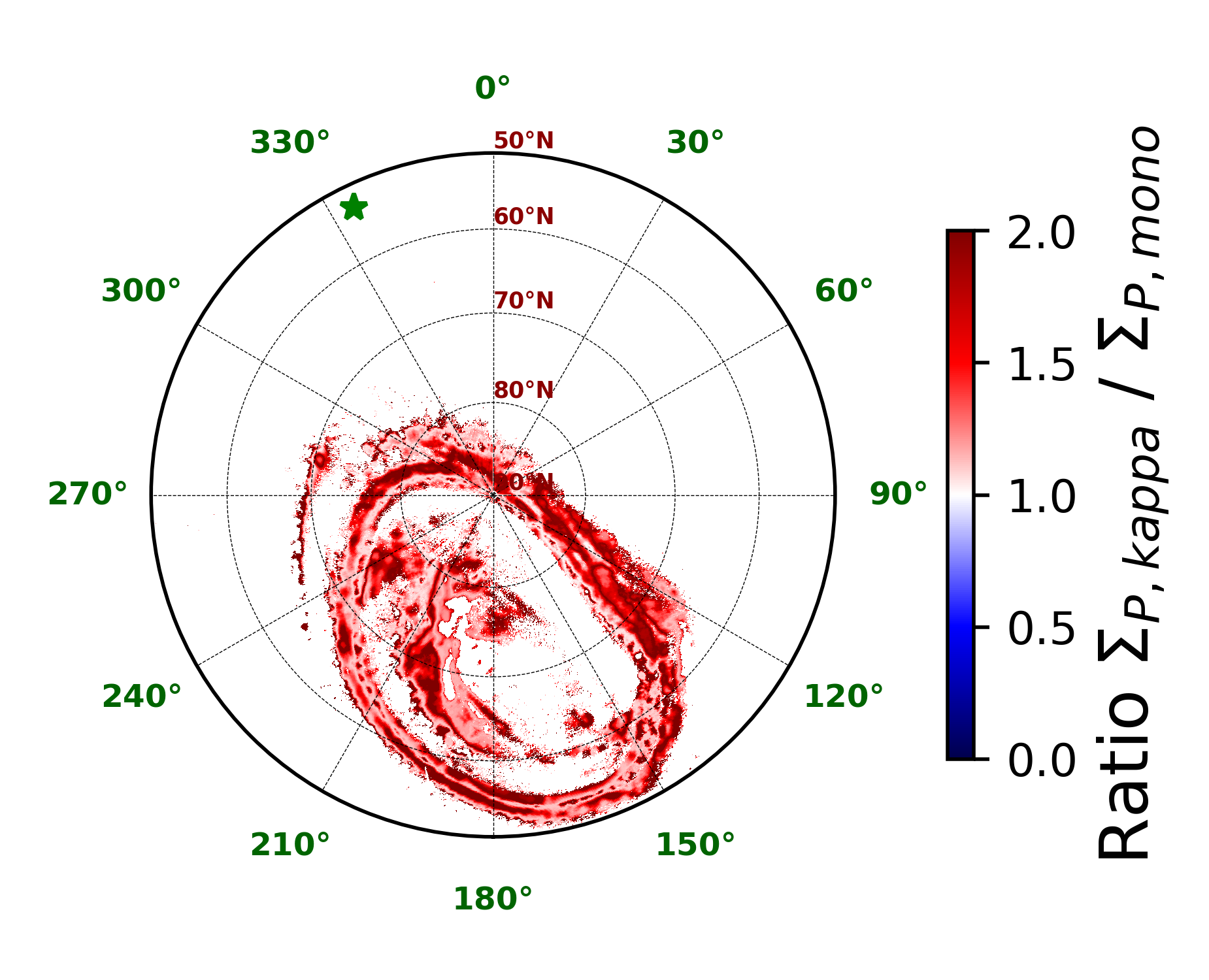}
        }
        
        \subfloat[ Ratio H - PJ22 north.]
        {\includegraphics[width=0.7\linewidth]{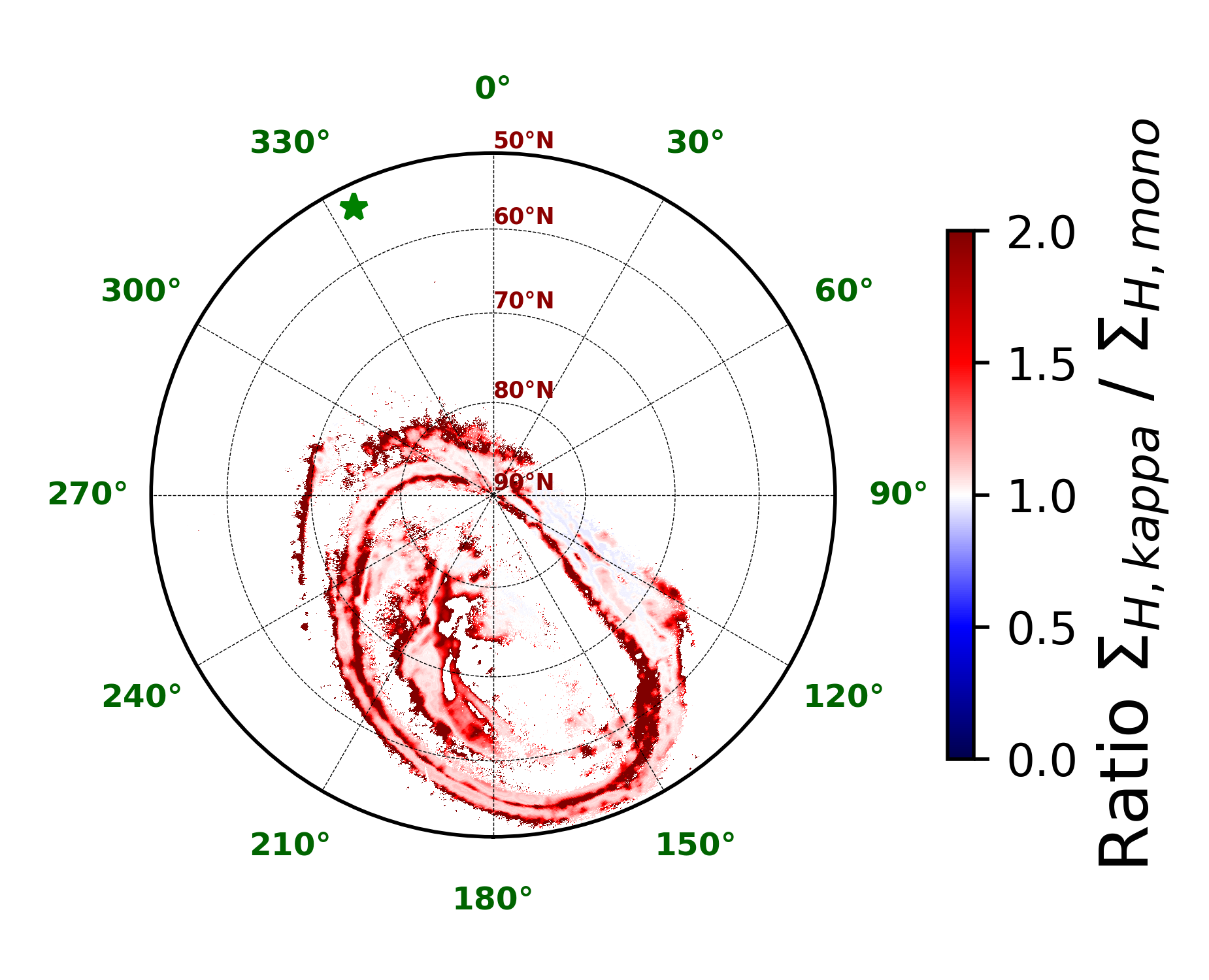}
        }
    \caption{
    Same description as Fig. \ref{fig:appendix_maps_conductnorth_pj1} for PJ22 north.
    }
    \label{fig:appendix_maps_conductnorth_pj22}
    \end{figure}

    \begin{figure} % One-column figure
    \centering
    \captionsetup[subfigure]{width=1\textwidth}
        \subfloat[ $\Sigma_P$ - PJ15 south.]
        {\includegraphics[width=0.7\linewidth]{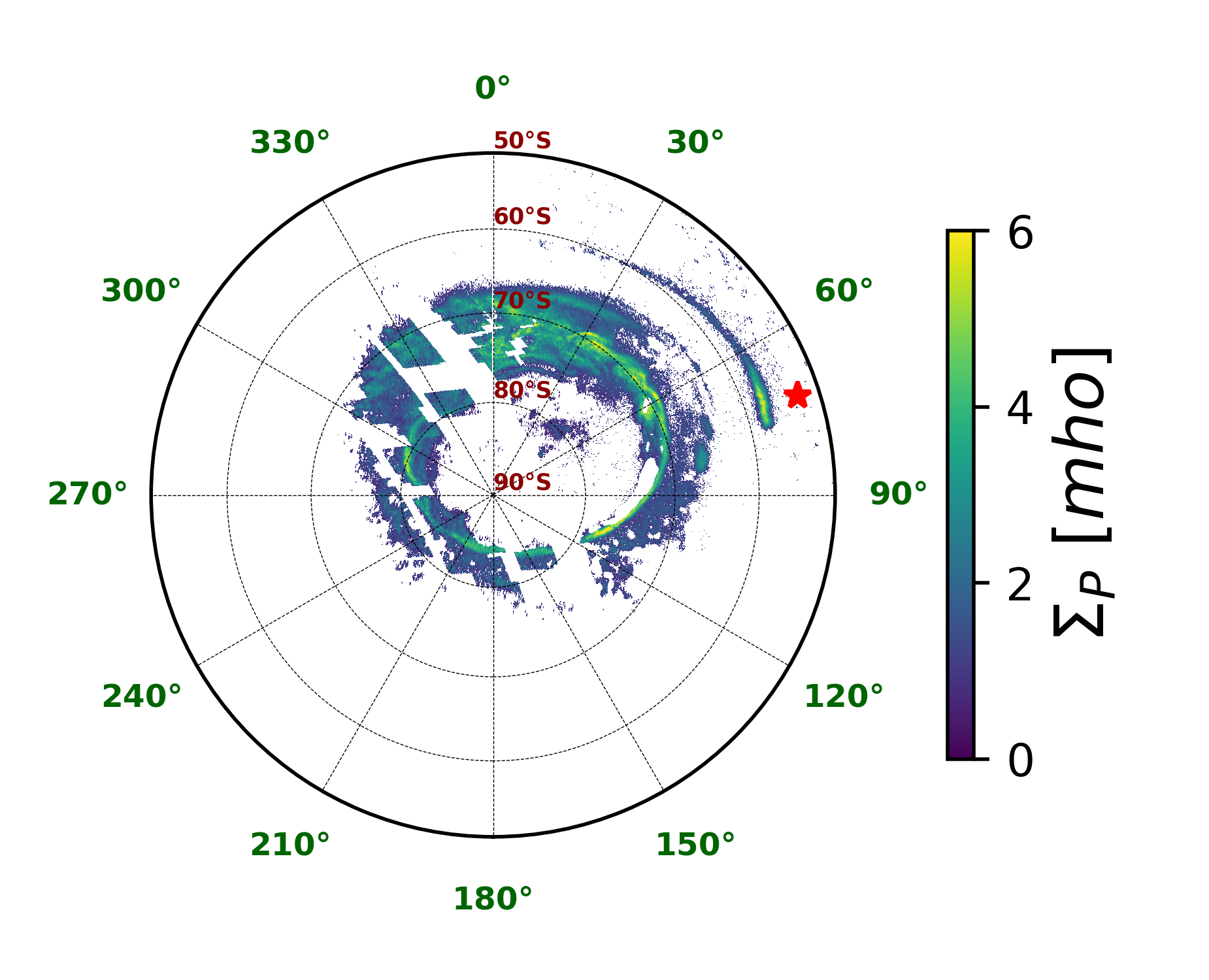}
        }
        
        \subfloat[ $\Sigma_H$ - PJ15 south.]
        {\includegraphics[width=0.7\linewidth]{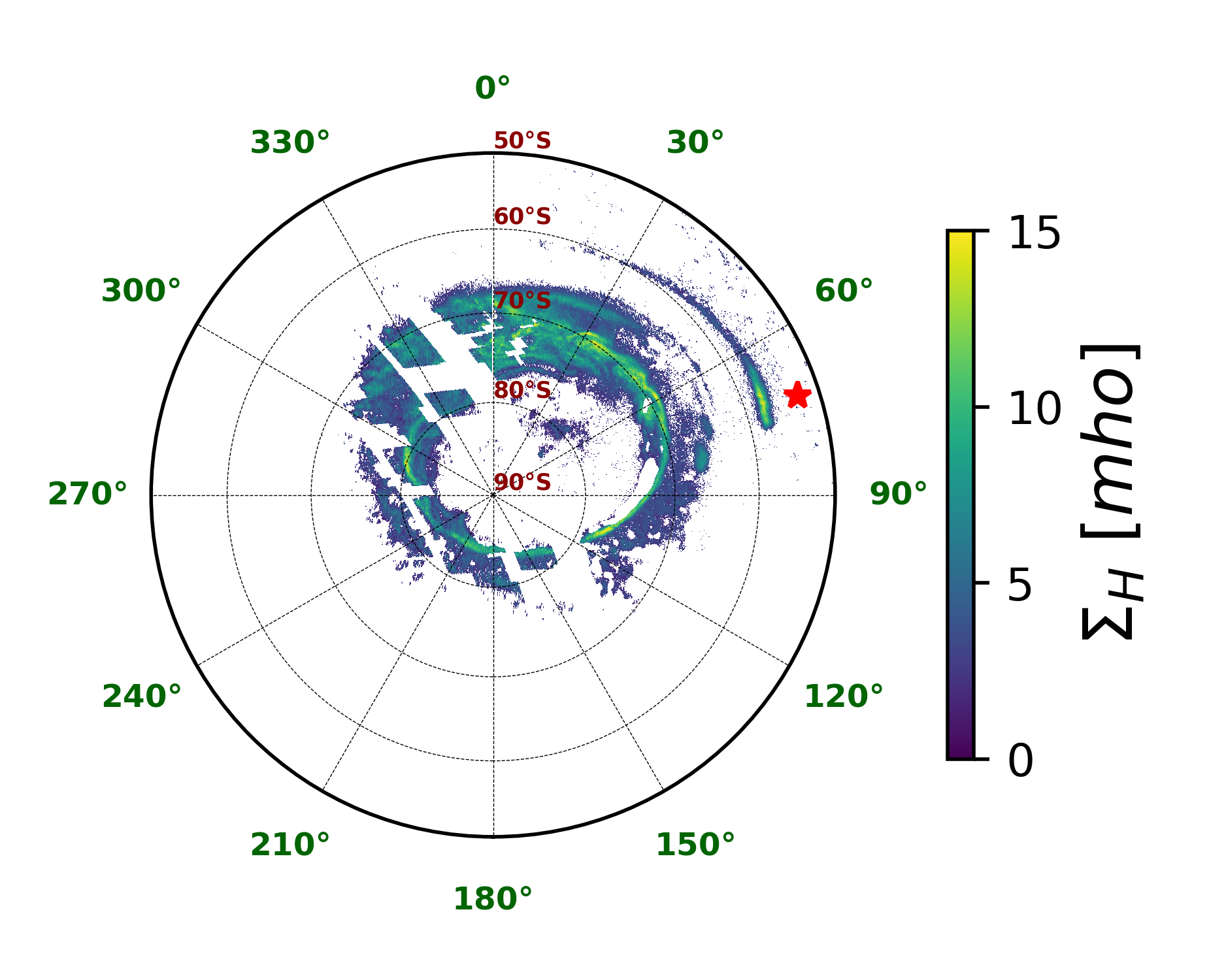}
        }
        
        \subfloat[ Ratio P - PJ15 south.]
        {\includegraphics[width=0.7\linewidth]{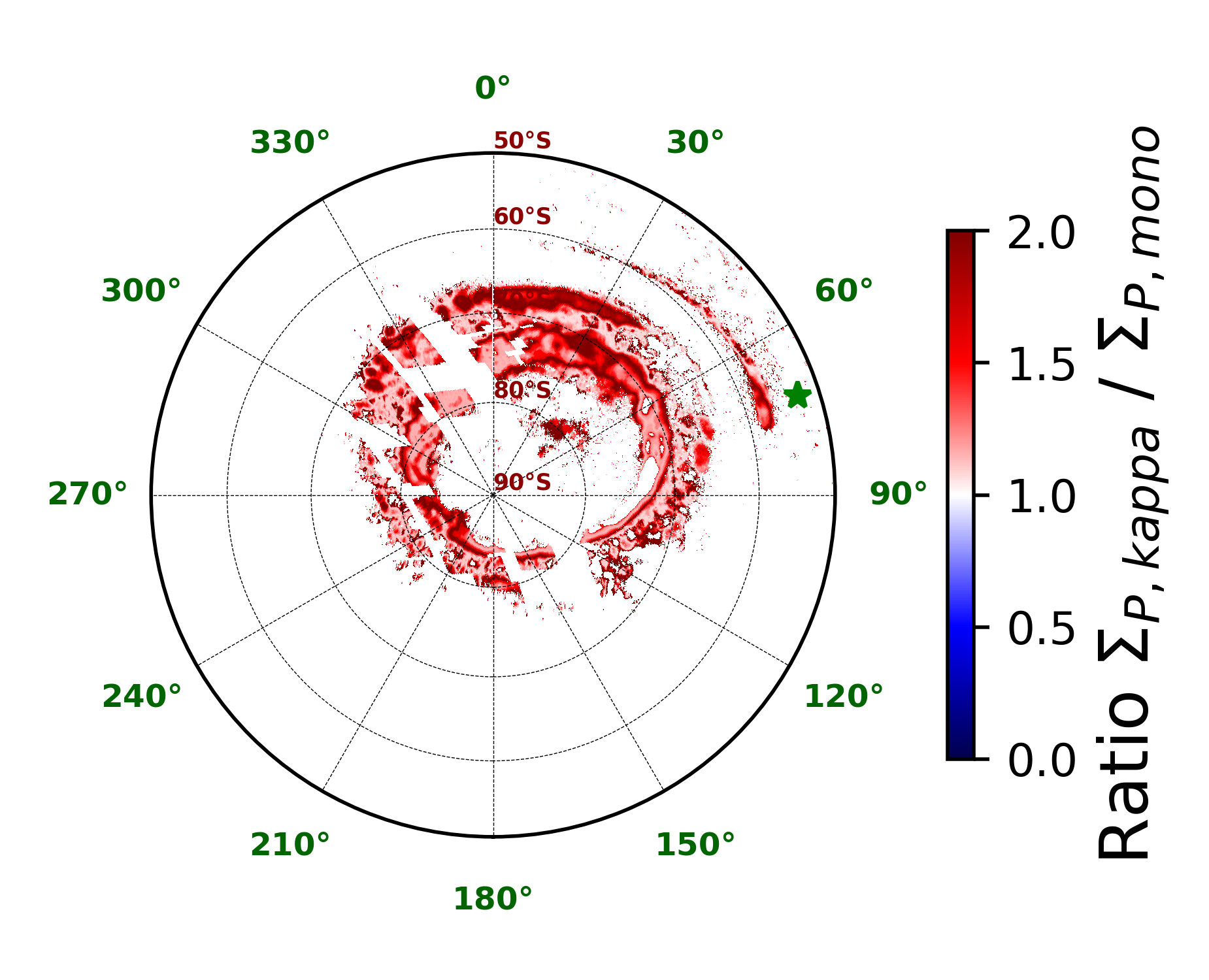}
        }
        
        \subfloat[ Ratio H - PJ15 south.]
        {\includegraphics[width=0.7\linewidth]{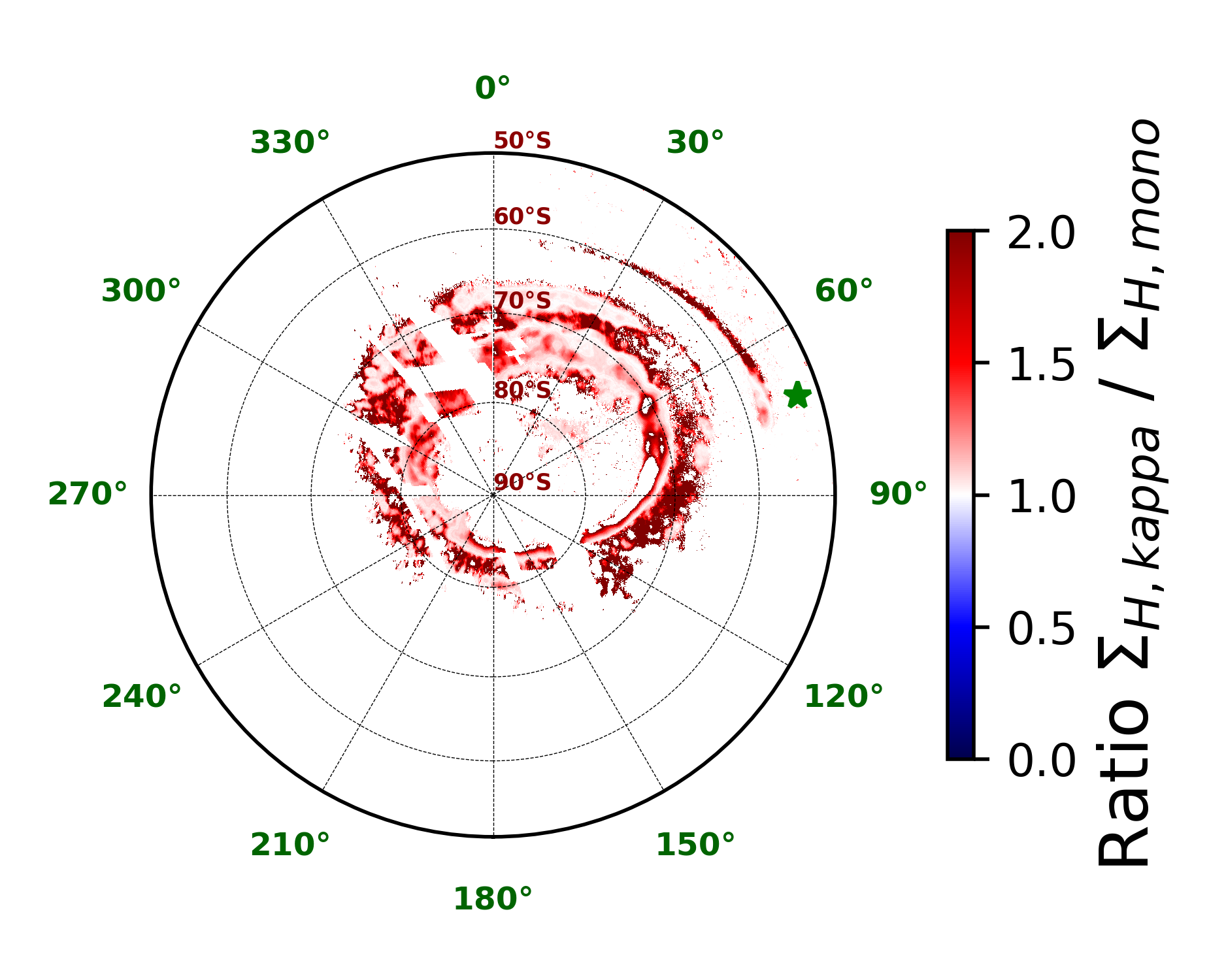}
        }
    \caption{
    Same description as Fig. \ref{fig:appendix_maps_conductnorth_pj1} for PJ15 south.
    }
    \label{fig:appendix_maps_conductsouth_pj15}
    \end{figure}

    \begin{figure} % One-column figure
    \centering
    \captionsetup[subfigure]{width=1\textwidth}
        \subfloat[ $\Sigma_P$ - PJ27 south.]
        {\includegraphics[width=0.7\linewidth]{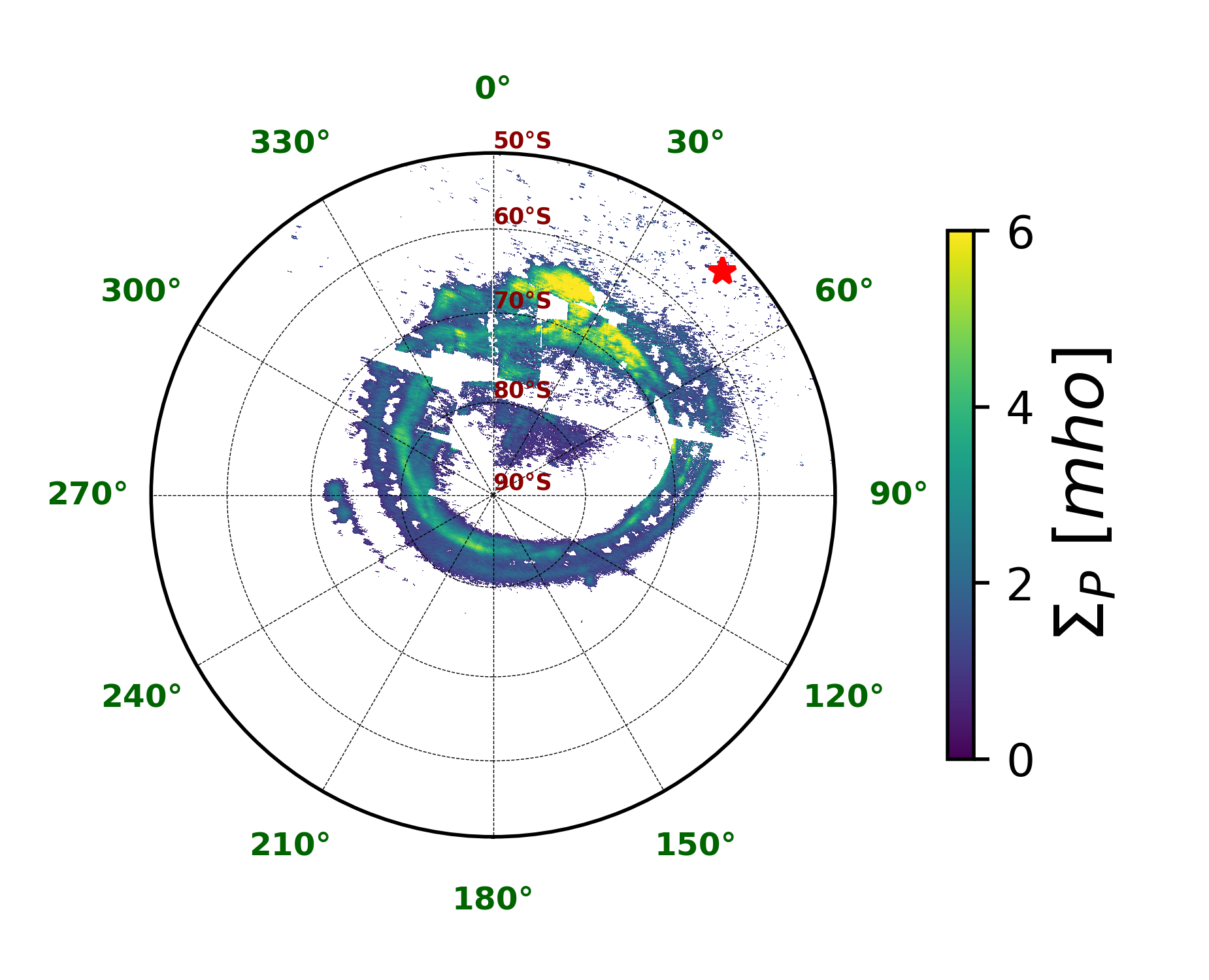}
        }
        
        \subfloat[ $\Sigma_H$ - PJ27 south.]
        {\includegraphics[width=0.7\linewidth]{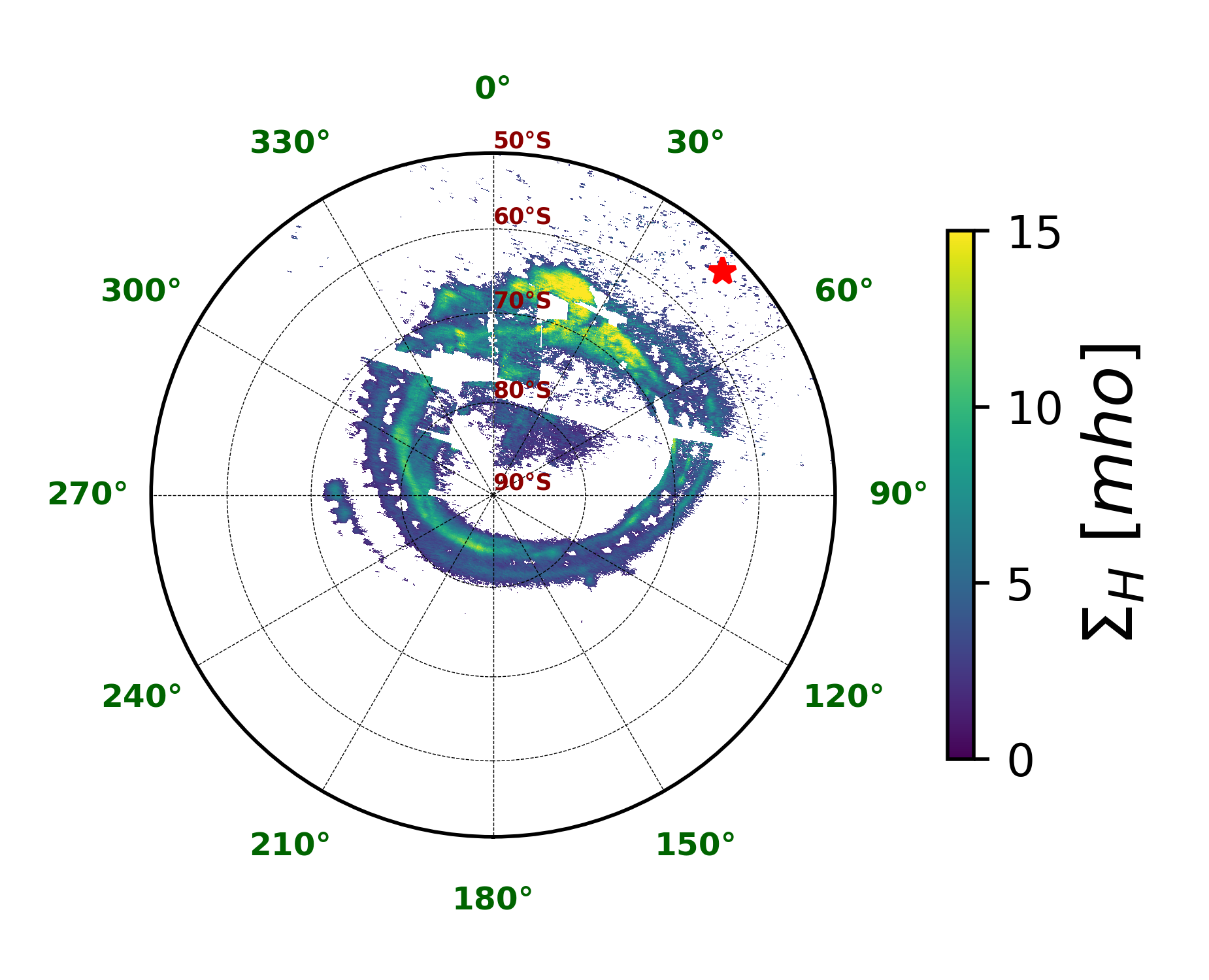}
        }
        
        \subfloat[ Ratio P - PJ27 south.]
        {\includegraphics[width=0.7\linewidth]{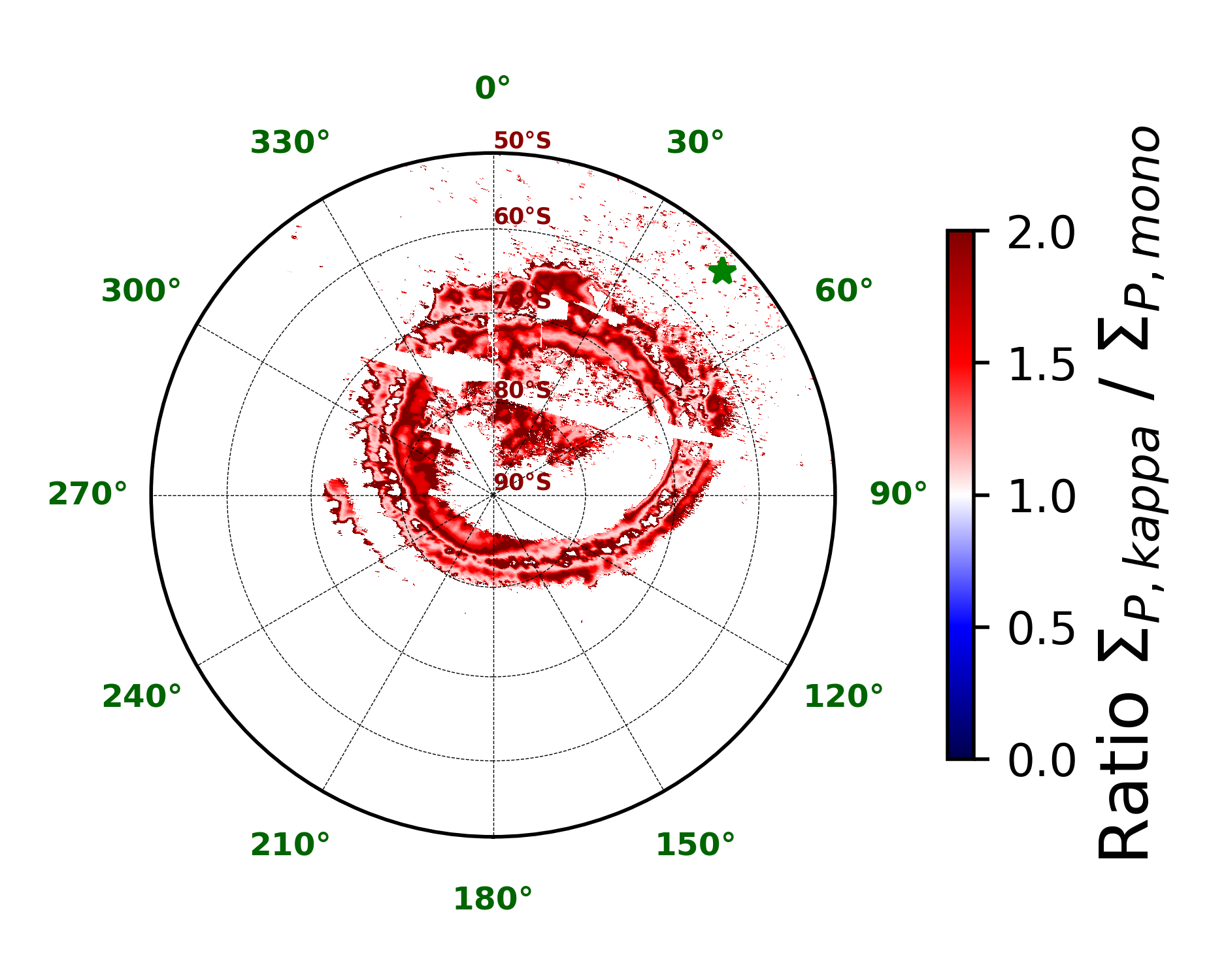}
        }
        
        \subfloat[ Ratio H - PJ27 south.]
        {\includegraphics[width=0.7\linewidth]{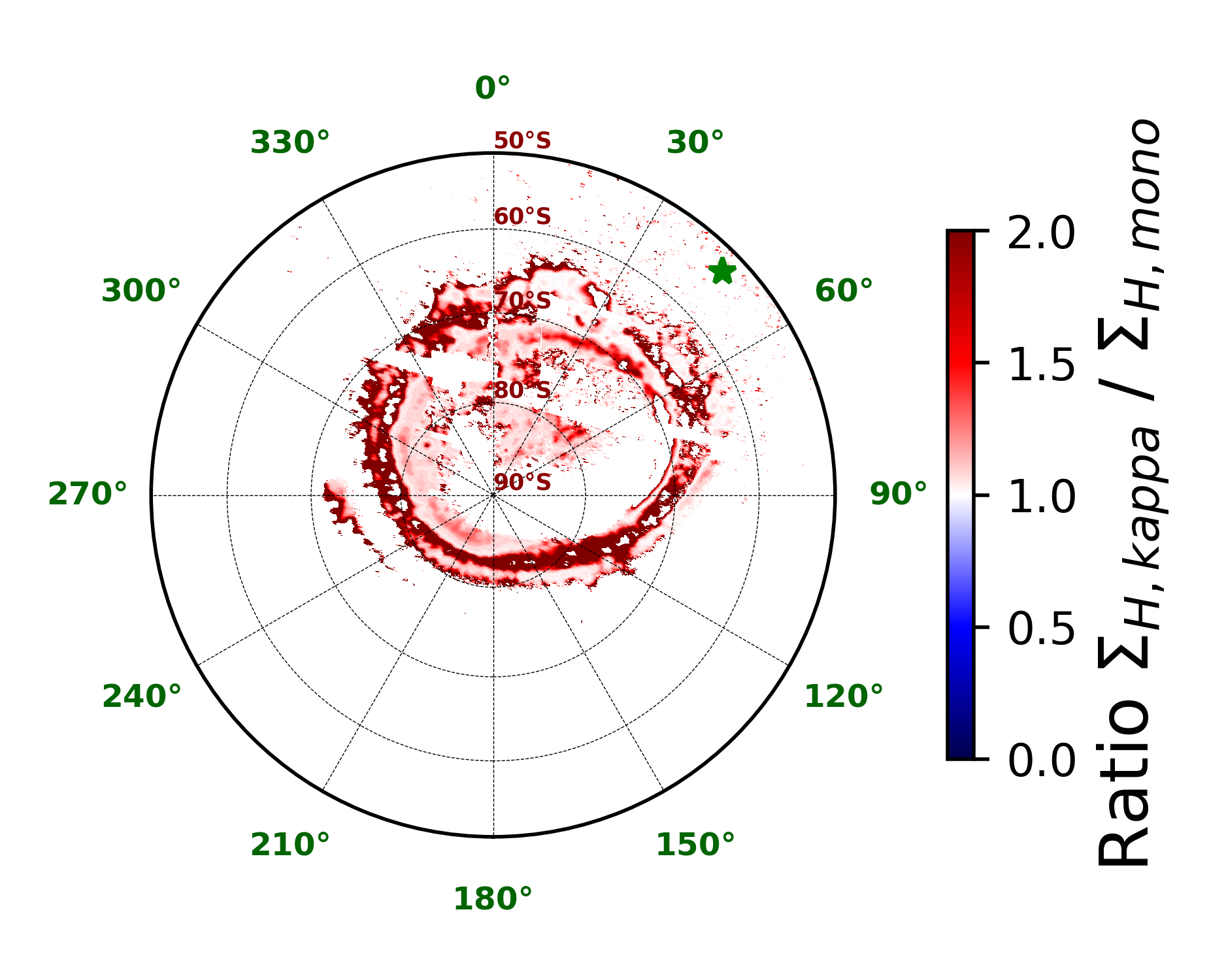}
        }
    \caption{
    Same description as Fig. \ref{fig:appendix_maps_conductnorth_pj1} for PJ27 south.
    }
    \label{fig:appendix_maps_conductsouth_pj27}
    \end{figure}

    \begin{figure} % One-column figure
    \centering
    \captionsetup[subfigure]{width=1\textwidth}
        \subfloat[ $\Sigma_P$ - PJ34 south.]
        {\includegraphics[width=0.7\linewidth]{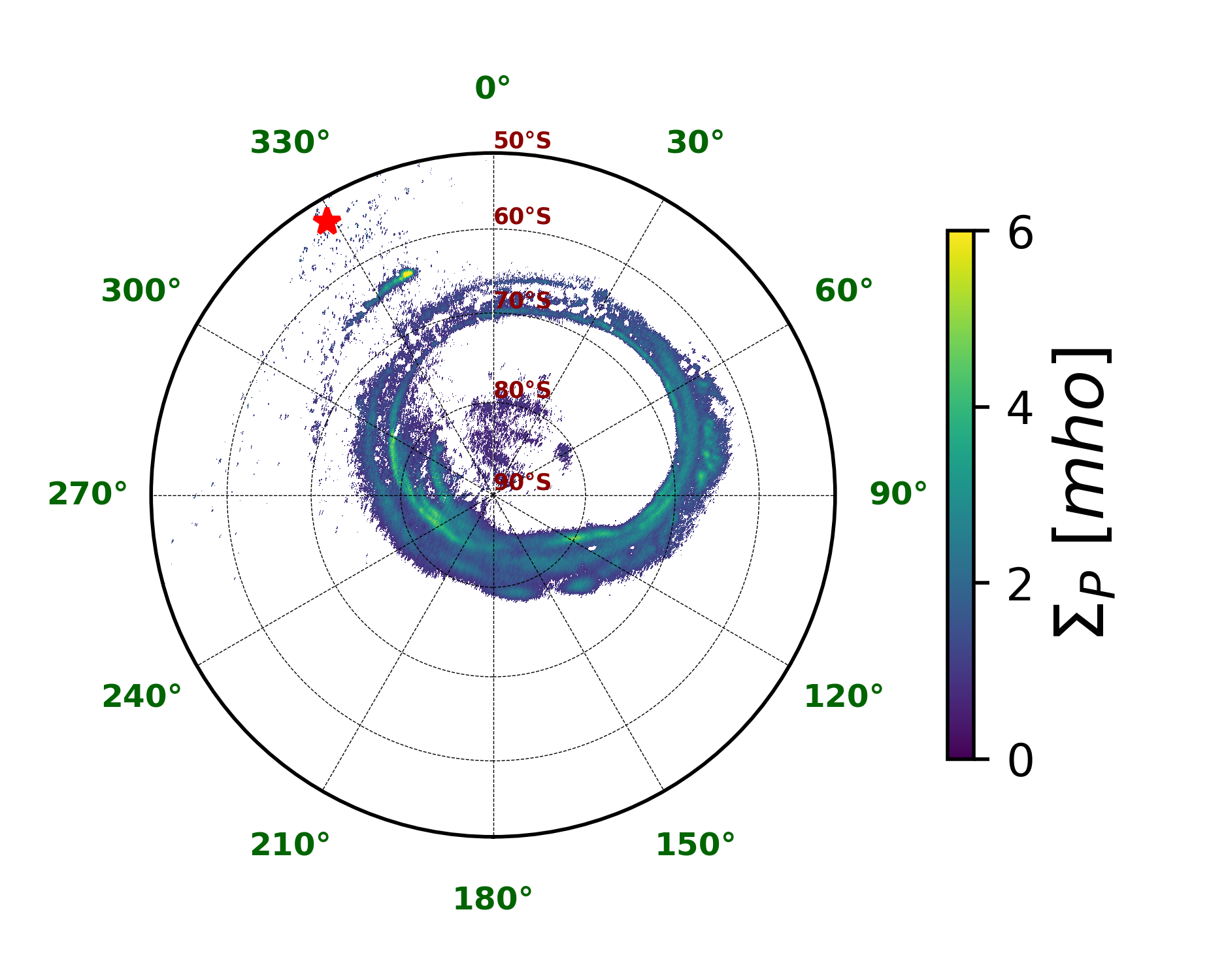}
        }
        
        \subfloat[ $\Sigma_H$ - PJ34 south.]
        {\includegraphics[width=0.7\linewidth]{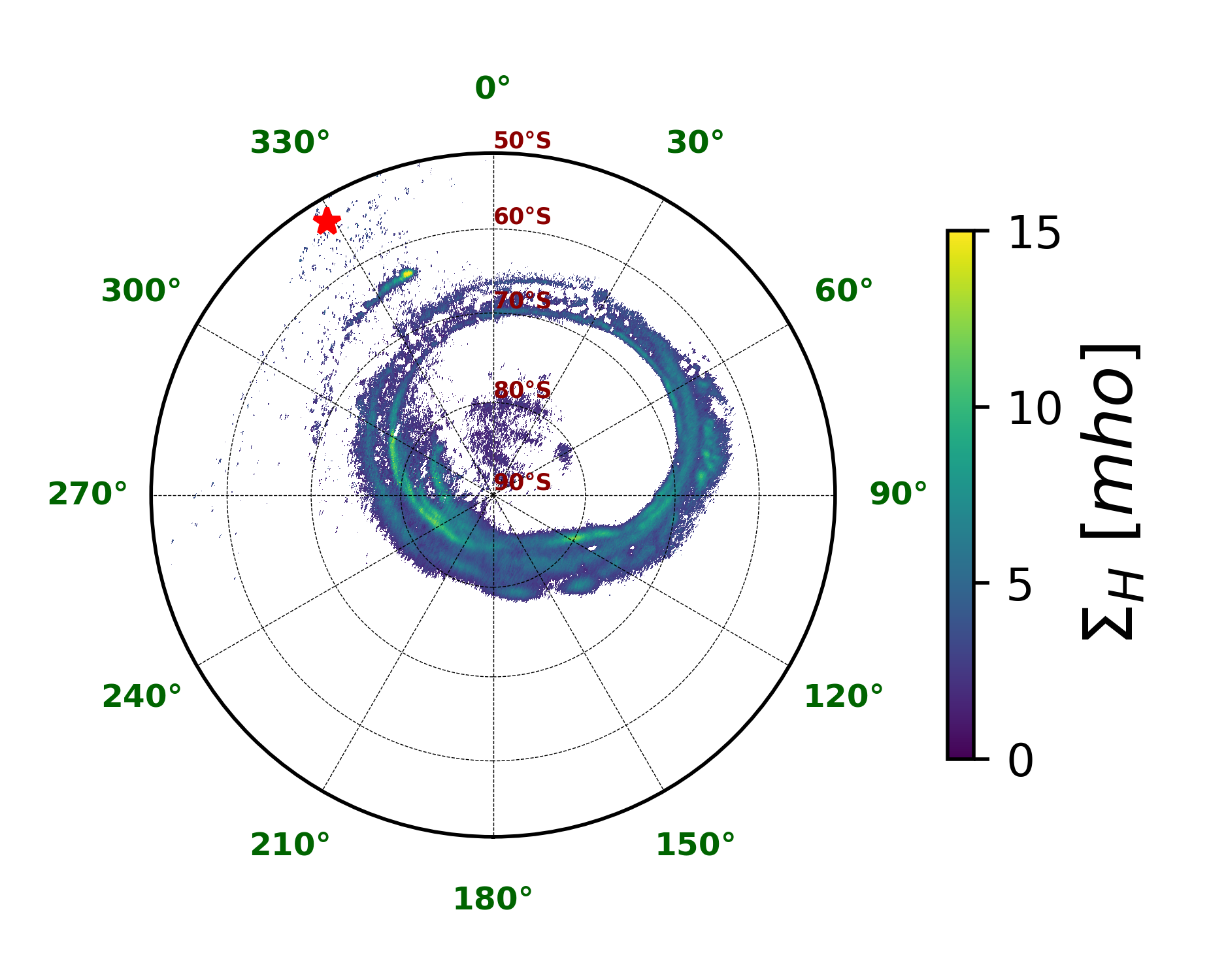}
        }
        
        \subfloat[ Ratio P - PJ34 south.]
        {\includegraphics[width=0.7\linewidth]{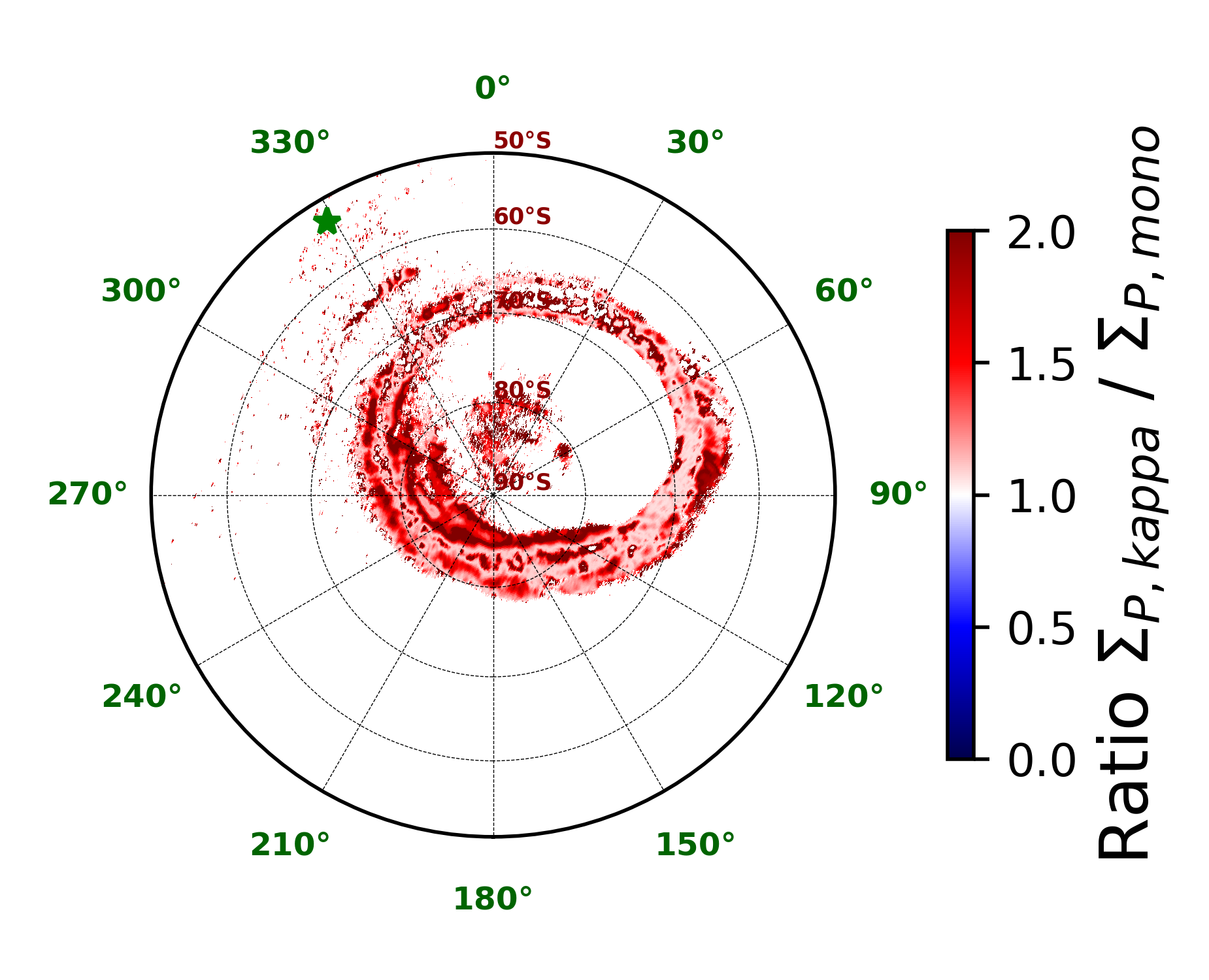}
        }
        
        \subfloat[ Ratio H - PJ34 south.]
        {\includegraphics[width=0.7\linewidth]{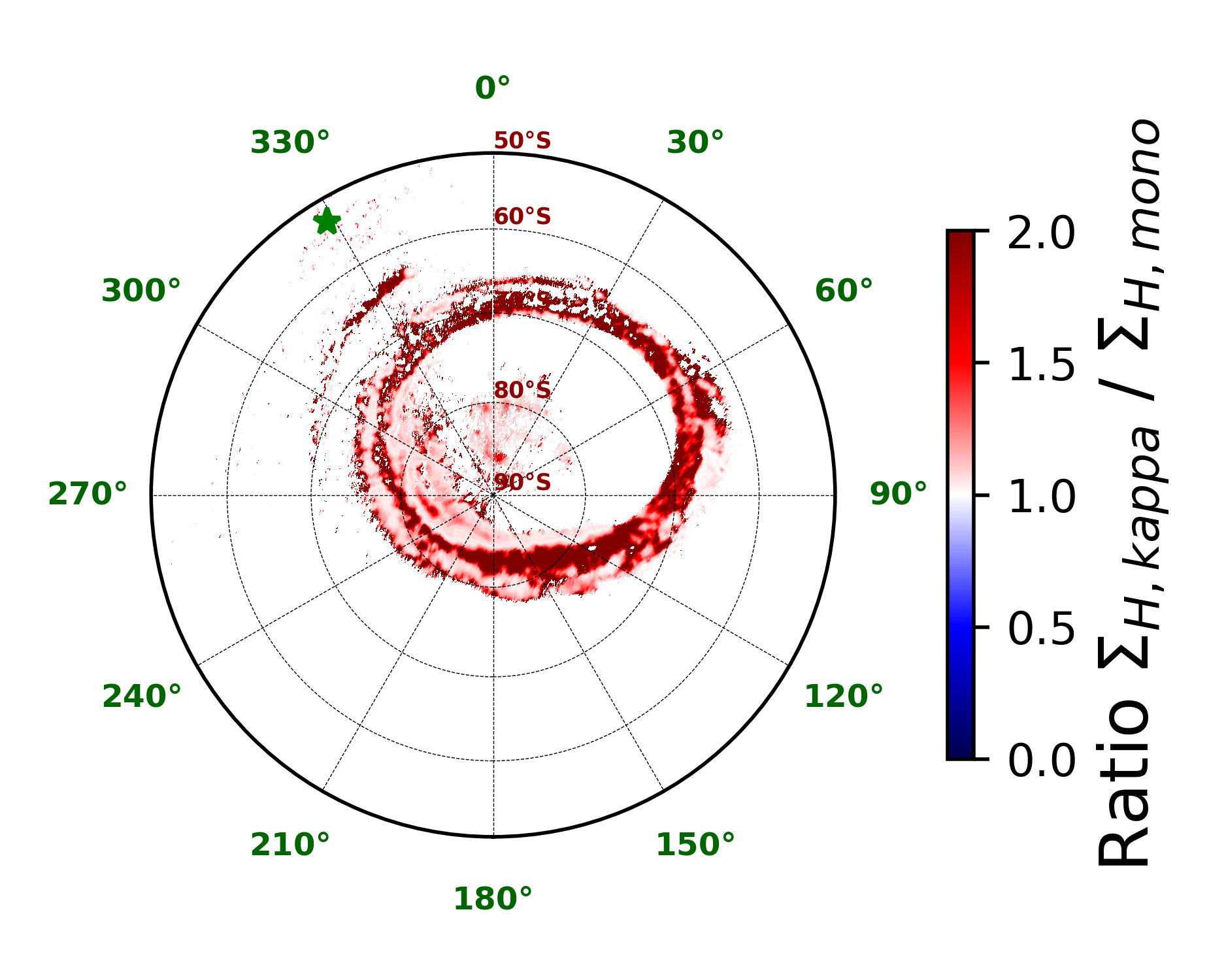}
        }
    \caption{
    Same description as Fig. \ref{fig:appendix_maps_conductnorth_pj1} for PJ34 south.
    }
    \label{fig:appendix_maps_conductsouth_pj34}
    \end{figure}
    
    \begin{figure} % One-column figure
    \centering
    \captionsetup[subfigure]{width=1\textwidth}
        \subfloat[ $\Sigma_P$ - PJ35 south.]
        {\includegraphics[width=0.7\linewidth]{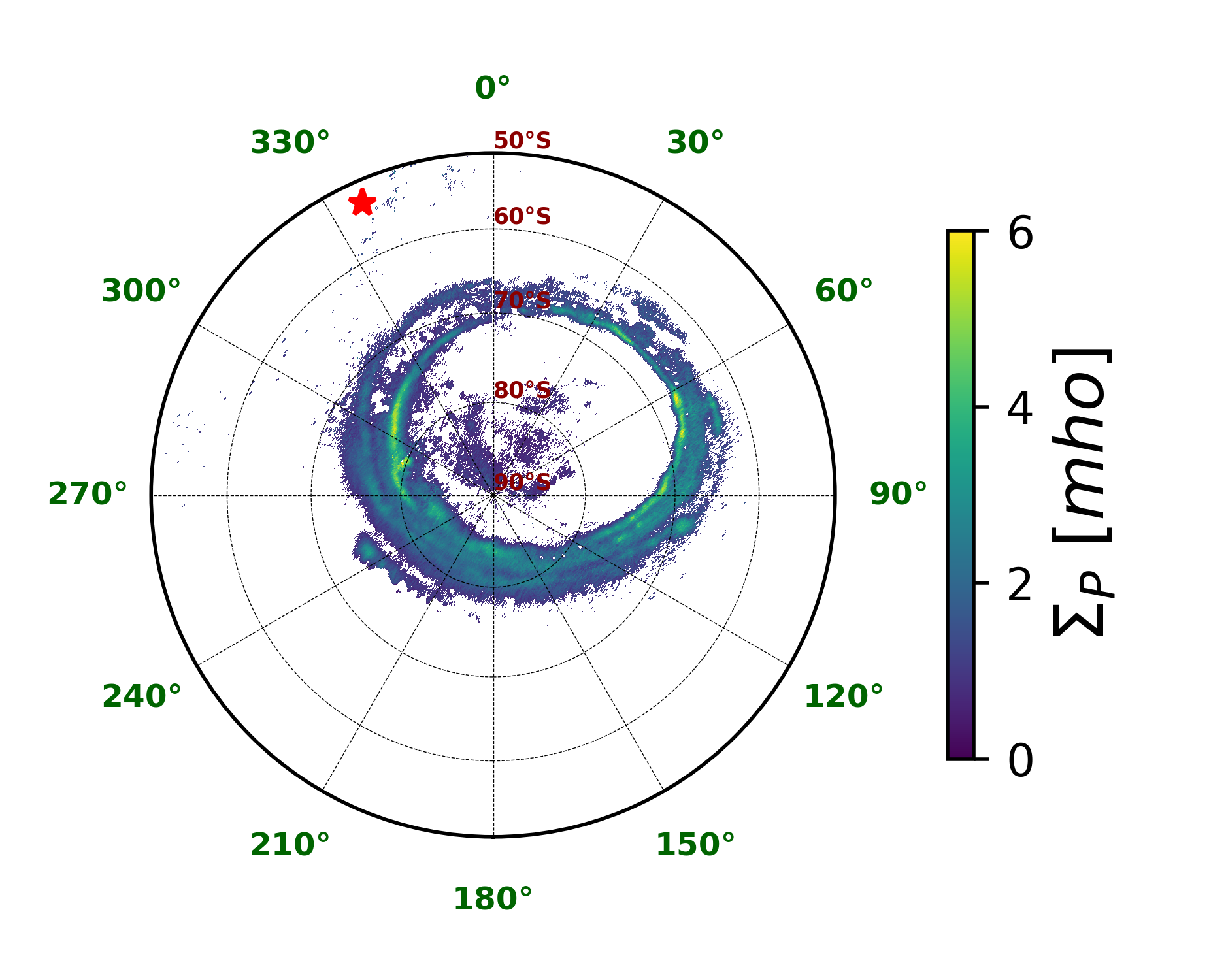}
        }
        
        \subfloat[ $\Sigma_H$ - PJ35 south.]
        {\includegraphics[width=0.7\linewidth]{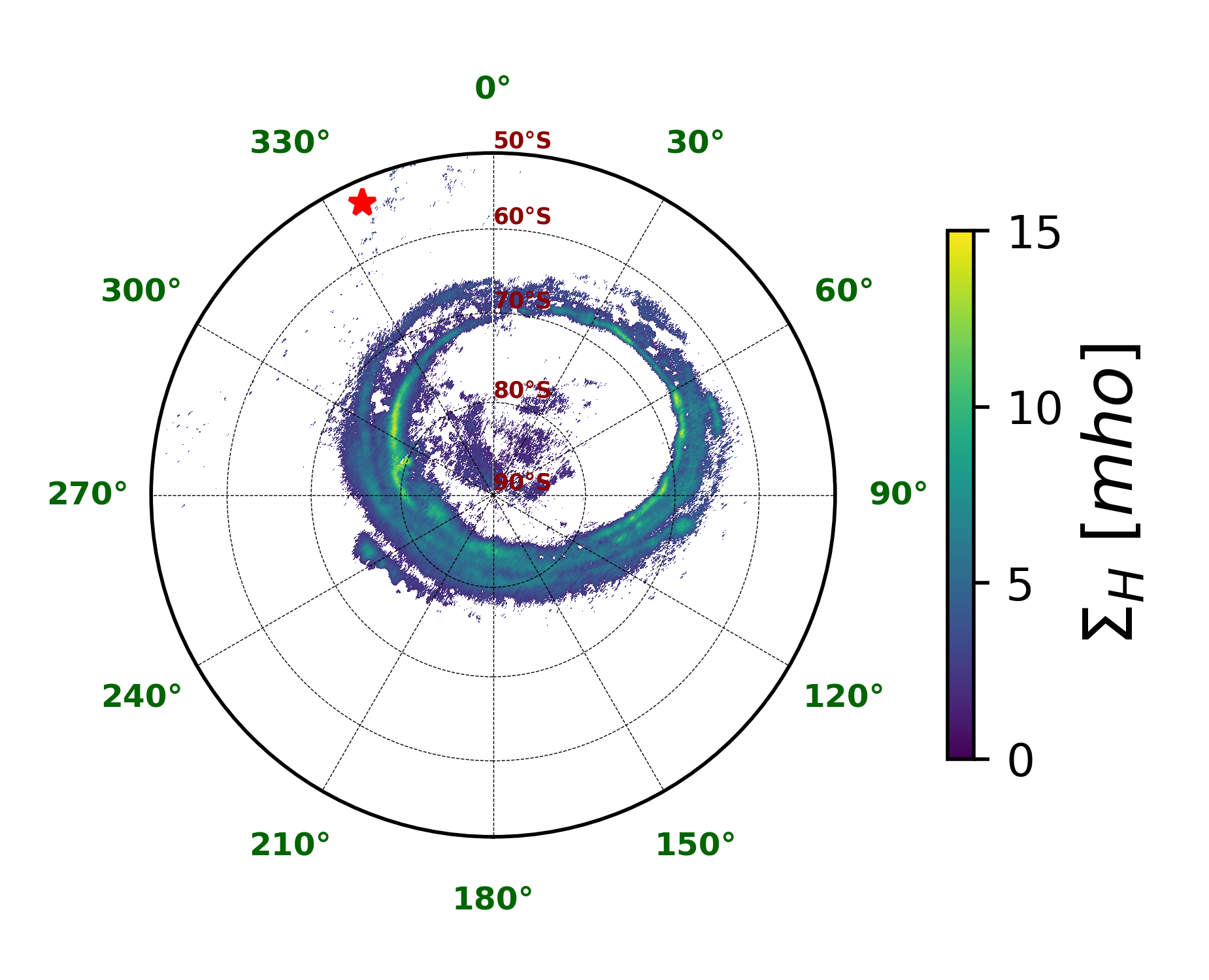}
        }
        
        \subfloat[ Ratio P - PJ35 south.]
        {\includegraphics[width=0.7\linewidth]{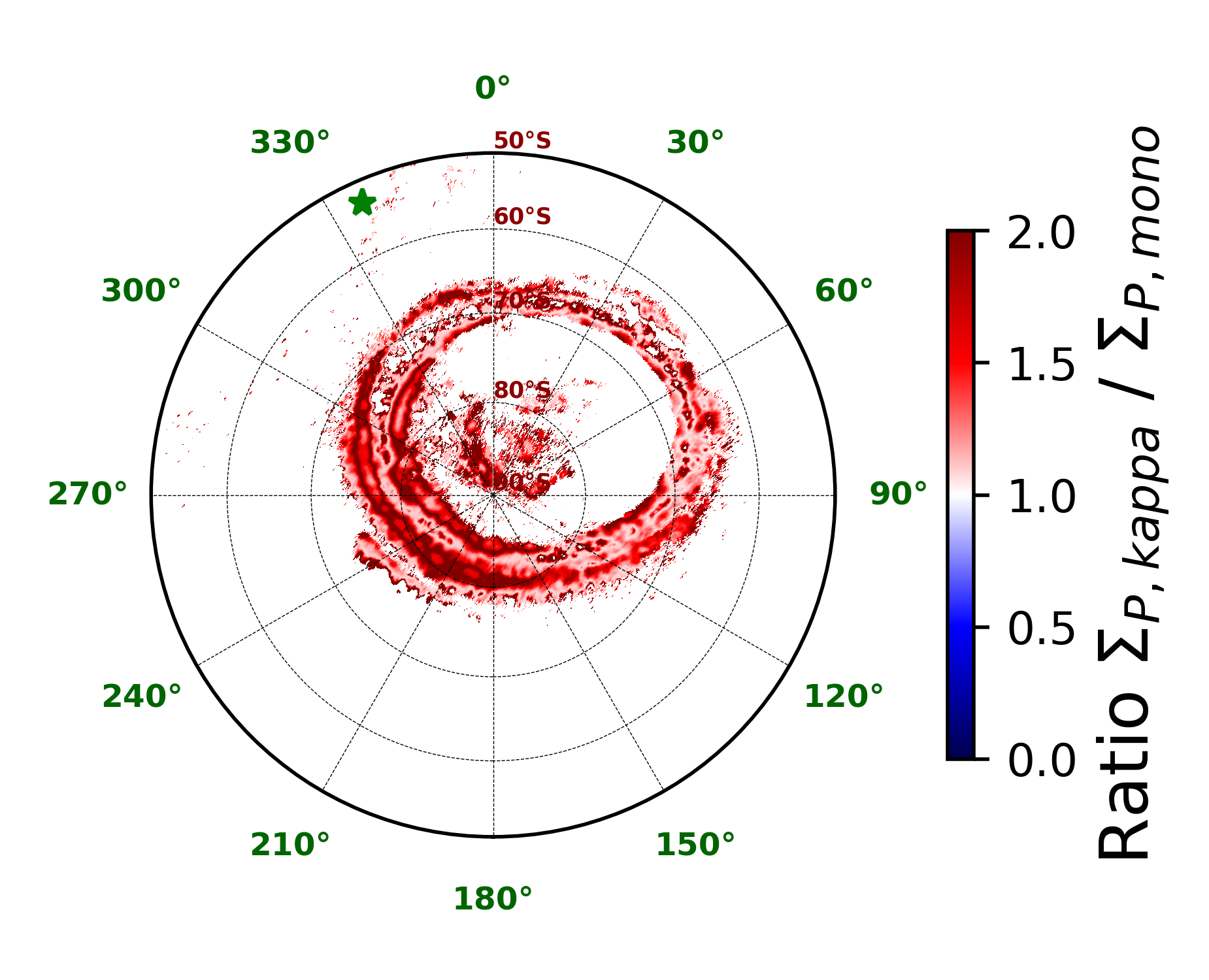}
        }
        
        \subfloat[ Ratio H - PJ35 south.]
        {\includegraphics[width=0.7\linewidth]{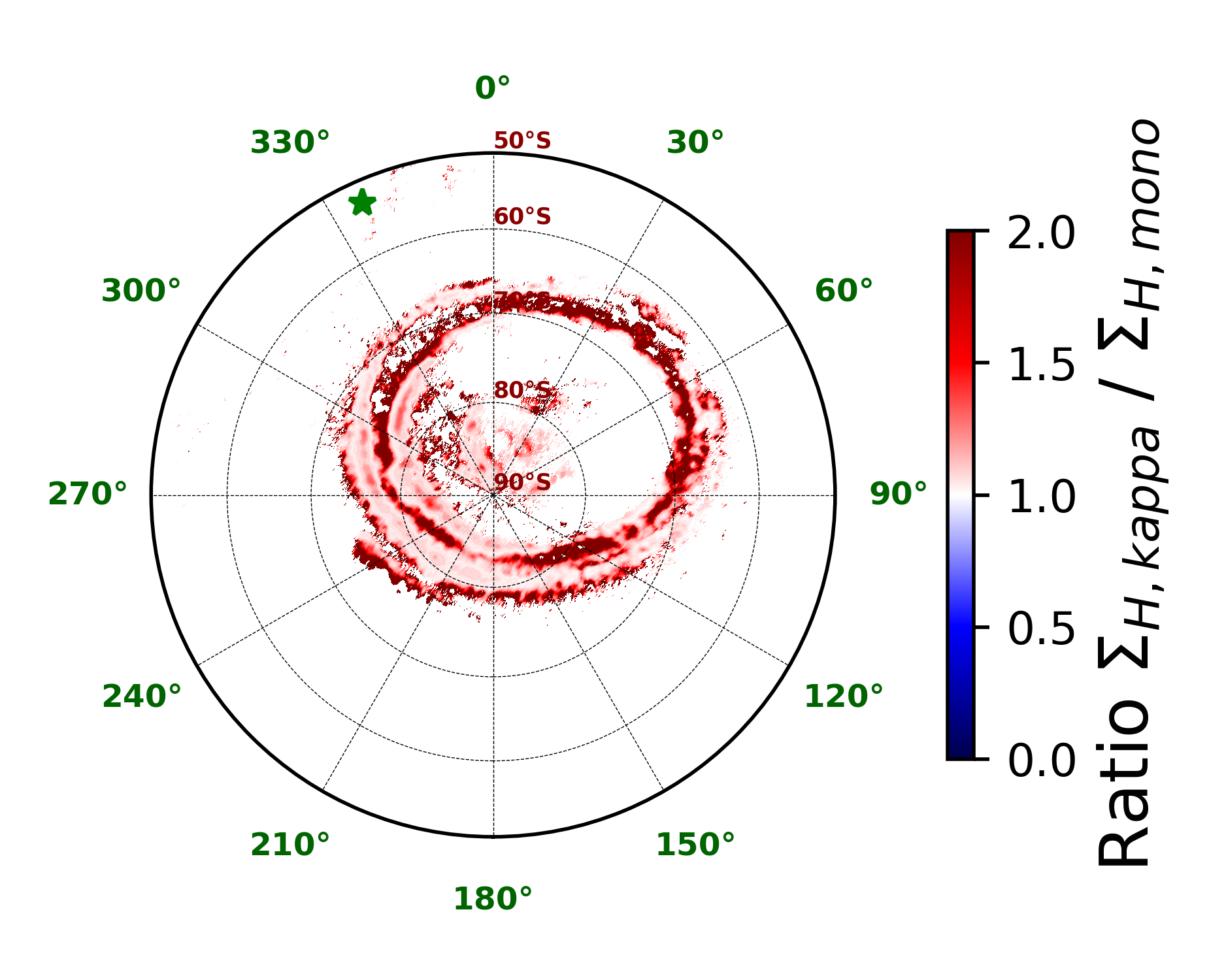}
        }
    \caption{
    Same description as Fig. \ref{fig:appendix_maps_conductnorth_pj1} for PJ35 south.
    }
    \label{fig:appendix_maps_conductsouth_pj35}
    \end{figure}

\end{appendix}

\end{document}